\documentclass[prb,aps,superscriptaddress,floatfix,tightenlines,showpacs,twocolumn]{revtex4-1}
\usepackage{graphicx}
\usepackage{dcolumn}   % needed for some tables
\usepackage{bm}        % for math
\usepackage{amssymb}   % for math
\usepackage{amsmath}
\usepackage{url}
\usepackage{layout}
\usepackage{epsfig}
\usepackage{graphicx}
\usepackage{booktabs}
\usepackage{float}
\usepackage[hyperindex,breaklinks]{hyperref}
\begin{document}

\title{Fate of a multiple-band Fermi liquid that is coupled with critical $\phi^4$ bosons}

\author{Zhiming Pan}
\affiliation{International Center for Quantum Materials, School of Physics, Peking University, Beijing 100871, China}
\affiliation{Collaborative Innovation Center of Quantum Matter, Beijing 100871, China}%
\author{Ryuichi Shindou}%
\email{rshindou@pku.edu.cn}
\affiliation{International Center for Quantum Materials, School of Physics, Peking University, Beijing 100871, China}%
\affiliation{Collaborative Innovation Center of Quantum Matter, Beijing 100871, China}%

\date{\today}

\begin{abstract}
Multiple-band nature of electronic energy bands leads to novel physical effects in solids. In this 
paper, we clarify physical properties of a Fermi system with a pair of electron and hole Fermi surfaces (FSs), 
whose coupling is mediated by a critical U(1) boson field. By using a one-loop renormalization group  
analysis, we show that when the boson field undergoes a quantum phase transition with broken 
U(1) symmetry, the multiple-band Fermi system shows a non-Fermi liquid (non-FL) behaviour in its 
thermodynamic and magnetic properties. At a quantum critical point (QCP), Fermi velocities 
of the two FSs are renormalized into a same critical velocity as a boson velocity, and the 
fermion's density of states (DOS) shows a pseudo-gap behaviour with a logarithmic energy dependence 
at the QCP. 
\end{abstract}

%\pacs{4444}

\maketitle
\section{introduction}
Experimental discoveries of graphene~\cite{novoselov05,zhang05} and 
topological insulators~\cite{koenig07,hsieh08} promote studies on 
unique physical effects that come from a multiple-band nature of electronic 
energy bands in solids~\cite{xiao10,qi11,hasan10}. The key 
ingredient in the recent studies is an interplay effect between the 
multiple-band nature and the other factors 
in solids, such as electron correlation, interaction with 
environments, disorder and so 
on~\cite{neto09,gross74,herbut06,assaad13,chubukov08,
fernandes14,sun09,moon13,savary14,dzero10,wolgast13,neupane13,
jiang13,li09,groth09,jiang09,meier18,stutzer18,raghu08,rachel10,haldane04,
shindou06,shindou08,wang10,wang12,chen17}. During the last 
decade, theoretical works revealed novel interplay effects in Dirac 
and Weyl semimetal~\cite{goswami11,isobe12,fradkin86,syzranov18}, together 
with experimental developement of relevant materials~\cite{armitage18}. 
Nonetheless, these fermions have zero (or tiny) density of states around 
the Fermi levels, where bulk electric transports are quantitatively tiny. 

\begin{figure}[t]
	\centering
	\includegraphics[width=0.9\linewidth]{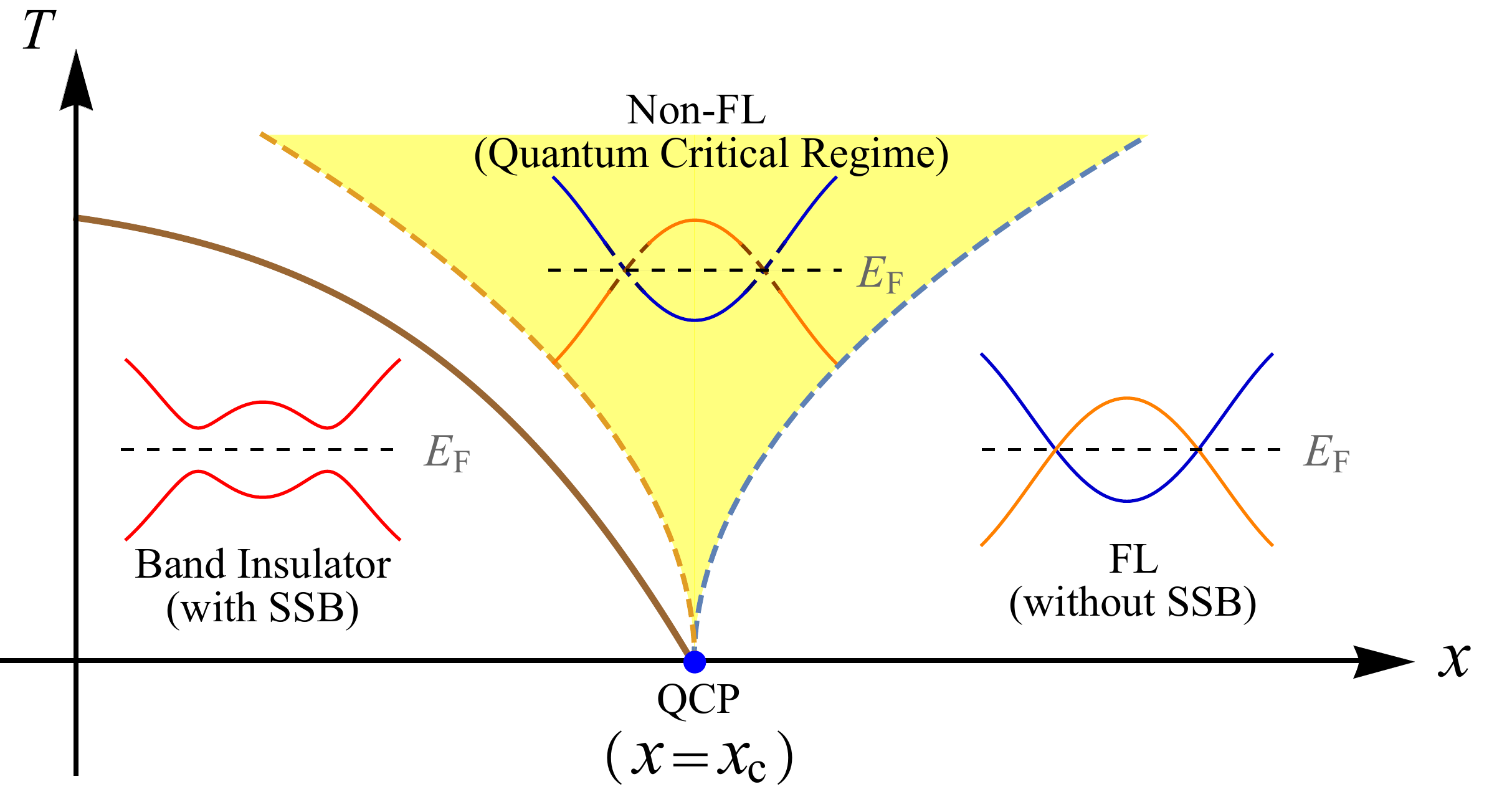}
	\caption{A phase diagram of the two-band fermion system coupled with U(1) symmetric $\phi^4$ boson: 
Eqs.~(\ref{boson-Eq1},\ref{fermion-Eq1a},\ref{f-b-Eq1a}). 
The vertical axis is temperature $T$, and horizontal axis $(x \sim m^2_0)$ is a 
parameter that induces a quantum phase 
transition with the SSB of the U(1) symmetry. In the SSB phase, the Fermi system acquires a finite band 
gap, while in the U(1) symmetric phase, the Fermi system belongs to a multiple-band Fermi liquid (m-FL) phase. At 
the QCP and its high-$T$ side (QCR), the Fermi system shows a non-FL behaviour (See Table.~\ref{table:1}).}
	\label{fig:1}
\end{figure}

In this paper, we study a multiple-band Fermi system with a finite volume of FS, 
that is coupled with a U(1) boson field in a $\phi^4$ type action in (3+1)-dimension. 
We uncover that an interplay between the multiple-band nature and the many-body 
effect leads to a non-Fermi liquid (non-FL) quantum critical 
regime~\cite{varma89,yakovenko93,lee89,altshuler94,kim94,metzner03,
rech06,fradkin10,lee09,abanov03,metlitski10a,metlitski10b,fitzpatrick13,fitzpatrick15b,mahajan13,fitzpatrick14,dalidovich13,ipsita15,ipsita16,pimenov18} 
in its electronic phase diagram (Fig.~\ref{fig:1}). The boson system is described by 
the (3+1)-dimensional U(1) symmetric $\phi^4$ action with the 
dynamical exponent $z=1$, 
\begin{align}
%\left\{\begin{array}{l}
S_{\phi} = \int d^{4}x  \!\ \big[m^2_{0} |\phi|^2 + |\partial_{\tau}\phi|^2 
+ c^2_0 |{\nabla_{\bm x}}\phi|^2 + \frac{\lambda_{0}}{4} (|\phi|^2)^2\big], \label{boson-Eq1} 
\end{align}
with $x \equiv (\tau,{\bm x})$, a bare boson mass $m^2_0$, boson velocity $c_0$, and 
bare $\phi^4$ coupling $\lambda_0$. The Fermi system has an electron-type 
fermion ($\psi_{+}$) with an isotropic FS and hole-type fermion $(\psi_{-})$ with the same 
size of isotropic FS;
\begin{table}[b]
  \centering
    \begin{tabular}{p{25mm} |c|c|c|c }
\hline 
%&$\rho(0)$ & $\sigma_z(0)$  & $\sigma_{\perp}(0)$   \\
& $C_{\rm el}/T $ & $\chi_{\rm mag}$ & $1/(T_1 T)$ & $\rho(\omega,T=0)$ 
 \\ \hline \hline 
non-Fermi liquid  & $\frac{1}{\sqrt{|\ln T|}}$ & $\frac{1}{\sqrt{|\ln T|}}$ 
& $\frac{1}{|\ln T|}$ & $\frac{1}{\sqrt{|\ln |\omega||}}$ \\  \hline 
 \end{tabular} 
 \caption{$T$-dependences of specific heat $C_{\rm el}$, 
and magnetic susceptibility $\chi_{\rm mag}$, and NMR relaxation time $T_1$ in the QCR 
and DOS near the Fermi level ($\omega=0$) at the QCP. 
For the magnetic response, we consider a trivial SU(2) generalization 
of Eqs.~(\ref{boson-Eq1},\ref{fermion-Eq1a},\ref{f-b-Eq1a}); 
$\psi_{\sigma} \rightarrow \psi_{\sigma,s}$, $\phi \psi^{\dagger}_{-}\psi_{+} 
\rightarrow \phi \psi^{\dagger}_{-,s}\psi_{+,s}$ with spin-1/2 $s=\uparrow,\downarrow$.}
\label{table:1}
\end{table}  
\begin{align}
S_{\psi} &= \sum_{\sigma=\pm }\int_{k} \psi^{\dagger}_{\sigma} (k)  
\big[i\omega - \sigma v_{\sigma0} l\big] \psi_{\sigma}(k), \label{fermion-Eq1a} \\
\int_k & \equiv k^{2}_{F} \int \frac{d^{2}\hat{\Omega}}{(2\pi)^{2}} 
\int^{\overline{\Lambda}_F}_{-\overline{\Lambda}_F} \frac{d\omega}{2\pi} 
\int^{\Lambda_F}_{-\Lambda_F} \frac{dl}{2\pi}. \label{fermion-Eq1b}
\end{align}
$k$ is fermion frequency and momentum; $k\equiv (\omega, {\bm k})$, ${\bm k} \equiv 
(k_F+l) \hat{\Omega}$. A unit vector $\hat{\Omega}$ defines a direction of ${\bm k}$ and a 
perpendicular component $l$ defines a vertical distance of ${\bm k}$ from the FS.
$v_{+0}$ and $v_{-0}$ are bare fermion velocities of electron-type and 
hole-type; $v_{+0}>0$, $v_{-0}>0$. The free fermion part is described by a 
spatially isotropic free theory with the dynamical exponent $z=1$. 
$\Lambda_F$ and $\overline{\Lambda}_F$ are ultraviolet (UV) cutoff for $l$, 
and $\omega$. The boson 
field $\phi$ mediates a coupling between the two FSs in a U(1) symmetric way, 
$\phi \rightarrow \phi e^{+2i\theta}, 
\psi^{\dagger}_{\pm} \rightarrow \psi^{\dagger}_{\pm} e^{\pm i\theta}$;    
\begin{align}
S_{\rm int} &=  g_0 \int_k \int_q \big[\phi^{\dagger}(q) \psi^{\dagger}_{+}(k) \psi_{-}(k+q) 
+ {\rm h.c.}\big], \label{f-b-Eq1a} \\
\int_q &\equiv \int_{|{\bm q}|<\Lambda_B}  \frac{d^{3}{\bm q}}{(2\pi)^{3}} 
\int^{+\overline{\Lambda}_B}_{-\overline{\Lambda}_B} \frac{d\varepsilon}{2\pi}. \label{f-b-Eq1b}
\end{align}
$g_0$ is nothing but a (bare) Yukawa coupling. In this paper, $k$ is always for 
the fermion ferquency and momentum and 
$q$ is boson frequency and momentum; $k \equiv (\omega,(k_F+l)\hat{\Omega})$, 
$q \equiv (\varepsilon, {\bm q})$. $\Lambda_B$ and $\overline{\Lambda}_B$ are 
UV cutoff for $q$ respectively.  

In the paper, we will clarify how the Fermi system acquires a band gap when the boson 
system undergoes a quantum phase transition with the spontaneous symmetry breaking 
(SSB) of the U(1) symmetry. We show that the fermion's DOS shows a pseudo-gap 
behaviour with a logarithmic energy dependence at the QCP and physical properties 
exhibit the logarithmic $T$-dependences in a high-temperature ($T$) side of the 
QCP: see Table.~\ref{table:1}.

\subsection{summary of this paper}
The structure of the paper is as follows. In Sec.~II, we introduce two concrete physical  
systems for which the effective action of Eqs.~(\ref{boson-Eq1},\ref{fermion-Eq1a},\ref{f-b-Eq1a}) 
is applicable. The first physical system is a 3-dimensional interacting Fermi 
system with a pair of electron and hole FSs with an excitonic instability. 
The intereacting Fermi system has a band inversion energy that measures an energy difference 
between the electron-band energy minimum and hole-band energy maximum. We discuss that when the 
band inversion energy is fine-tuned to a `frustrated' point where the excitonic instability is maximally supressed, 
a  quantum dielectric transition associated with the excitonic instability is characterized by the effective 
action given by Eqs.~(\ref{boson-Eq1},\ref{fermion-Eq1a},\ref{f-b-Eq1a}). 

The second physical system is introduced on a cubic lattice; a quantum rotor model (QRM) coupled 
with a two-band free-fermion model. The QRM comprises of a U(1) phase variable $\theta_{\bm j}$ and its 
canonical conjugate momentum $L_{\bm j}$, defined on the cubic lattice site ${\bm j}$. 
For a heavier mass of the quantum rotor, the boson model undergoes a quantum phase 
transition to a long-range ordering of the U(1) phase variable. The free-fermion model is 
a tight-binding model on the cubic lattice, that has $s$ orbital ($s_{\bm j}$) and $p_{+} \equiv p_x+ip_y$ 
orbital ($p_{\bm j}$) on the same lattice site ${\bm j}$. The tight-binding model has a pair of electron 
and hole pockets around high symmetric points. We discuss that when the rotor is regarded as an 
electric/magnetic dipole moment, the simplest symmetry-allowed coupling between the rotor and the 
fermions in the tight-binding model 
is given by $e^{i\theta_{\bm j}} s^{\dagger}_{\bm j} p_{\bm j}$.  Such an on-site coupling 
together with the QRM and the free-fermion model can be effectively described by 
Eqs.~(\ref{boson-Eq1},\ref{fermion-Eq1a},\ref{f-b-Eq1a}) near the quantum phase transition point.

In Sec.~III, we give a general consideration on ground-state phase diagram and 
renormalizability of the effective model. For Eqs.~(\ref{boson-Eq1},\ref{fermion-Eq1a},\ref{f-b-Eq1a}) 
with the UV cutoff Eqs.~(\ref{fermion-Eq1b},\ref{f-b-Eq1b}), vertex functions (amputated 1-particle-irreducible 
part of connected Green functions) 
have potentially UV divergences in the power of the large UV cutoff 
$\Lambda$ ($\Lambda=\Lambda_B, \overline{\Lambda}_B, \Lambda_F, \overline{\Lambda}_F$). 
The order of the UV divergence ($M$) of the vertex function $\Gamma^{(N_F,N_B)}$ with $N_F$ and $N_B$ 
external fermion and boson lines respectively can be superficially evaluated by dimensional countings 
and it is bounded from above by $M \leq 4 - \frac{3}{2} N_F - N_B$ in the three-spatial dimension. 
That says, the effective model 
in the (3+1)-dimension is renormalizable, where only the following four vertex functions have the 
UV divergences in $\Lambda$;
\begin{eqnarray}
\left\{\begin{array}{l}
\Gamma^{(2,0)}_{\sigma} (i\omega,{\bm k}) =  \ldots \Lambda + \ldots \ln \Lambda \!\ i\omega 
+ \ldots \ln \Lambda \!\ l,  \\
\Gamma^{(0,2)}(i\varepsilon,{\bm q}) = \ldots \Lambda^2 + \ldots 
\ln \Lambda \!\ \varepsilon^2 + \ldots \ln \Lambda \!\ {\bm q}^2,  \\
\Gamma^{(2,1)}_{\sigma}(\cdots) = \ldots \ln \Lambda + {\cal O}(\omega,l,q), \\
\Gamma^{(0,4)}(\cdots) = \ldots \ln \Lambda + {\cal O}(\omega,l,q). \\ 
\end{array}\right. \label{Eq3} 
\end{eqnarray}  
Here the subscript $\sigma=\pm$ is a band index, that distinguishes different functions for the same $N_F$ and 
$N_B$. Thereby, we use the standard renormalized perturbation theory, and eliminate all the UV divergences 
in Eq.~(\ref{Eq3}) by absorbing them into renormalizations of field operator amplitudes, and physical measurable 
quantities. 

The effective model in Eqs.~(\ref{boson-Eq1},\ref{fermion-Eq1a},\ref{f-b-Eq1a}) assumes 
that $q, \omega, l \ll \Lambda \ll k_F$.  
By the renormalized perturbation theory, the set of renormalized vertex functions shall be free from 
the UV cutoff $\Lambda$; the $\Lambda$-divergent terms in Eq.~(\ref{Eq3}) are absorbed into the renormalizations 
of the physical quantities and field operator amplitudes, while $q/\Lambda$, $\omega/\Lambda$, $l/\Lambda$ 
are set to zero in the renormalized functions. On the one hand, the renormalized vertex functions thus obtained 
depend not only on $q$, $\omega$, $l$ but also on $k_F$. In Sec.~III, we shall also argue 
that such an exceptional treatement of $k_F$ in spite of $q,\omega,l \ll \Lambda \ll k_F$ 
becomes possible, because all the vertex functions in Eq.~(\ref{Eq3}) can be expanded in the power of 
$k^2_F$. To be more specific, we show that the vertex functions for $N_F=0,2$ can be expanded in the power of 
$k^2_F$; $\Gamma^{(N_F,N_B)} = \Gamma^{(N_F,N_B;0)} + k^{2}_F \Gamma^{(N_F,N_B;1)} + \cdots$, 
where $k^{2n}_F \Gamma^{(N_F,N_B;n)}$ is a sum of all those amputated 1-particle irreducible (1PI)
Feynman diagrams for $\Gamma^{(N_F,N_B)}$ with $n$ numbers of internal closed 
fermion loops. Importantly, $\Gamma^{(N_F,N_B;n)}$ has no $k_F$-dependence and 
their $k_F$-dependence appear only through the overall factor $k^{2n}_F$. 
Thanks to this analytic nature as functions of $k^2_F$, 
the UV divergences in Eq.~(\ref{Eq3}) can be removed 
at every order in $k^2_F$. In $\Gamma^{(N_F,N_B;n)}$ thus given, 
the superficial degree of the UV divergence in the power of $\Lambda$ is smaller than that of 
$\Gamma^{(N_F,N_B;0)}$ by $2n$.  Accordingly, we have only to keep track of 
the first order in $k^2_F$ of $\Gamma^{(0,2)}$. %; see Eq.~(\ref{Eq3}).  

In Sec.~IV, we shall eliminate the UV divergences in all the vertex functions in Eq.~(\ref{Eq3}), while 
including the UV cutoff dependences into the renormalization of the field operators amplitudes 
and physical quantities. To this end, we impose the following renormalization conditions 
on the renormalied vertex function $\overline{\Gamma}^{(N_F,N_B)}$; 
\begin{eqnarray}
\left\{\begin{array}{l}
\overline{\Gamma}^{(2,0)}_{\sigma}(\omega=l=0) = 0, \!\ 
\frac{\partial \overline{\Gamma}^{(2,0)}_{\sigma}(\omega,l)}{\partial (i\omega)}
\Big|_{v_{\sigma} l=0,\omega=\kappa} = 1, \\
\frac{\partial \overline{\Gamma}^{(2,0)}_{\sigma}(\omega,l)}{\partial (v_{\sigma} l)}\Big|_{v_{\sigma} l=\kappa,\omega=0} = 
-\sigma, \\ 
\overline{\Gamma}^{(0,2)}(\varepsilon=\kappa,{\bm q} = 0) = m^2 + \kappa^2, \\ 
\frac{\partial \overline{\Gamma}^{(0,2)}(q)}{\partial \varepsilon}\Big|_{{\bm q}=0,\varepsilon=\kappa} = 
 2\kappa + \frac{g^2}{\pi^2 (v_{+}+v_{-})} \frac{k^2_F}{\kappa} \\
\frac{\partial \overline{\Gamma}^{(0,2)}(q)}{\partial (c |{\bm q}|)}\Big|_{c|{\bm q}|=\kappa,\varepsilon=0} 
= 2\kappa + \frac{g^2}{\pi^2 (v_{+}+v_{-})} \frac{k^2_F}{\kappa}, \\   
\overline{\Gamma}^{(2,1)}_{\sigma}(\cdots) = - g, \!\ 
\overline{\Gamma}^{(0,4)}(\cdots) = - \lambda,   \\
\end{array}\right. \label{Eq4}
\end{eqnarray} 
with $\sigma=\pm$. 
$\kappa$ in Eq.~(\ref{Eq4}) is an external momentum and 
frequency of the functions, at which the conditions are imposed. The 
right hand sides of Eq.~(\ref{Eq4}) are free from the UV cutoff $\Lambda$, and 
they depend only on physical measurable quantities, such as renormalized boson 
mass $m$, renormalized Fermi velocities $v_{\pm}$, boson velocity $c$, and renormalized 
coupling constants $g$, $\lambda$. When the renormalized mass is set to the zero 
(at the critical point), a finite `$\kappa$' controls infrared behaviours of the vertex functions. 
%In accordance with the expansion of 
%$\Gamma^{(0,2)}$ in $k^2_F$, the conditions for $\overline{\Gamma}^{(0,2)}$ 
%are expanded in the power of $k^2_F$ up to the first order.  

In sec. IV, we employ the minimal substraction scheme~\cite{peskin-schroeder,amit}, to 
make the renormalized vertex functions to satisfy the renormalization conditions 
perturbatively in the coupling constants, $g$ and $\lambda$. The perturbative treatment 
will be a posteriori justified by $\beta$-functions of $g$ and $\lambda$ (see below). 
A general relation among numbers of internal integral variables, $g$ and $\lambda$ in 
a vertex function suggests that $g^2$ and $\lambda$ 
should be treated as the same order (Eq.~(\ref{Lexp}) in Sec.~III). Under Eq.~(\ref{Eq4}), 
all the UV divergences in the vertex functions in Eq.~(\ref{Eq3}) are absorbed into 
bare quantities ($v_{\sigma0}$, $m^2_0$, $c_0$, $g_0$, $\lambda_0$) 
and renormalizations of field operator amplitudes ($Z_{\sigma}$, $Z_{\phi}$). At 
the quantum critical point with $m=0$, the bare quantities thus determined 
are given by
\begin{align}
v_{\sigma0} &= v_{\sigma} + \frac{g^2 (v_{\sigma}-c)}{4\pi^2 c^2 (c+v_{\overline{\sigma}})} 
\log\Big(\frac{\Lambda_B}{\kappa}\Big) 
+ \cdots, \label{Eq5} \\
Z_{\sigma} & = 1 - \frac{g^2}{4\pi^2 c^2 (c+v_{\overline{\sigma}})}\log\Big(\frac{\Lambda_B}{\kappa}\Big) + \cdots, \label{Eq6} \\
m^2_0 & = \frac{g^2 k^2_F}{\pi^2(v_{+}+v_{-})} \log \Big(\frac{\Lambda_F}{\kappa}\Big) 
- \frac{\lambda \Lambda^2_B}{16 \pi^2 c^3} + \cdots, \label{Eq7} \\
g_0 &= g + \frac{g^3}{8\pi^2 c^2} 
\frac{2c+v_{+}+v_{-}}{(c+v_{+})(c+v_{-})}
%\Big\{\frac{1}{c+v_{+}}+ \frac{1}{c+v_{-}}\Big\} 
\log\Big(\frac{\Lambda_B}{\kappa}\Big)
 + \cdots, \label{Eq8}  \\
\lambda_0 &= \lambda +\frac{5\lambda^2}{16\pi^2 c^3} 
\log\Big(\frac{\Lambda_B}{\kappa}\Big) + \cdots,  \label{Eq9} 
\end{align}
with $c_0=c + \cdots$, $Z_{\phi} = 1+ \cdots$,  $\overline{\sigma} = \mp$ for $\sigma=\pm$.
Here only the leading order terms in $g^2$ and $\lambda$ that depend on either 
$\Lambda$ or $\kappa$ are shown explicitly in the right hand sides, while the others 
are omitted by `$\cdots$'. The renormalized vertex 
functions defined by renormalized field operators, 
$\overline{\psi}_{\sigma} \equiv \psi_{\sigma}/\sqrt{Z_{\sigma}}$, 
$\overline{\phi} \equiv \phi/\sqrt{Z_{\phi}}$, are free from the UV cutoff as 
functions of the renormalized physical quantities. Note that 
Eq.~(\ref{Eq7}) defines a subspace of $m=0$ (quantum critical `point') in a multiple-parameter space 
subtended by the bare quantities and $\ln \kappa$. When $m=0$, the external momenta $\kappa$ 
plays role of a renormalization group (RG) scale variable and all the $\log \Lambda$ in the above 
equations come in pairs with $-\log \kappa$. 

In Sec.~V, we clarify non-FL features in the renormalized two-point fermion function 
around quantum critical point. To this end, we first derive a Callan-Symmanzik (CS) equation for
the renormalized two-point fermion vertex function;
%\begin{widetext}
\begin{align} 
&\Big\{\frac{\partial}{\partial \ln \kappa} 
+ \sum_{\sigma^{\prime}=\pm}\beta_{v_{\sigma^{\prime}}} \frac{\partial}{\partial v_{\sigma^{\prime}}}  
+ \beta_{g} \frac{\partial}{\partial g} + \beta_{\lambda} 
\frac{\partial}{\partial \lambda}  \nonumber \\
& \ \ \ \   - \gamma_{\sigma} \Big\} \overline{\Gamma}^{(2,0)}_{\sigma} (l,\omega;c,\cdots,\lambda,\kappa)  
= 0, \label{Eq10}
\end{align}
%\end{widetext}
together with $\beta$ functions of Fermi velocities, and coupling constants, 
\begin{eqnarray}
\left\{\begin{array}{l}
\beta_{v_{\sigma}} \equiv 
\frac{\partial v_{\sigma}}{\partial \ln \kappa}\Big|_{c_0,\cdots,\Lambda}  
 = \frac{g^2}{4\pi^2 c(c+v_{\overline{\sigma}})} \Big(\frac{v_{\sigma}}{c} - 1\Big) + 
\cdots, \\  
\beta_g \equiv \frac{\partial g}{\partial \ln \kappa} \Big|_{c_0,\cdots,\Lambda}  
=  \frac{g^3}{8 \pi^2 c^2} \Big(\frac{1}{c+v_{+}} + \frac{1}{c+v_{-}} \Big) + \cdots \\
\beta_{\lambda} \equiv 
\frac{\partial \lambda}{\partial \ln \kappa} \Big|_{c_0,\cdots,\Lambda}   
= \frac{5\lambda^2}{16\pi^2 c^3} + \cdots,  \\
\end{array}\right. \label{Eq11}
\end{eqnarray}
and $\gamma$ functions of the two fermion bands ($\sigma=\pm$),
\begin{align}
\gamma_{\sigma} \equiv \frac{\partial  \ln Z_{\sigma}}{\partial \ln \kappa}\Big|_{c_0,\cdots,\Lambda} = 
\frac{g^2}{4\pi^2 c^2 (c+v_{\overline{\sigma}})} +\cdots.   \label{Eq12}
\end{align}
The $\beta$ functions of the velocities show that two Fermi velocities $v_{\pm}$ 
are renormalized into the boson velocity $c$ in the infrared (IR) limit ($\kappa \rightarrow 0$). 
This suggests that $v_{\pm}$ in the CS equation could be set to $c$ in the low-energy 
limit. $\beta$ functions of the coupling constants dictate that the two coupling constants are 
marginally irrelevant in the IR limit. The marginal irrelevance justifies a posteriori the 
perturbative treatment of the coupling constants. Besides, the $\lambda$-dependence 
of $\Gamma^{(2,0)}$ must start from the order of ${\cal O}(g^2 \lambda)$, 
so that $\partial_{\lambda}$ in Eq.~(\ref{Eq10}) leads to the higher-order terms, 
${\cal O}(g^2 \lambda^2)$, and can be omitted. These simplifications lead to the 
homogeneous CS equation with a one-parameter scaling. The solution of such 
CS equation is given in the form of the renormalized 
Green's function with 
$\Gamma^{(2,0)} G^{(2,0)}=1$; 
\begin{align}
\overline{G}^{(2,0)}_{\sigma}(l,\omega;g,\kappa) &= 
\frac{\exp \Big[-\int^g_{\overline{g}}\frac{\gamma_{\sigma}(g^{\prime})}{\beta_g(g^{\prime})} 
dg^{\prime}\Big]}{i\omega - \sigma cl} 
=\frac{\frac{\overline{g}(t)}{g}}{i\omega - \sigma cl}. \label{Eq16}
\end{align}
$\overline{g}(t)$ is a solution of 
the one-parameter scaling equation; $\partial \overline{g}/\partial t = \beta_g(\overline{g})$, 
$\overline{g}(t=0)=g$, 
\begin{align}
\overline{g}(t) = \frac{g}{\sqrt{1- 2 \alpha t}}, \!\ \!\ \!\ \!\  t \equiv \log \bigg[\frac{\sqrt{\omega^2 + (cl)^2}}{\kappa}\bigg], 
\label{Eq17}
\end{align}
with $\alpha \equiv g^2/(8\pi^2 c^3)$. 
%In Eq.~(\ref{Eq16}), $\gamma_{\sigma}(g)/\beta_g(g) = 1/g$ is used from 
%Eqs.~(\ref{Eq11},\ref{Eq12}) with $v_{\pm}=c$. 

\begin{figure}[t]
	\centering
	\includegraphics[width=0.9\linewidth]{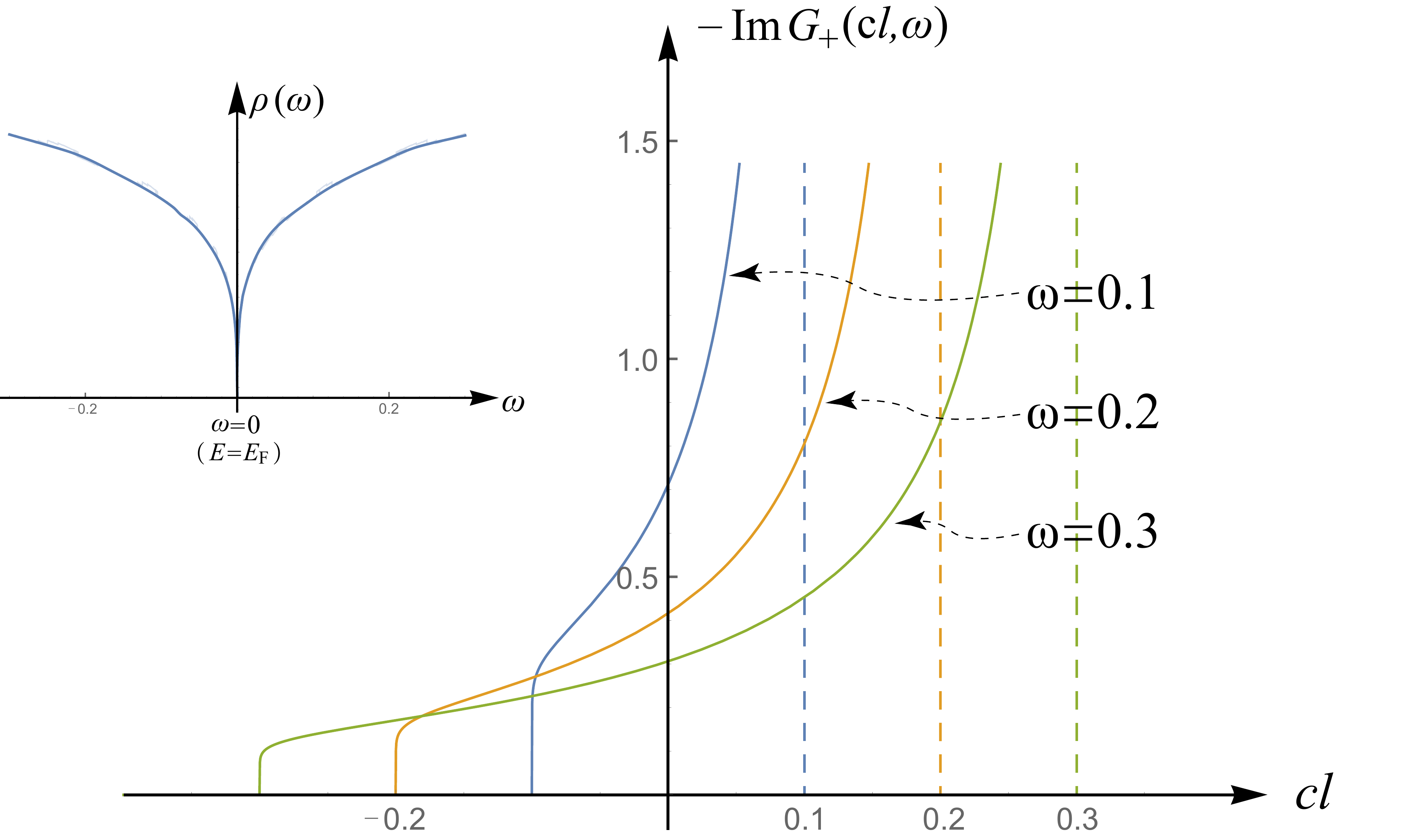}
	\caption{(a) Renormalized spectral function for the electron-type band at 
the QCP: $-{\rm Im}G^{R}_{+}(cl,\omega)$ is plotted as a function of $cl$ 
for $\omega=0.1, \!\ 0.2,\!\ 0.3$ with the RG scale $\kappa=1$. (inset) DOS at the QCP: $\rho(\omega)$. 
}
	\label{fig:3}
\end{figure}

In Sec.~VI, we obtain fermion's two-point retarded Green's function, spectral function and DOS at the QCP 
(Fig.~\ref{fig:3}). After an analytic continuation, $i\omega \rightarrow \omega+i\delta$, 
Eq.~(\ref{Eq16}) gives the retarded Green's function. Thereby, it can be clearly seen from 
Eqs.~(\ref{Eq16},\ref{Eq17}) with $i\omega \rightarrow \omega+i\delta$ that 
the quasi-particle spectral weight at $T=0$ is zero 
because $\overline{g}/g$ in Eq.~(\ref{Eq16}) goes to zero whenever $\omega \rightarrow \sigma cl$; 
${\rm Re} \!\ t \rightarrow -\infty$. The vanishing quasi-particle spectral weight at the 
quantum critical point is the cornerstone of the non-Fermi liquid property in this paper. 
The spectral function comprises only of continuum spectrum, that shows an asymptotic 
form of $1/[\eta |\ln \eta|^{\frac{3}{2}}]$ when $\omega = \sigma (cl +\eta)$ 
with small positive $\eta$ (Fig.~\ref{fig:3}). 
The asymptotic form around $\omega = \sigma cl$ plays major role in 
the pseudo-gap behaviour of the DOS at the QCP; $\rho(\omega)\propto 1/|\ln| \omega||^{\frac{1}{2}}$ 
(inset of Fig.~\ref{fig:3}). 

In Sec.~VII, we conclude the paper with discussion on thermodynamic and 
magnetic properties of the two-band fermion system coupled with the critical $\phi^4$ bosons. 
In the appendix, we derive and solve the Callan-Symanzik equation for two-point boson 
Green functions and discuss the boson spectral function at the QCP. 

\section{Physical Systems} 
As for conrete physical systems to which the effective model of 
Eqs.~(\ref{boson-Eq1},\ref{fermion-Eq1a},\ref{f-b-Eq1a}) is applicable, we consider the two 
systems: (i) a three-dimensional (3D) interacting two-bands fermion system with an excitonic instability, and 
(ii) multiple-band electron model coupled with quantum 
rotor model with/without an $Z_n$ anisotropy term ($n\ge 4$).   

\subsection{excitonic condensations in interacting two-bands fermion systems 
at a `frustrated' point}
\begin{figure}[t]
	\centering
	\includegraphics[width=1.0\linewidth]{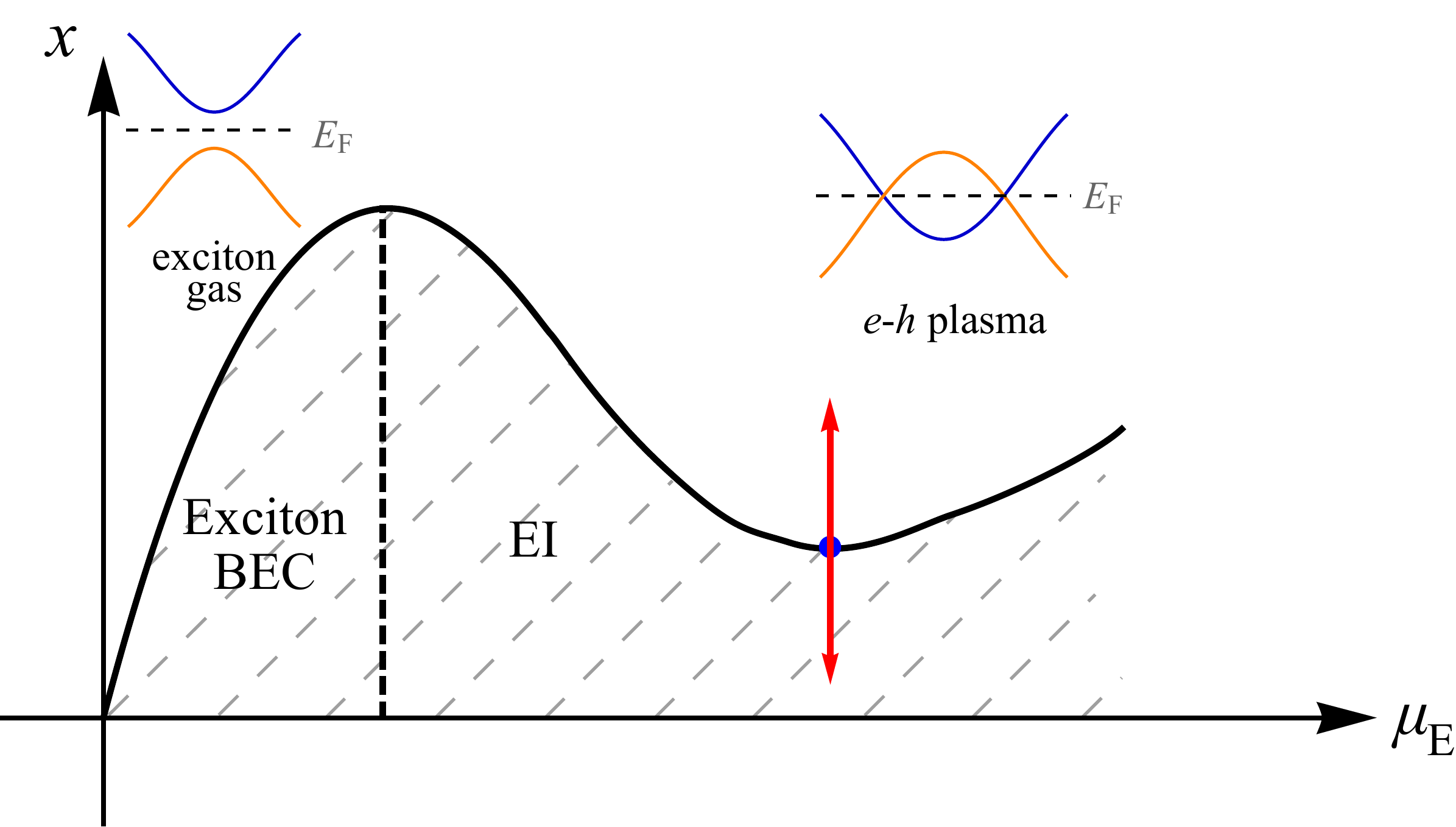}
	\includegraphics[width=1.0\linewidth]{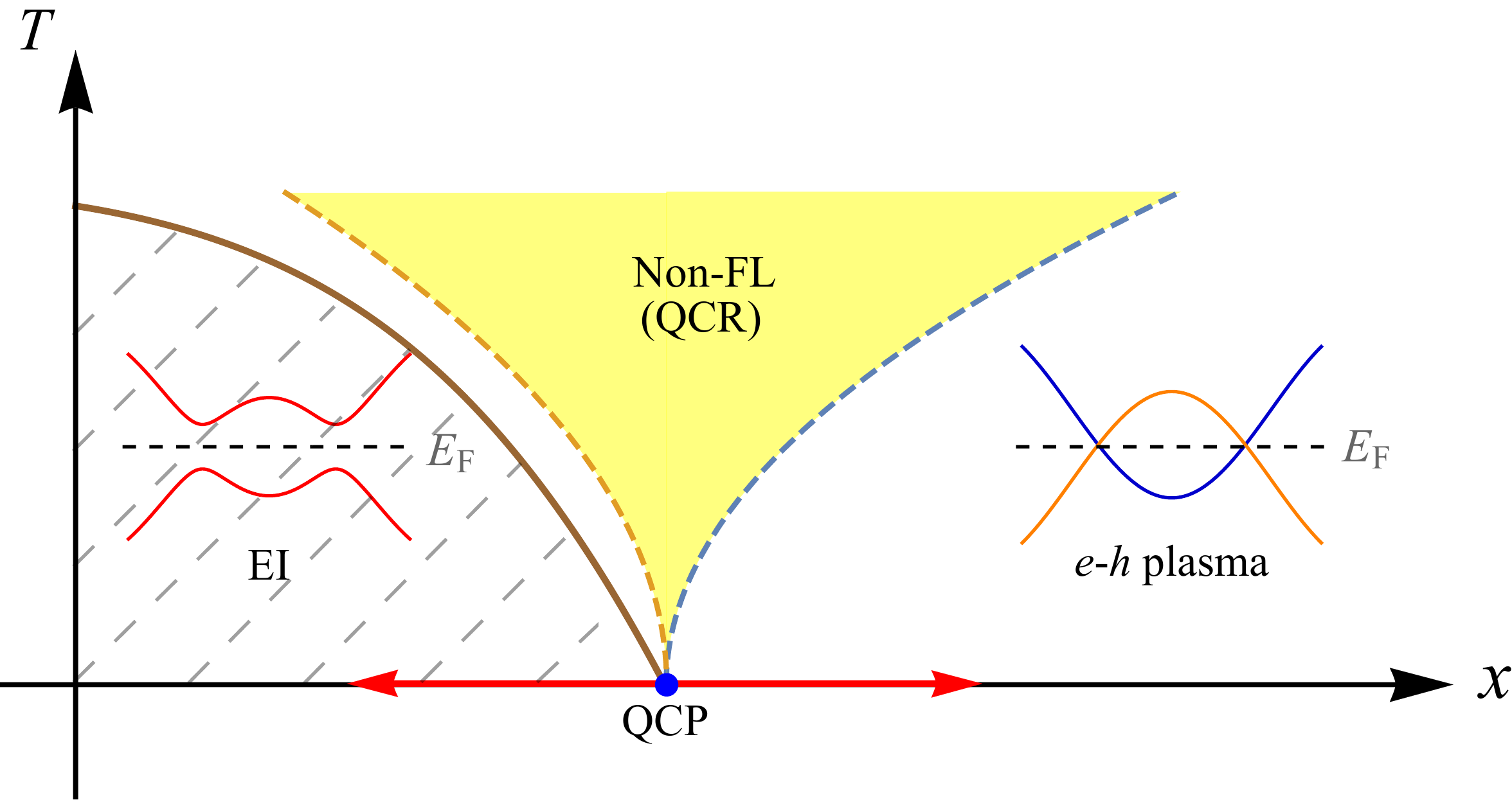}
	\caption{a: (upper) A schematic ground state phase diagram of 
an interacting two-bands fermion model in the 3 spatial dimension of Eq.~(\ref{ef0}). 
The phase diagram is subtended by a band inversion parameter $\mu_E$ and other 
parameter $x$. The excitonic phase with the broken U(1) symmetry is  
identified by a region of $A<0$ in Eq.~(\ref{ef1}), while the U(1) symmetric normal phase 
can be identified by the other region of $A>0$. The phase boundary between these 
two regions is defined by $x=X(\mu_E)$. We fine-tune $\mu_E$ at those values of 
$\mu_E$ where $X$ have local minima, say $\mu_E=\mu_{E,f}$. A quantum phase
 transition driven by the other parameter $x$ at $\mu_E=\mu_{E,f}$ can be well 
described by the effective model studied in this paper. b: (lower) A schematic $T$-$x$
phase diagram associated with the quantum phase transition at $\mu_E=\mu_{E,f}$. The 
non-FL region is expected in a yellow colored region. }
	\label{sfig:6}
\end{figure}

Consider a 3D two-band interacting spinless fermion model with an excitonic instability. 
A partition function for the excitonic-pairing order parameter can be given by  
\begin{align}
Z &= \int {\cal D} a {\cal D} b {\cal D} \Phi \!\ {\rm exp}\Big[-\int^{\beta}_0 d\tau \int 
d^3{\bm x} 
\Big\{ a^{\dagger}\partial_{\tau} a + b^{\dagger} \partial_{\tau} b \nonumber \\
& \hspace{1cm} - \mu_{E} (a^{\dagger} a - b^{\dagger} b) - \mu_0 (a^{\dagger} a + b^{\dagger} b)  \nonumber \\
& \hspace{1cm} 
+ H_{el}(a,b) + a^{\dagger} b \Phi + b^{\dagger} a \Phi^{*} -\frac{1}{g} |\Phi|^2 \Big\} \Big] \label{ef0} \\
& = \int {\cal D} \Phi \!\ {\rm exp} \Big[-\int^{\beta}_{0} d\tau 
\int d^3{\bm x} \Big\{ u \Phi^* \partial_{\tau} \Phi + v|\partial_{\tau} \Phi|^2 \nonumber \\
& \hspace{0.8cm}  
+ w |\partial_{\bm x} \Phi|^2 + A|\Phi|^2 + B(|\Phi|^2)^2  + \cdots \Big\} \Big].  
\label{ef1}
\end{align}
Here $H_{el}(a,b)$ is a free (kinetic energy) part of the two-bands electron systems. $a$ 
and $b$ are field operators of the $a$-band (electron-type) fermion and $b$-band 
(hole-type) fermion respectively. $\mu_E$ and $\mu_{0}$ are `staggered' and 
uniform chemical potential. The staggered chemical potential $\mu_E$ is nothing but 
an energy difference between the electron-band energy minimum and the hole-band 
energy maximum (Fig.~\ref{sfig:6}(a)). We call this as `band inversion energy'. 

A repulsive interaction between the two fermions is decomposed into a 
coupling between an excitonic pairing field $\Phi$, and the two fermion 
field operators. After an integration of the fermion fields, 
one would obtain the $(3+1)$D $\phi^4$-type effective 
action for the excitonic field, that varies slowly in space and time 
(see Eq.~(\ref{ef1})).  In the absence of the Berry phase term 
$\Phi^{*}\partial_{\tau}\Phi$ ($u=0$), such an effective action becomes 
identical to the boson part of the effective model with $z=1$ studied 
in this paper; Eq.~(\ref{boson-Eq1}). A coefficient for the Berry phase term, $u$,  as well 
as others ($v$, $A$, $w$, $\cdots$) depend on microscopic parameters such as $\mu_E$ and other 
parameter $x$;
\begin{align}
u(\mu_E,x, \cdots), \ \ v(\mu_E, \cdots), \ \ A(\mu_E, \cdots), \cdots. \nonumber
\end{align}   
In general, $u$ is finite, and thus the dynamical exponent $z$ of the boson part of the 
action is $2$. Nonetheless, the linear coefficient $u$ can be exactly zero when $\mu_E$ is fine-tuned 
at a certain point, say $\mu_E=\mu_{E,f}$,
\begin{align}
u(\mu_E = \mu_{E,f},\cdots) = 0,  \ \ v(\mu_E=\mu_{E,f},\cdots) >0, 
\end{align} 
where $z=1$. A quantum phase transition at such fine-tuned point ($\mu_E=\mu_{E,f}$), 
that is driven by the other parameter $x$, is effectively well described by the 
fermion-boson coupled model studied in this paper, 
Eqs.~(\ref{boson-Eq1},\ref{fermion-Eq1a},\ref{f-b-Eq1a}).      

To identify such fine-tuned critical point for $\mu_E$, 
we follow an argument based on a local gauge symmetry~\cite{fisher89}. 
To this end, note first that Eq.~(\ref{ef0}) is invariant under the following 
local gauge transformation;
\begin{align}
& a^{\dagger} \rightarrow a^{\dagger} e^{i\frac{\theta(\tau)}{2}}, 
\ b^{\dagger} \rightarrow b^{\dagger} e^{-i\frac{\theta(\tau)}{2}},  \nonumber \\ 
& \Phi \rightarrow \Phi e^{-i\theta(\tau)}, \  
\mu_E \rightarrow \mu_E - i \frac{\partial_{\tau}\theta}{2}. 
\end{align} 
So is Eq.~(\ref{ef1}). Thus, for sufficiently small $\partial_{\tau}\theta(\tau)$, we could 
expand Eq.~(\ref{ef1}) and $u$, $v$ and $A$ in the power of $\partial_{\tau}\theta$, where  
the variation must be zero. This gives the following general relations,  
\begin{align}
\frac{1}{2} \frac{\partial A}{\partial \mu_E} = u, \ \ \frac{1}{2} \frac{\partial u}{\partial \mu_E}  = - 2v.   
\label{relation1}
\end{align}

With Eq.~(\ref{relation1}) in mind, consider a $T=0$ phase diagram of the 
interacting two-bands fermion system. For simplicity, we consider a two-dimensional 
phase diagram subtended by $\mu_E$ and $x$ (Fig.~\ref{sfig:6}). 
In the $T=0$ phase diagram, a region with $A<0$ could be 
regarded as an excitonic phase with the SSB of the global 
U(1) symmetry ($|\Phi| \ne 0$), while the other region with $A>0$ can be 
regarded as a U(1) symmetric normal phase ($|\Phi| =0$). 
A phase boundary between these two regions is denoted by $x=X(\mu_E)$. 
From Eq.~(\ref{relation1}), one can see the minima of $X$ as a function of $\mu_E$ 
are the desired fine-tuned critical points with $u=0$ and $v>0$; 
\begin{eqnarray}
\left\{\begin{array}{lcc} 
u = 0, & v <0, &  \!\ \!\  {\rm for} \!\  \!\ \frac{dX}{d\mu_E} = 0 \!\ \cap  \!\  \frac{d^2X}{d\mu^2_E}  
<0, \\
u = 0, & v > 0, & \!\ \!\ {\rm for} \!\ \!\ \frac{dX}{d\mu_E} = 0  \!\ \cap  \!\ \frac{d^2X}{d\mu^2_E}  
> 0. \\
\end{array}\right. 
\end{eqnarray} 
Since the excitonic phase is suppressed as a function of $x$ at the fine-tuned points, 
we call such fine-tuned values of $\mu_{E,f}$ as `frustrated' points with maximally 
suppressed excitonic instability. 
  
\subsection{$sp$ model coupled with a quantum rotor model}
The second physical system is a quantum rotor model coupled with two-orbital spinless fermion model. 
The fermion part is introduced on a 3D cubic lattice. Each lattice points has $s$-orbital 
and $p_{+} \equiv p_x+ip_y$ orbital degree of freedom of the spinless fermion. 
The fermion Hamiltonian is described by a two-orbital tight-binding model;
\begin{align}
&H_{\rm sp}  \equiv  \sum_{{\bm j},{\bm m}} 
\sum_{\alpha,\beta = s,p}
[\mathbb{H}]_{({\bm j},\alpha|{\bm m},\beta)} c^{\dagger}_{{\bm j},\alpha} c_{{\bm m},\beta} \nonumber \\
& \equiv \sum_{\bm j} \bigg\{ 
\Delta_0 \Big(- s^{\dagger}_{\bm j} s_{\bm j} 
+ p^{\dagger}_{\bm j} p_{\bm j}\Big) - t_s \sum_{\mu=x,y,z}  
\Big(s^{\dagger}_{{\bm j}+\mu} s_{\bm j} + {\rm h.c.} \Big) \nonumber \\
&  + t_p  \Big( p^{\dagger}_{{\bm j}+x} p_{\bm j} + p^{\dagger}_{{\bm j}+y} p_{\bm j} 
+ {\rm h.c.} \Big) 
- t^{\prime}_{p} \Big( p^{\dagger}_{{\bm j}+z} p_{\bm j} + {\rm h.c.} \Big) \nonumber \\ 
& \hspace{-0.4cm} + t_{sp} \Big((s^{\dagger}_{{\bm j}+x} + i s^{\dagger}_{{\bm j}+y} 
- s^{\dagger}_{{\bm j}-x} - i s^{\dagger}_{{\bm j}-y}) p_{\bm j} + {\rm h.c.} \Big) \bigg\}, 
\label{free-fermi}
\end{align}   
with ${\bm j} \equiv (j_x,j_y,j_z)$, $x \equiv (1,0,0)$ and so on.
Here all the intra-orbital hopping integrals $t_s$, $t_{p}$, $t^{\prime}_{p}$ as well as charge energy $\Delta_0$ 
are positive. When $t^{\prime}_p$ is chosen to be a negative value and the others are positive, the model 
becomes a prototype model for a magnetic Weyl semimetal and layered Chern insulator~\cite{yang11,chen15,liu16}. 

In this work, we take $t_s=t^{\prime}_{p}=t_{p}=t>0$ and $t_{sp}=t^{\prime}$ for simplicity. 
The two energy bands are always separated by finite direct band gap; 
\begin{align}
E_{\pm}({\bm k}) = -2tc_z \pm \sqrt{(\Delta_0 + 2t(c_x+c_y))^2 + 4{t^{\prime}}^2 (s^2_x+s^2_y)}, \nonumber  
\end{align} 
with $c_x \equiv \cos k_x$, $s_x \equiv \sin k_x$ and so on.
When $\Delta_0-4t >0$,  an upper energy band $E_{+}$ has an energy miminum at $(k_x,k_y,k_z)=(\pi,\pi,0)$ and 
the lower energy band $E_{-}$ has an energy maximum at $(k_x,k_y,k_z)=(\pi,\pi,\pi)$. When the Fermi level is set 
to zero, the two bands form an electron pocket and hole pocket around these extremes respectively, 
\begin{align}
E_{\pm } = &\pm \Big(-2t + (\Delta_0-4t) + t(k^2_x+k^2_y+k^2_z)  \nonumber \\
& \ \  + \frac{2t^{\prime}}{\Delta_0-4t} (k^2_x+k^2_y)\Big) 
+ \cdots. \nonumber  
\end{align}
Note that a Fermi surface (FS) of the electron pocket around $(\pi,\pi,0)$ and  that of the hole pocket around  $(\pi,\pi,\pi)$ 
are identical in shape and size; $\mu=0$ is the charge-neutrality point. 
When $\Delta_0, t \gg t^{\prime}$, the FS can be approximated by an isotropic 
Fermi surface and the two-orbital spinless fermion model can be well described by Eq.~(\ref{fermion-Eq1a}). Thereby, 
the electron and hole pockets are primarily 
composed of $p$ orbital and $s$ orbital respectively. In Eq.~(\ref{free-fermi}), 
a mixing is induced only by small $t^{\prime}$.

The boson part takes a form of quantum rotor model on the same cubic lattice with a $Z_4$ anisotropy~\cite{sachdev}, 
\begin{align}
& H_{\rm qrm} =  \frac{\hat{L}^2_{\bm j}}{2M} - J_0 \sum_{\mu=x,y} \sum_{\bm j} \cos(\hat{\theta}_{\bm j} 
- \hat{\theta}_{{\bm j}+\mu}) \nonumber \\
& \  \ + J_{\perp} \sum_{\bm j} \cos(\hat{\theta}_{\bm j} 
- \hat{\theta}_{{\bm j}+z}) - \Delta_4 \sum_{\bm j} \cos(4 \hat{\theta}_{\bm j})
\end{align} 
with $J_0$, $J_{\perp}$, $\Delta_4>0$. $\theta_{\bm j}$ represents a rotor degree of freedom defined 
on the cubic lattice site, taking a form of the 
U(1) phase variable. $\Delta_4$ is the $Z_4$ anisotorpy term that locks the rotor into four directions within 
a plane; $\theta_{{\bm j}}$ to $0, \pm \frac{\pi}{2}, \pi$. The positive $J_0$ and $J_{\perp}$ favor an 
antiferro-type order of the U(1) phase variable; $\theta_{\bm j}$ orders in the ferro-type way within a  
$xy$ plane of the cubic lattice, and it orders in the antiferro-type way along the $z$ axis. 
$\hat{L}_{\bm j}$ is a momentum canonically conjugate to $\theta_{\bm j}$;
\begin{align}
\big[ \hat{\theta}_{\bm j}, \hat{L}_{\bm m}\big] = i\hbar \delta_{{\bm j},{\bm m}}. 
\end{align} 
$M$ stands for a mass of the rotor. Larger $M$ reduces a kinetic energy of the rotor, favoring the 
ordering of the U(1) phase variable, while smaller $M$ induces a quantum 
phase transition from the antiferro-type order to a quantum disorder phase. In the quantum disorder phase, the 
momentum takes the definite integer value at every site, $L_{\bm j}=0$. Since the $Z_4$ anisotropy term 
is (dangerously) irrelevant at the quantum critical point~\cite{jose77}, the quantum phase transition 
is well described by the $(3+1)$-dimensional $\phi^4$ action with $z=1$ 
as in Eq.~(\ref{boson-Eq1}), where  $e^{i\hat{\theta}_{\bm j}}(-1)^{j_z}$ plays 
role of a slowly-varying $\phi ({\bm x})$ in Eq.~(\ref{boson-Eq1}) 
with ${\bm j}\equiv (j_x,j_y,j_z)$ being identified with ${\bm x}$~\cite{sachdev}.  

\subsubsection{coupling between rotor and $sp$ model} 
In this subsection, we shall introduce the most natural symmetry-allowed on-site coupling between 
the rotor and two-orbital spinless fermions. 
To this end, note first that the fermion model respects a magnetic point group symmetry 
of $4/mm^{\prime}m^{\prime}$, whose group elements are  
\begin{align}
C^z_{4}, R\sigma_x, R\sigma_y, I, \label{magnetic} 
\end{align} 
and their combinations. Here $C^{z}_4$ is $\pi/2$ rotation around the $z$-axis, $R$ is the time-reversal operation, 
$\sigma_x$, $\sigma_y$ are mirror with respect to the $x=0$ and $y=0$ plane respectively, $I$ is the spatial 
inversion. Namely, the tight-binding Hamiltonian respects the following symmetries;  
\begin{align}
&[\mathbb{H}]_{(C^z_4({\bm j})|C^z_4({\bm m}))} = 
\left(\begin{array}{cc} 
1 & 0 \\
0 & -i \\ 
\end{array}\right) [\mathbb{H}]_{({\bm j}|{\bm m})} \left(\begin{array}{cc} 
1 & 0 \\
0 & +i \\ 
\end{array}\right) \ : \!\ C^z_4,  \nonumber \\ 
& [\mathbb{H}^T]_{(\sigma_x({\bm j})|\sigma_x({\bm m}))} = 
\left(\begin{array}{cc} 
1 & 0 \\
0 & -1 \\ 
\end{array}\right) [\mathbb{H}]_{({\bm j}|{\bm m})} \left(\begin{array}{cc} 
1 & 0 \\
0 & -1 \\ 
\end{array}\right) \  : \!\ R\sigma_x, \nonumber \\ 
&[\mathbb{H}^T]_{(\sigma_y({\bm j})|\sigma_y({\bm m}))} = 
[\mathbb{H}]_{({\bm j}|{\bm m})} \  : \!\ R\sigma_y, \nonumber \\  
& [\mathbb{H}]_{(I({\bm j})|I({\bm m}))} = 
\left(\begin{array}{cc} 
1 & 0 \\
0 & -1 \\ 
\end{array}\right) [\mathbb{H}]_{({\bm j}|{\bm m})} \left(\begin{array}{cc} 
1 & 0 \\
0 & -1 \\ 
\end{array}\right) \  : \!\ I. \label{symmetry}
\end{align} 
In other words, Eq.~(\ref{free-fermi}) is invariant under the following transformations of 
the creation and annihilation operators of $s$ and $p$ orbitals,
\begin{align}
& \left(\begin{array}{c} 
s \\
p \\ 
\end{array}\right) \rightarrow \left(\begin{array}{cc} 
1 & 0 \\
0 & +i \\ 
\end{array}\right) \left(\begin{array}{c} 
s \\
p \\ 
\end{array}\right) \ : \!\ C^z_{4}, \nonumber \\
& \left(\begin{array}{c} 
s \\
p \\ 
\end{array}\right) \rightarrow \left(\begin{array}{cc} 
s^{\dagger} & p^{\dagger} \\
\end{array}\right) 
\left(\begin{array}{cc} 
1 & 0 \\
0 & -1 \\ 
\end{array}\right) \ : \!\ R\sigma_x, \nonumber \\
& \left(\begin{array}{c} 
s \\
p \\ 
\end{array}\right) \rightarrow \left(\begin{array}{cc} 
s^{\dagger} & p^{\dagger} \\
\end{array}\right) \ : \!\ R\sigma_y, \nonumber \\
&\left(\begin{array}{c} 
s \\
p \\ 
\end{array}\right) \rightarrow \left(\begin{array}{cc} 
1 & 0 \\
0 & -1 \\ 
\end{array}\right) \left(\begin{array}{c} 
s \\
p \\ 
\end{array}\right) \ : \!\ I
\end{align}
with $C^z_4( A_{({\bm j},\alpha|{\bm m},\beta)} c^{\dagger}_{{\bm j},\alpha} c_{{\bm m},\beta}) = 
A_{\cdots} C^z_{4}(c^{\dagger}_{{\bm j},\alpha}) C^z_4(c_{{\bm m},\beta})$, 
$I(A c^{\dagger}_{{\bm j},\alpha} c_{{\bm m},\beta}) = A 
I(c^{\dagger}_{{\bm j},\alpha}) I(c_{{\bm m},\beta})$, and 
$R\sigma_{\mu}( A c^{\dagger}_{{\bm j},\alpha} c_{{\bm m},\beta}) = A 
R\sigma_{\mu}(c_{{\bm m},\beta}) \!\ R\sigma_{\mu}(c^{\dagger}_{{\bm j},\alpha})$. 

In the following, we introduce an on-site coupling between the two-orbital fermions and the 
quantum rotor, that respects all (or part of) these magnetic point group symmetry operations. To do so, 
we consider two physical cases, (i) when the rotor on each site has the same symmetry as $x$ and 
$y$-components of electric dipole;  $e^{i\theta_{\bm j}} \sim E_x({\bm j})+ i E_{y}({\bm j})$, 
and (ii) when the rotor has the same symmetry 
as the $x$ and $y$-components of magnetic dipole;  
$e^{i\theta_{\bm j}} \sim B_x({\bm j})+ i B_{y}({\bm j})$. 

\subsubsection{when rotor is electric dipole}
Consider the first case, where real and imaginary part of $e^{i\theta_{\bm j}}$ has the same symmetry as 
the $x$ and $y$ components of the electric dipole moment defined at ${\bm j}$ respectively. Under 
the symmetry operations of $4/mm^{\prime}m^{\prime}$, the rotor degree of freedom will be 
transformed as follows,
\begin{align}
&e^{i\theta_{\bm j}}  \rightarrow e^{i(\theta_{\bm j}+\frac{\pi}{2})} \ : \!\ C^z_{4}, \ \ 
e^{i\theta_{\bm j}}  \rightarrow e^{i\theta_{\bm j}} \ : \!\ R, \nonumber \\
&e^{i\theta_{\bm j}}  \rightarrow e^{-i\theta_{\bm j}} \ : \!\ \sigma_y, \  \ 
e^{i\theta_{\bm j}}  \rightarrow e^{i(\pi-\theta_{\bm j})} \ : \!\ \sigma_x, \nonumber \\
&e^{i\theta_{\bm j}}  \rightarrow  - e^{i\theta_{\bm j}} \ : \!\ I, \ \ 
e^{i\theta_{\bm j}}  \rightarrow - e^{-i\theta_{\bm j}} \ : \!\ R\sigma_x, \nonumber \\ 
& e^{i\theta_{\bm j}}  \rightarrow e^{-i\theta_{\bm j}} \ : \!\ R\sigma_y. \ 
\end{align}
Thus, the simplest on-site coupling between the electric dipole and the fermions that respects all the 
magnetic point group symmetry of $4/mm^{\prime}m^{\prime}$ is as follows,
\begin{eqnarray}
H_{f-b} = \sum_{\bm j} \Big( 
e^{-i\hat{\theta}_{\bm j}} s^{\dagger}_{\bm j} p_{\bm j} + e^{i\hat{\theta}_{\bm j}} 
p^{\dagger}_{\bm j} s_{\bm j} \Big). \label{f-b-coupling} 
\end{eqnarray}
In the presence of the on-site fermion-boson coupling, the 
slowly varying $\phi({\bm x} \simeq {\bm j}) = e^{-i\hat{\theta}_{\bm j}}(-1)^{j_z}$ couples 
between the electron and hole pockets in the same form as in Eq.~(\ref{f-b-Eq1a}).       

\subsubsection{when rotor is magnetic dipole}
Consider the second case, where real and imaginary part of $e^{i\theta_{\bm j}}$ has the same symmetry as 
the $x$ and $y$ components of the magnetic dipole moment defined at ${\bm j}$. Under the 
magnetic point group symmetry operations, the rotor degree of freedom will be transformed 
as follows,
\begin{align}
&e^{i\theta_{\bm j}}  \rightarrow e^{i(\theta_{\bm j}+\frac{\pi}{2})} \ : \!\ C^z_{4}, \ \ 
e^{i\theta_{\bm j}}  \rightarrow - e^{i\theta_{\bm j}} \ : \!\ R, \nonumber \\ 
& e^{i\theta_{\bm j}}  \rightarrow e^{i(\pi-\theta_{\bm j})} \ : \!\ \sigma_y, \  \ 
e^{i\theta_{\bm j}}  \rightarrow e^{-i\theta_{\bm j}} \ : \!\ \sigma_x, \nonumber \\
& e^{i\theta_{\bm j}}  \rightarrow  e^{i\theta_{\bm j}} \ : \!\ I, \ \ 
e^{i\theta_{\bm j}}  \rightarrow - e^{-i\theta_{\bm j}} \ : \!\ R\sigma_x, \nonumber \\
& e^{i\theta_{\bm j}}  \rightarrow e^{-i\theta_{\bm j}} \ : \!\ R\sigma_y. \ 
\end{align}         
Thus the on-site coupling in Eq.~(\ref{f-b-coupling}) is symmetrically allowed by $C^z_4$, $R\sigma_x$ 
and $R \sigma_y$, while it is disallowed by $I$. In other words, the coupling of Eq.~(\ref{f-b-Eq1a}) 
is allowed by a subgroup of $4/mm^{\prime}m^{\prime}$: $4m^{\prime}m^{\prime}$. The group 
elements of  the magnetic point group of $4m^{\prime}m^{\prime}$ are 
\begin{align}
C^z_{4}, R \sigma_x, R\sigma_y, 
\end{align} 
and their combinations. 

%\begin{figure}[t]
%	\centering%
%	\includegraphics[width=0.9\linewidth]{fig2.pdf}
%	\caption{(a-c) One-loop corrections included in the perturbative renormalization theory analysis. Solid/broken 
%line represent fermion/boson lines respectively.}
%	\label{fig:2}
%\end{figure} 

\section{Effective boson-fermion coupled model and its renormalizability}
\subsection{$(d+1)$-dimensional effective model and its ground-state phase diagram}
The two physical systems discussed in the previous section can be described by 
Eqs.~(\ref{boson-Eq1},\ref{fermion-Eq1a},\ref{f-b-Eq1a}).  To put it generally, 
consider the fermion-boson coupled system in the $(d+1)$-dimension,
\begin{align}
&S_{t} = S_{\psi} + S_{\phi} + S_{\psi-\phi}, \label{model1} \\
&S_{\psi} = \sum_{\sigma=\pm}\int \psi^{\dagger}_{\sigma}({\bm x},\tau) 
\big[\partial_{\tau} + \mu - E_{\sigma}(i\nabla)\big] \psi_{\sigma}({\bm x},\tau), \label{model2} \\
&S_{\phi} =  \!\ \int \big[m^2_{0} |\phi|^2 + |\partial_{\tau}\phi|^2 
+ c^2_0 |{\bm \nabla}\phi|^2 + \frac{\lambda_{0}}{4} \big(|\phi|^2\big)^2\big],  \label{model3} \\
&S_{\psi-\phi} = g_0 \int \!\ \big[\phi^{\dagger} \psi^{\dagger}_{+}\psi_{-} 
+ \phi \psi^{\dagger}_{-} \psi_{+}\big],  \label{model4} \\
&\int \equiv \int d\tau \int d^d{\bm x}. \nonumber
\end{align}
The coupled system has  global U(1) symmetries;
\begin{eqnarray}
\left\{\begin{array}{l}
{\rm (i)} \!\ \!\ \!\ \psi^{\dagger}_{\pm} \rightarrow \psi^{\dagger}_{\pm} e^{\pm i\theta}, \ \ 
 \psi_{\pm} \rightarrow \psi_{\pm} e^{\mp i\theta}, 
 \ \ \phi^{\dagger} \rightarrow \phi^{\dagger} e^{-2i\theta}. \\
{\rm (ii)} \!\ \!\ \!\  \psi^{\dagger}_{\pm} \rightarrow \psi^{\dagger}_{\pm} e^{i\theta}, \ \   
\psi_{\pm} \rightarrow \psi_{\pm} e^{-i\theta}. \\ 
\end{array}\right. 
 \label{global-gauge-symmetry}
\end{eqnarray}
The kinetic energy of the free fermion part is isotropic in the $d$-dimensional space, and an 
momentum-energy dispersion for the two fermion bands depend only on a norm of 
momentum, $E_{\sigma}(|{\bm k}|)$ ($\sigma=\pm$). The two fermion's energy 
bands are an electron-type band ($\sigma=+$) and hole-type band ($\sigma=-$) respectively. 
The chemical potential for the fermions is fine-tuned to a charge neutrality point (CNP), 
where the isotropic Fermi surface of the electron and that of the hole 
become identical to each other. The two momentum-energy dispersions are linearized 
around the isotropic Fermi surface. The fermion part is described by a free theory with 
the dynamical exponent $z=1$;
\begin{align}
S_{\psi} &= \int \frac{d^{d+1}k}{(2\pi)^{d+1}} 
\psi^{\dagger}_{+}(k) \big[i\omega - v_{+0} l \big] \psi_{+}(k) \nonumber \\
& \hspace{1cm} +  \int \frac{d^{d+1}k}{(2\pi)^{d+1}}
\psi^{\dagger}_{-}(k) \big[i\omega + v_{-0} l \big] \psi_{-}(k), \nonumber \\
& \equiv   \sum_{\sigma=\pm} \int_{\Omega,\omega,l} \psi^{\dagger}_{\sigma} (k) 
\big[i\omega - \sigma v_{\sigma 0} l\big] \psi_{\sigma}(k), 
\label{model5}
\end{align}  
and 
\begin{align}
\int_k \equiv k^{d-1}_{F} \int \frac{d^{d-1}\hat{\Omega}}{(2\pi)^{d-1}} 
\int^{\overline{\Lambda}_F}_{-\overline{\Lambda}_F} \frac{d\omega}{2\pi} 
\int^{\Lambda_F}_{-\Lambda_F} \frac{dl}{2\pi},  \label{k-int}
\end{align} 
with $v_{+0} > 0$, $v_{-0}>0$, $k\equiv (\omega,{\bm k})$ and 
${\bm k} = (k_F+l) \hat{\Omega}$. $k_F$ 
is the Fermi wavelength of the isotropic Fermi surface and $l$ is the perpendicular 
component of the $d$-dimensional momentum, and $\hat{\Omega}$ is a unit vector in the 
$d$-dimensional momentum space. 
In the presence of the isotropic Fermi surface, the $d$-dimensional momentum integral is decomposed 
into one-dimensional integral over $l$ and $(d-1)$-dimensional integral over 
$\hat{\Omega}$. $\Lambda_F$ and $\overline{\Lambda}_F$ 
are high energy (ultraviolet) cutoff for the one-dimensional $l$-integral and 
the frequency ($\omega$) integral respectively. 

The boson part and the coupling part of the action are given in the momentum-frequency space,
\begin{align}
S_{\phi} &= \int_{q} \big(m^2_{0} + c^2_0 {\bm q}^2 + \varepsilon^2 \big) 
\phi^{\dagger}(q) \phi(q) \nonumber \\
& \hspace{-0.9cm} + \frac{\lambda_{0}}{4} \int_{q_1,q_2,q_3} 
\big(\phi^{\dagger}(q_1) \phi(q_2)\big)\big(\phi^{\dagger}(q_3) \phi(q_1+q_3-q_2)\big),  \label{model6} \\
S_{\psi-\phi} &= g_0 \int_{k,q}  \big[\phi^{\dagger}(q) \psi^{\dagger}_{+}(k)\psi_{-}(k+q) 
+ {\rm h.c.} \big], \label{model7}
\end{align}
with $q \equiv (\varepsilon,{\bm q})$,  $q_j \equiv (\varepsilon_j,{\bm q}_j)$ ($j=1,2,\cdots$), and 
\begin{align} 
&\int_q \equiv \int_{|{\bm q}|<\Lambda_B} \frac{d^d{\bm q}}{(2\pi)^d}
\int^{\overline{\Lambda}_B}_{-\overline{\Lambda}_B}\frac{d\varepsilon}{2\pi}, \label{q-int} \\
&\int_{k,q} \equiv \int_{k} \int_{q}, \ \  
\int_{q_1,q_2,q_3} \equiv \int_{q_1} \int_{q_2} \int_{q_3}. \label{kq-int}
\end{align} 
Here $\Lambda_B$ and $\overline{\Lambda}_B$ are the ultraviolet (UV) cutoff for the boson momentum 
and frequency. 

\begin{figure}[t]
	\centering
	\includegraphics[width=0.8\linewidth]{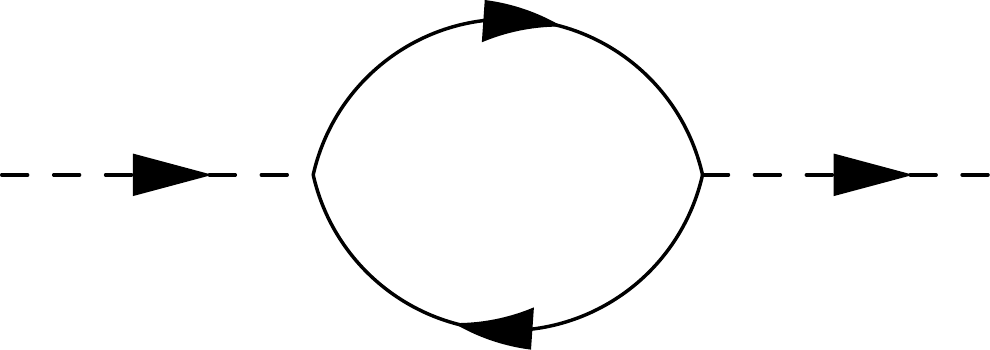}
	\caption{Polarization function in the interband channel. Two solid lines with arrow 
stand for the fermion Green's function in particle and hole channel respectively. The dotted line 
stand for the boson Green's function.}
	\label{sfig:0}
\end{figure}

A ground-state phase diagram of the coupled system possibly 
comprises of two phases; the SSB phase with $\langle \psi^{\dagger}_{+}\psi_{-}\rangle \ne 0$ 
(band insulator phase) and U(1) symmetric phase 
(multiple-band FL phase). At the exact charge neutrality point (CNP), a polarization function 
in an interband channel (Fig.~\ref{sfig:0}) has an infrared (IR) logarithmic singularity. 
The singularity in combination with any small Yukawa coupling $g_0$ could result 
in the low-$T$ band insulator phase with the SSB. A transition temperature $T_c$ is exponentially 
small for small $g_0$; $\log[T_c/\overline{\Lambda}_F]  \propto - 1/g^2_{0}$. In actual physical systems, 
however, the IR logarithmic singularity is weak enough that it is easily regularized by a tiny effect of 
other degree of freedom in low-energy scale. Such effects are disorder, an uncontrolled deviation 
from the CNP and/or spatially inhomegeneity of the carrier densities. Assuming these IR regularizations 
implicitly, we take it for granted the presence of the multiple-band FL phase and the quantum 
critical point (QCP) between the FL and the band insulator phases (Fig.~\ref{fig:1}).  Thereby,  
the phase transition between the two is primarily driven by a change of 
the boson mass. In the following, we will uncover universal critical properties at 
such QCP and in a high-$T$ side of the QCP (quantum critical regime; QCR). 

The Green's and vertex functions play the central role in the characterizations of quantum 
criticality. Let us call as $\Gamma^{(N_F,N_B)}$ and $G^{(N_F,N_B)}$ the 
vertex and Green's function respectively, that have $N_F$ external fermion lines and $N_B$ external 
boson lines. The Green functions 
are introduced together with the effective action $S_t$ as,
\begin{align}
&(2\pi)^{d+1} \delta^{d+1}(k-k^{\prime}) \delta_{\sigma,\sigma^{\prime}} 
G^{(2,0)}_{\sigma}(k) \equiv  \langle 
\psi_{\sigma} (k) \psi^{\dagger}_{\sigma^{\prime}}(k^{\prime}) \rangle, \nonumber \\
&(2\pi)^{d+1} \delta^{d+1}(q-q^{\prime})  G^{(0,2)}(q)  
\equiv \langle \phi(q) \phi^{\dagger}(q^{\prime}) \rangle, \nonumber  \\
&(2\pi)^{d+1} \delta^{d+1}(k^{\prime}-k-q) G^{(2,1)}_{+}(k,k+q;q) \nonumber \\
&\hspace{3.8cm} \equiv \langle \psi_{+}(k) \psi^{\dagger}_{-}(k^{\prime})\phi(q) \rangle, \nonumber \\ 
& (2\pi)^{d+1} \delta^{d+1}(k^{\prime}-k+q) G^{(2,1)}_{-}(k,k-q;q) \nonumber \\
& \hspace{3.8cm} \equiv \langle \psi_{-}(k) \psi^{\dagger}_{+}(k^{\prime})\phi^{\dagger}(q) \rangle, \nonumber \\
& (2\pi)^{d+1} \delta^{d+1}(q_4+q_3-q_2-q_1) G^{(0,4)}(q_1,q_2,q_3) \nonumber \\
&\hspace{3.2cm} \equiv \langle \phi(q_4) \phi(q_3)\phi^{\dagger}(q_2) \phi^{\dagger}(q_1) \rangle, \label{Green} 
\end{align} 
and 
\begin{align}
\langle \cdots \rangle &\equiv \frac{1}{Z} 
\int D\phi^{\dagger} D\phi D\psi^{\dagger} D\psi \!\ e^{-S_t} \cdots, \nonumber \\
Z &\equiv \int D \phi^{\dagger} D\phi D\psi^{\dagger} D\psi  \!\ e^{-S_t}. \label{average}
\end{align}
Here the band index $\sigma,\sigma^{\prime}=\pm$ in 
Eq.~(\ref{Green}) is an additional subscript, that distinguishes 
different functions for same $N_F$ and $N_B$. Because of the U(1) gauge symmetry 
Eq.~(\ref{global-gauge-symmetry}), $G^{(2,1)}$ has only two 
different functions; $G^{(2,1)}_{\pm}$. 
$G^{(0,4)}$ comprises of connected and disconnected 
parts. The disconnected part is given by products of 
two $G^{(0,2)}$. The vertex functions are obtained from an amputation of 
the one-particle irreducible (1PI) parts of the connected Green's functions;
\begin{align}
&G^{(2,0)}_{\sigma}(k) \Gamma^{(2,0)}_{\sigma}(k) \equiv 1, \ \ 
G^{(0,2)}(q) \Gamma^{(0,2)}(q)  \equiv 1, \nonumber \\
&G^{(2,1)}_{\pm}(k,k\pm q;q)  \equiv \nonumber \\
& \hspace{1cm} G^{(2,0)}_{\pm}(k) G^{(2,0)}_{\mp}(k\pm q) 
G^{(0,2)}(q) \Gamma^{(2,1)}_{\pm}(k,k\pm q;q), \nonumber \\
&G^{(0,4)}(q_1,q_2,q_3)  \equiv G^{(0,2)}(q_1) G^{(0,2)}(q_2) G^{(0,2)}(q_3)  \nonumber \\
& \hspace{0.8cm} 
 \times  G^{(0,2)}(q_1+q_2-q_3) \Gamma^{(0,4)}(q_1,q_2,q_3) 
+ \cdots.  \label{amputated1PI}
\end{align}
Here `$\cdots$' in the last line stands for the disconnected part of $G^{(0,4)}$.

\subsection{renormalizability}
For the effective model of Eqs.~(\ref{model5},\ref{model6},\ref{model7}), the vertex function  
$\Gamma^{(N_F,N_B)}$ has potentially an ultraviolet (UV) divergence in the power of a large UV cutoff $\Lambda$, 
which is either $\Lambda_B$, $\overline{\Lambda}_B$, $\Lambda_F$ or $\overline{\Lambda}_F$ defined in 
Eqs.~(\ref{k-int},\ref{q-int},\ref{kq-int}). 
The degree of the UV divergence depends on the number of the external lines, $N_F$ and $N_B$, and 
the spatial dimension $d$. By the field theory, 
$\Gamma^{(N_F,N_B)}$ is given by a sum of amputated Feynman diagrams for the 1PI connected 
$G^{(N_F,N_B)}$. Each amputated Feynman diagram is given by an 
integral over $L$-number of internal $D$-dimensional momenta (frequency and momentum) with 
$D \equiv d+1$. The integrand is a product among $V_F$-number of internal fermion lines (fermion Green's functions), 
$V_B$ internal boson lines (boson Green's functions), $V_1$ vertices of the Yukawa couplings ($g_0$) and 
$V_2$ vertices of the $\phi^4$ coupling ($\lambda_{0}$). Such an integral can have a 
UV divergence with respect to the large UV cutoff $\Lambda$. The degree of the divergence can 
be superficially evaluated by a dimensional counting of the integral~\cite{peskin-schroeder,amit}. 
From the dimensional counting, the superficial degree of the UV divergence, $M$, is 
bounded from above by $D_{\rm max}= DL - V_F-2V_B$; $M\le D_{\rm max}$. 
Here the equality holds true for those amputated Feynman diagrams which do not have any integrals over 
internal fermion momenta. The inequality applies for those Feynman diagrams which 
have the integrals over the internal fermion momenta (see also below). A sum of the internal and 
external boson lines in the connected $G^{(N_F,N_B)}$ is given by 
$V_1/2+2V_2+N_B/2 \equiv V_B+N_B$, while a sum 
of the internal and external fermion lines is by $V_1+N_F/2 \equiv V_F+N_F$. In terms of  these two 
identities together with 
$L=V_F+V_B-V_1-V_2+1$, the dependences on $V_F$ and $V_B$ in $D_{\rm max}$ and $L$ can be eliminated,   
\begin{align}
D_{\rm max} = & D + \frac{D-4}{2} V_1 + (D-4) V_2 \nonumber \\
& \ \ - \frac{D-1}{2} N_F - \frac{D-2}{2} N_B,  \label{dmax} 
\end{align} 
and 
\begin{align}
L = \frac{V_1}{2} + V_2 - \frac{N_F}{2} - \frac{N_B}{2} + 1. \label{Lexp}
\end{align}
Eq.~(\ref{dmax}) suggests that $D=4$ ($d=3$) is an upper critical dimension, 
where the effective model of Eqs.~(\ref{model5},\ref{model6},\ref{model7}) is renormalizable. 
Eq.~(\ref{Lexp}) shows that for fixed $N_E$ and $N_B$, the number of the internal integral variables 
$L$ increases with a unit of either $V_1=2$ or $V_2=1$.

This paper focuses on the upper critical dimension of the effective model, 
where only the following four vertex functions have potentially the UV divergences at $D=4$, 
\begin{align}
\Gamma^{(2,0)}, \!\ \Gamma^{(0,2)}, \!\ \Gamma^{(2,1)}, \!\  \Gamma^{(0,4)}.  \label{four-vertex}
\end{align} 
From Eq.~(\ref{dmax}), their superficial degrees of the UV divergences in the power of the UV cutoff $\Lambda$ 
are evaluated (at most) $1$, $2$, $0$ and $0$ respectively. Note that both $\Gamma^{(1,0)}$, 
$\Gamma^{(0,1)}$ and $\Gamma^{(1,1)}$ are zero because of the gauge 
symmetries; Eq.~(\ref{global-gauge-symmetry}).  
In the next section, we will use a renormalized perturbation theory and eliminate the dependences of 
the UV cutoff from these vertex functions by way of renormalizations of field operator amplitudes, 
boson mass, fermion velocities, boson velocity, Yukawa coupling and the $\phi^4$ coupling. 

Before closing this section, let us explain how the Fermi 
wavelength $k_F$ is treated in this paper. The effective model implicitly assumes that the 
Fermi wavelength $k_F$ is much larger than the UV cutoff $\Lambda$ and the UV cutoff 
is much larger than fermion and boson momenta in Eqs.~(\ref{model5},\ref{model6},\ref{model7});
\begin{align}
|{\bm q}|,\varepsilon, \omega, l \ll \Lambda_B, \overline{\Lambda}_B, \Lambda_F,\overline{\Lambda}_F \ll k_F. \label{large-kF}
\end{align}
The renormalized theory shall be free from the UV cutoff $\Lambda$ (being $\Lambda_B$, $\overline{\Lambda}_B$, 
$\Lambda_F$ or  
$\overline{\Lambda}_F$), where $|{\bm q}|/\Lambda$, $\varepsilon/\Lambda$, $l/\Lambda$, $\omega/\Lambda$ 
are regarded as infinitesimally small and set to zero in the renormalized vertex functions. 
Meanwhile, we will allow the renormalized functions to 
depend not only on the boson and fermion momenta ${\bm q}$, $\varepsilon$, $\omega$, $l$ 
but also on the Fermi wavelength $k_F$. That says, the Fermi wavelength 
is treated as a finite (although the largest) measurable physical quantity. 
%while the UV cutoff $\Lambda$ 
%are not treated as measurable physical quantities. This is because the UV cutoff simply defines an energy and 
%momentum scale below which the effective continuum theory of Eqs.~(\ref{model5},\ref{model6},\ref{model7}) 
%becomes a valid description for a certain lattice model. 

To enable such an exceptional treatment of $k_F$ in spite of 
$|{\bm q}|$, $\varepsilon$, $\omega$, $l$ $\ll \Lambda$ $\ll k_F$, 
it is important to note that all the vertex functions in Eq.~(\ref{four-vertex}) 
can be expanded in the power of $k^{d-1}_F$; 
\begin{align}
\Gamma^{(N_F,N_B)} = & \Gamma^{(N_F,N_B;0)} + k^{d-1}_F \Gamma^{(N_F,N_B;1)} \nonumber \\
& \hspace{-0.8cm} + k^{2(d-1)}_F 
\Gamma^{(N_F,N_B;2)} + \cdots,  \!\ \!\ \!\ \!\ (N_F=0,2). \label{ex-g}
\end{align}
Here $ k^{n(d-1)}_F \Gamma^{(N_F,N_B;n)}$ in the right hand side is a sum of all those amputated 1PI Feynman diagrams for 
$\Gamma^{(N_F,N_B)}$ that have $n$ numbers of internal closed loops formed by fermion lines (`closed fermion 
loops'). Importantly, {\it it has no $k_F$-dependence other than the overall factor, $k^{n(d-1)}_F$, for $N_F=0,2$}. 
To see this, let us consider how to assign the $L$-number of the internal integral variables to the 
$(V_F+V_B)$-number of the internal lines in an amputated 1PI Feynman diagram with $n$-number 
of the closed fermion loops. The most natural way of doing this 
in the case of $\Lambda \ll k_F$  
is as follows. For each closed fermion loop, say the $j$-th closed fermion loop with $j=1,\cdots,n$, 
assign an integral variable of fermion momentum to one and {\it only} one internal fermion line in the loop, 
say $(\omega_j,{\bm k}_j)$ with ${\bm k}_j=(k_F+l_j) \hat{\Omega}_j$. Meanwhile, assign the 
other $(L-n)$-number of 
the integral variables to the internal boson lines, such that momentum and frequency conservations  
are preserved at every vertex. The momenta of the other internal fermion lines in the $j$-th 
closed fermion loop shall be given by a sum of $(\omega_j,{\bm k}_j)$ and 
the boson momenta (either internal or external). 
For $N_F=2$ with an open fermion line with two external fermion points, one of the two external 
fermions is given by $(\omega,{\bm k})$, while the other external 
fermion is given by a sum of $(\omega,{\bm k})$ and external boson momenta (if any). 
Thereby, the momenta of the internal fermions in the open fermion line shall be given by a sum of the external 
fermion momentum $(\omega,{\bm k})$ and (either internal or external) boson momenta. With this way of 
the assignment of the integral variables, the integrand clearly has no $k_F$-dependence in the case of 
$\Lambda \ll k_F$. The only $k_F$-dependence in the integral appears through 
the overall factor, $k^{n(d-1)}_F$;
\begin{widetext}
\begin{align}
k^{n(d-1)}_{F}\Gamma^{(N_F \le 2,N_B;n)} =& k^{n(d-1)}_F\sum_{L}  \int \frac{d^{d-1}\hat{\Omega}_1}{(2\pi)^{d-1}} 
\int_{\Lambda} \frac{d\omega_1dl_1}{(2\pi)^2}
\cdots   \int \frac{d^{d-1}\hat{\Omega}_n}{(2\pi)^{d-1}}  \int_{\Lambda} \frac{d\omega_ndl_n}{(2\pi)^2} \nonumber \\
& \hspace{2.8cm} 
\int_{\Lambda} \frac{d^{d+1} q_1}{(2\pi)^{d+1}} \cdots \int_{\Lambda} \frac{d^{d+1} q_{L-n}}{(2\pi)^{d+1}} 
\!\ \big({\rm integrand} \!\ \!\ {\rm that} \!\ \!\ {\rm is} \!\  \!\ {\rm free} \!\ \!\  {\rm from} \!\ \!\ k_F \big).  \label{Gamma-loop-exp} 
\end{align}
\end{widetext}
%Here $(\omega_j,{\bm k}_j)$ with ${\bm k}_j=(k_F+l_j) \hat{\Omega}_j$ is for an internal fermion line in the $j$-th closed 
%fermion loop ($j=1,\cdots,n$). 
In $\Gamma^{(N_F,N_B;n)}$ thus given, the superficial degree of the UV divergence with 
respect to the power in $\Lambda$ is smaller than that of $\Gamma^{(N_F,N_B;0)}$ by $n(d-1)$.  
\begin{eqnarray}
\left\{\begin{array}{l}
\big[ \Gamma^{(N_F,N_B;0)} \big] = DL - V_F-2V_B \equiv D_{\rm max}, \\ 
\big[ \Gamma^{(N_F,N_B;1)} \big] = D_{\rm max} - (d-1),  \\
\big[ \Gamma^{(N_F,N_B;2)} \big] = D_{\rm max} - 2(d-1), \ \ \cdots, \ \  \\  
\big[ \Gamma^{(N_F,N_B;n)} \big] = D_{\rm max} - n(d-1), \ \  \cdots. \\
\end{array}\right. \label{UV-divergence-n}
\end{eqnarray} 
The integer in the right hand side refers to the superficial degree of the UV divergence in the left hand side. 
When the integer in the right hand side 
is less than $0$, the left hand side should be regarded as on the order of ${\cal O}(1)$ in the power of 
the UV cutoff $\Lambda$.  

Thanks to the analytic feature of the vertex functions as functions of $k_F$, Eq.~(\ref{ex-g}), the dependence 
on the UV cutoff $\Lambda$ in the vertex functions can be removed at every order in $k^{d-1}_F$. 
Generally speaking, higher orders in $k^{d-1}_F$ mean smaller numbers of the superficial 
degree of the UV divergences in $\Lambda$; Eq.~(\ref{UV-divergence-n}). Thus, it is usually the case 
that only a few low-order powers in $k^{d-1}_F$ need to be considered in the expansion. 
In this paper, we consider $d=3$ and 
carry out the renormalizations of the zeroth order of $\Gamma^{(2,0)}$ whose UV divergence degree 
in $\Lambda$ is 1 (Secs.~IV,V), and the zeroth and first order of $\Gamma^{(0,2)}$ whose UV 
divergence degrees in $\Lambda$ are $2$ and $0$ respectively (Sec.~IV, Appendix).
To this end, renormalization conditions for the vertex functions will be expanded in the 
powers of $k^{2}_F$ ($d=3$). See Eq.~(\ref{boundary}), where a renormalization condition on 
the boson self-energy part is given up to the first order in $k^2_F$. 

A dimensional analysis suggests that renormalized 
$\Gamma^{(N_F,N_B;n)}$ thus obtained has a different scaling form for different $n$; 
see Eq.~(\ref{scaling-form}) for example. In each of these scaling forms, the boson 
and fermion momenta, $q$, $\varepsilon$, $\omega$, $l$, must be treated as much smaller 
than $k_F$; Eq.~(\ref{large-kF}). In Sec.~V and Appendix, asymptotic behaviours of the boson and 
fermion spectral functions will be clarified for such small momentum and frequency region. 
It turns out that the asymptotic behaviours with respect to the small momenta are free from $k_F$, 
while overall factors of the spectral weights can have an explicit dependence on $k_F$ (Appendix).        
         
\section{Renormalization}
In this section, we will eliminate the UV divergences in all the vertex functions in Eq.~(\ref{four-vertex}), 
while include the UV cutoff dependences into the renormalizations of the field operator amplitudes 
and physical quantities. To see how many physical 
quantities are needed for this purpose, let us first expand their 
superficial degrees of the UV divergence in terms of the external momentum and 
frequency, 
\begin{align}
\Gamma^{(2,0)}_{\sigma} (i\omega,{\bm k}) &=  \ldots \Lambda + \ldots \ln \Lambda \!\ i\omega 
+ \ldots \ln \Lambda \!\ l,  \nonumber \\
\Gamma^{(0,2)}(i\varepsilon,{\bm q}) &= \ldots \Lambda^2 + \ldots 
\ln \Lambda \!\ \varepsilon^2 + \ldots \ln \Lambda \!\ {\bm q}^2,  \nonumber \\
\Gamma^{(2,1)}_{\sigma}(\cdots) &= \ldots \ln \Lambda + {\cal O}(\omega,l,q), \nonumber \\
\Gamma^{(0,4)}(\cdots) &= \ldots \ln \Lambda + {\cal O}(\omega,l,q),  \nonumber 
\end{align}
with $\sigma=\pm$ and 
${\bm k} = (k_F+l) \hat\Omega$. Here the UV cutoff $\Lambda$ is either 
$\Lambda_F$, $\overline{\Lambda}_F$ 
$\Lambda_B$ or $\overline{\Lambda}_B$. 
For each of the two fermion bands ($\sigma=\pm$), 
$\Gamma^{(2,0)}_{\sigma}(i\omega,{\bm k})$ has a linear 
divergence in $\Lambda$, that corresponds to an energy shift of each fermion energy band. 
$\Gamma^{(2,0)}_{\sigma}(i\omega,{\bm k})$ also 
has the logarithmic UV divergences in its linear terms of its frequency $\omega$ 
and the one-dimensional momentum $l$. These logarithmic divergences correspond to the 
renormalizations of the fermion field operator amplitude and the Fermi velocity respectively. 
The two-point boson vertex function has the logarithmic divergences in the quadratic 
terms of its frequency and momentum. They correspond to the renormalizations of the boson field 
operator amplitude and boson velocity. In total, the theory in $d=3$ ($D=4$) has eleven kinds of the 
UV divergences with respect to the UV cutoff $\Lambda$. To renormalize all of them, one 
should use the shifts of the Fermi 
wavelengths, $\delta k_{F\sigma}$ ($\sigma=\pm$), renormalizations of the fermion field 
amplitudes, $Z_{\sigma}$ $(\sigma=\pm)$, the fermion velocities, $v_{\sigma}$ ($\sigma=\pm$), 
the boson field amplitude, $Z_{\phi}$, boson mass, $m^2$, boson velocity, $c^2$, 
the Yukawa coupling $g$ and the $\phi^4$ coupling $\lambda$. As shown below (Sec.~IVC),  
the $\Lambda$-linear terms in $\Gamma^{(2,0)}_{\pm}$ do not appear in the 
following perturbation theory calculation; $\delta k_{F\sigma}=0$. 
We thus consider only the renormalizations of 
$Z_{\sigma}$, $v_{\sigma}$ $(\sigma=\pm)$, $Z_{\phi}$, $m^2$, $c^2$, $g$ and $\lambda$, while taking the 
Fermi wavelength to be always $k_F$; $k_{F\sigma}=k_{F0}+\delta k_{F\sigma}\equiv k_F$ 
($\sigma=\pm$, $k_{F0}$ is a bare Fermi wavelength).     
 
\subsection{renormalized Green's and vertex functions, effective action and counterterms part}

Based on this observation, we employ the minimal substraction scheme~\cite{amit,peskin-schroeder},
and decompose the bare action $S_t$ into the effective 
action $S_E$ and counterterms part $S_c$, 
\begin{align}
S_t = & S_E + S_c \nonumber \\
S_E=& \sum_{\sigma=\pm} \int_{\Omega,\omega,l} \overline{\psi}^{\dagger}_{\sigma} (k) 
\big[i\omega - \sigma v_{\sigma} l\big] \overline{\psi}_{\sigma}(k) \nonumber \\ 
& + \int_{q} \big(m^2 + c^2 {\bm q}^2 + \varepsilon^2 \big) 
\overline{\phi}^{\dagger}(q) \overline{\phi}(q) \nonumber \\
&  + \frac{\lambda}{4} \int_{q_1,q_2,q_3} 
\big(\overline{\phi}^{\dagger}(q_1) \overline{\phi}(q_2)\big)\big(\overline{\phi}^{\dagger}(q_3) \overline{\phi}(q_1+q_3-q_2)\big) 
\nonumber \\
& +  g \int_{k,q}  \big[\overline{\phi}^{\dagger}(q) \overline{\psi}^{\dagger}_{+}(k)\overline{\psi}_{-}(k+q) 
+ {\rm c.c.}\big] \label{se} \\
S_c  = & \sum_{\sigma} \int_{\Omega,\omega,l} \overline{\psi}^{\dagger}_{\sigma}  
\big[i \delta_{\sigma} \!\ \omega - \sigma \delta v_{\sigma} \!\ l\big] \overline{\psi}_{\sigma} \nonumber \\
& + \int_{q} \big(\delta m^2 + \delta c^2 \!\ {\bm q}^2 + \delta_{\phi} \!\ \varepsilon^2 \big) 
\overline{\phi}^{\dagger} \overline{\phi} \nonumber \\
& \hspace{-0.5cm} + \frac{\delta \lambda}{4} \int_{q_1,q_2,q_3} 
\big(\overline{\phi}^{\dagger} \overline{\phi}\big)\big(\overline{\phi}^{\dagger} \overline{\phi}\big) 
 +  \delta g \int_{k,q}  \big[\overline{\phi}^{\dagger} \overline{\psi}^{\dagger}_{+}\overline{\psi}_{-} 
+ {\rm c.c.}\big]. \label{sc}
\end{align}
Here renormalized field operators ($\overline{\psi}_{+}$, $\overline{\psi}_{-}$, $\overline{\phi}$) and 
renormalized physical quantities ($c^2, v_{+}, v_{-}, m^2, g, \lambda$) are related with their bare and 
counterparts by,  
\begin{align}
&\psi_{\sigma} \equiv \sqrt{Z_{\sigma}} \overline{\psi}_{\sigma}, 
%\  \psi_{-} \equiv \sqrt{Z_{-}} \overline{\psi}_{-}, \ 
\phi \equiv \sqrt{Z_{\phi}} \overline{\phi},  \ c^2_0 \equiv 
Z_{c^2} c^2, \  v_{\sigma0} \equiv Z_{v_\sigma} v_\sigma,  \nonumber \\ 
%&v_{-0} \equiv Z_{v_{-}} v_{-}, 
& \ m^2_0 \equiv Z_m m^2,  \ g_0 \equiv \frac{Z_g}{\sqrt{Z_{\phi}Z_{+}Z_{-}}} g, \   \lambda_0 \equiv
\frac{Z_{\lambda}}{Z^2_{\phi}} \lambda,  \label{MSS1}
\end{align}
and 
\begin{align}
&Z_{\sigma} \equiv 1 + \delta_{\sigma}, \!\ %\!\ Z_{-} \equiv 1 + \delta_{-}, \!\ 
Z_{\phi} \equiv 1 + \delta_{\phi}, \!\ 
Z_{\sigma} Z_{v_{\sigma}} v_{\sigma} \equiv 
v_{\sigma} + \delta v_{\sigma}, \nonumber \\ %\!\ Z_{-} Z_{v_{-}} v_{-} \equiv v_{-} + \delta v_{-}, \nonumber \\
& Z_{\phi} Z_{c^2} c^2 \equiv c^2 + \delta c^2, \!\ Z_{\phi} Z_{m} m^2 \equiv m^2 + \delta m^2, \nonumber \\
& Z_{g} g \equiv g+ \delta g, \!\ Z_{\lambda} \lambda \equiv \lambda + \delta \lambda. \label{MSS2}
\end{align}
with $\sigma=\pm$.

The objective of the renormalization theory is to include all the UV divergences in the bare vertex functions 
into the renormalizations of the field operator amplitudes ($Z_{\pm}$, $Z_{\phi}$) and bare quantities 
($v_{\pm0}$, $c^2_0$, $m_0$, $g_0$, $\lambda_0$), while making renormalized vertex functions 
and renormalized physical quantities ($v_{\pm}$, $c^2$, $m$, $g$, $\lambda$) to be free from the UV cutoff. 
Thereby, the renormalized Green's and vertex functions are 
defined in the same way as Eqs.~(\ref{Green},\ref{amputated1PI}) with $\psi_{\pm}$ and $\phi$ being 
replaced by $\overline{\psi}_{\pm}$ and $\overline{\phi}$,  e.g. 
\begin{align}
&(2\pi)^4 \delta^4(k-k^{\prime}) \delta_{\sigma,\sigma^{\prime}} \overline{G}^{(2,0)}_{\sigma}(k) \equiv  \langle 
\overline{\psi}_\sigma (k) \overline{\psi}^{\dagger}_{\sigma^{\prime}}(k^{\prime}) \rangle, \nonumber \\
&  (2\pi)^4 \delta^4(q-q^{\prime})  
\overline{G}^{(0,2)}(q) \equiv \langle \overline{\phi}(q) \overline{\phi}^{\dagger}(q^{\prime}) \rangle, \nonumber \\
& \overline{G}^{(2,0)}_{\sigma}(k) 
\overline{\Gamma}^{(2,0)}_{\sigma}(k) = 1, \!\    
\overline{G}^{(0,2)}(q) \overline{\Gamma}^{(0,2)}(q) = 1, \nonumber  \\
& (2\pi)^4 \delta^4(k^{\prime}-k-q) \overline{G}^{(2,1)}_{+}(k,k+q;q) \equiv 
\langle \overline{\psi}_{+}(k) \overline{\psi}^{\dagger}_{-}(k^{\prime})\overline{\phi}(q) \rangle \nonumber \\
& = \overline{G}^{(2,0)}_{+}(k) \overline{G}^{(2,0)}_{-}(k+q) \overline{G}^{(0,2)}(q) 
\overline{\Gamma}^{(2,1)}_{+}(k,k+q;q), \nonumber \\
& (2\pi)^4 \delta^4(q_4+q_3-q_2-q_1) \overline{G}^{(0,4)}(q_1,q_2,q_3) \nonumber \\
& \hspace{3cm} \equiv \langle \overline{\phi}(q_4) \overline{\phi}(q_3)\overline{\phi}^{\dagger}(q_2) 
\overline{\phi}^{\dagger}(q_1) \rangle, \nonumber \\  
&\overline{G}^{(0,4)}(q_1,q_2,q_3) = \overline{G}^{(0,2)}(q_1) 
\overline{G}^{(0,2)}(q_2) \overline{G}^{(0,2)}(q_3) \nonumber \\
& \ \  \times  \overline{G}^{(0,2)}(q_1+q_2-q_3) 
\overline{\Gamma}^{(0,4)}(q_1,q_2,q_3)  + \cdots \label{renormalizedGreen}
\end{align} 
Note that $\langle \cdots \rangle$ in the right hand sides is defined in 
Eq.~(\ref{average}) with the same $S_t$ as in Eq.~(\ref{model1}) with $d=3$. 
Thus, the renormalized Green's and vertex functions differ from 
their bare functions only by $Z_{\phi}$ and/or $Z_{\pm}$,  
\begin{align}
&G^{(2,0)}_{\sigma}(k) = Z_{\sigma} \overline{G}^{(2,0)}_{\sigma}(k), \ \ 
G^{(0,2)}(q) = Z_{\phi} \overline{G}^{(0,2)}(q), \nonumber \\ 
&\overline{\Gamma}^{(2,1)}(k,k+q:q)= Z^{\frac{1}{2}}_{+} Z^{\frac{1}{2}}_{-} Z^{\frac12}_{\phi} 
\Gamma^{(2,1)}(k,k+q;q), \nonumber \\
&\overline{\Gamma}^{(0,4)}(q_1,q_2,q_3) = Z^2_{\phi} \Gamma^{(0,4)}(q_1,q_2,q_3). 
\label{relation} 
\end{align} 
In the following, we will omit the superscripts of 
$(N_F,N_B)$ from the Green's and vertex functions and use the following 
simplified notations,
\begin{align} 
&\overline{G}^{(2,0)}_{\sigma} \rightarrow \overline{G}_{\sigma},  \ \ 
\overline{G}^{(0,2)} \rightarrow \overline{G}_{\phi}, \ \  \overline{\Gamma}^{(2,0)}_{\sigma} 
\rightarrow \overline{G}^{-1}_{\sigma},  \nonumber \\ 
&\overline{\Gamma}^{(0,2)} \rightarrow \overline{G}^{-1}_{\phi}, \ \  
\overline{\Gamma}^{(2,1)}_{+} \rightarrow \overline{\Gamma}_g, \ \ 
\overline{\Gamma}^{(0,4)} \rightarrow \overline{\Gamma}_{\lambda},  
\end{align}  
with $\sigma=\pm$. 

\subsection{renormalization conditions}
To make the renormalized vertex functions to be free from the UV cutoff $\Lambda$, 
we impose the following conditions on the vertex functions, 
\begin{widetext}
\begin{eqnarray}
\left\{\begin{array}{l}
\overline{G}^{-1}_{\sigma}(\omega=l=0) = 0, \!\ 
\frac{\partial \overline{G}^{-1}_{\sigma}(\omega,l)}{\partial (i\omega)}\Big|_{v_{\sigma} l=0,\omega=K} = 1 \!\  \!\ 
(\sigma=\pm), \!\ \!\ 

\frac{\partial \overline{G}^{-1}_{+}(\omega,l)}{\partial (-v_{+} l)}\Big|_{v_{+} l=K,\omega=0} = 
\frac{\partial \overline{G}^{-1}_{-}(\omega,l)}{\partial (v_{-} l)}\Big|_{v_{-} l=K,\omega=0} = 1 \\ 
\overline{G}^{-1}_{\phi}(\varepsilon=Q,|{\bm q}| = 0) = m^2 + Q^2, \!\ 
\frac{\partial \overline{G}^{-1}_{\phi}(q)}{\partial \varepsilon}\Big|_{{\bm q}=0,\varepsilon=Q} = 
\frac{\partial \overline{G}^{-1}_{\phi}(q)}{\partial (c |{\bm q}|)}\Big|_{c|{\bm q}|=Q,\varepsilon=0} 
= 2Q + \frac{g^2}{\pi^2 (v_{+}+v_{-})} \frac{k^2_F}{Q}, \\  
%+ {\cal O}(g^4, g^2 \lambda, \lambda^2), \\   
%
\overline{\Gamma}_g(k,k+q;q)\Big|_{q=0,\omega=l=0} = - g, \!\ 
\overline{\Gamma}_{\lambda}(q_1,q_2,q_3)\Big|_{q_1=(P,0),q_2=(P/3,0),q_3=(2P/3,0)} = - \lambda. \\  
%+ {\cal O}(g^6,g^4\lambda,g^2 \lambda^2,\lambda^3),   \\
\end{array}\right. \label{boundary}
\end{eqnarray}
\end{widetext} 
%with $k=(\omega,{\bm k})$, ${\bm k}=(k_F+l)\hat{\Omega}$, and 
%$q=(\varepsilon,{\bm q})$. 
Importantly, the right hand sides of the conditions are free from the UV 
cutoff $\Lambda$. They depend only on physical quantities such as the Fermi wavelength $k_F$, 
renormalized boson mass, $m$, renormalized velocities, $v_{\pm}$, $c$, and renormalized 
coupling constants, $g$, $\lambda$. 

\begin{figure}[t]
	\centering
	\includegraphics[width=1.0\linewidth]{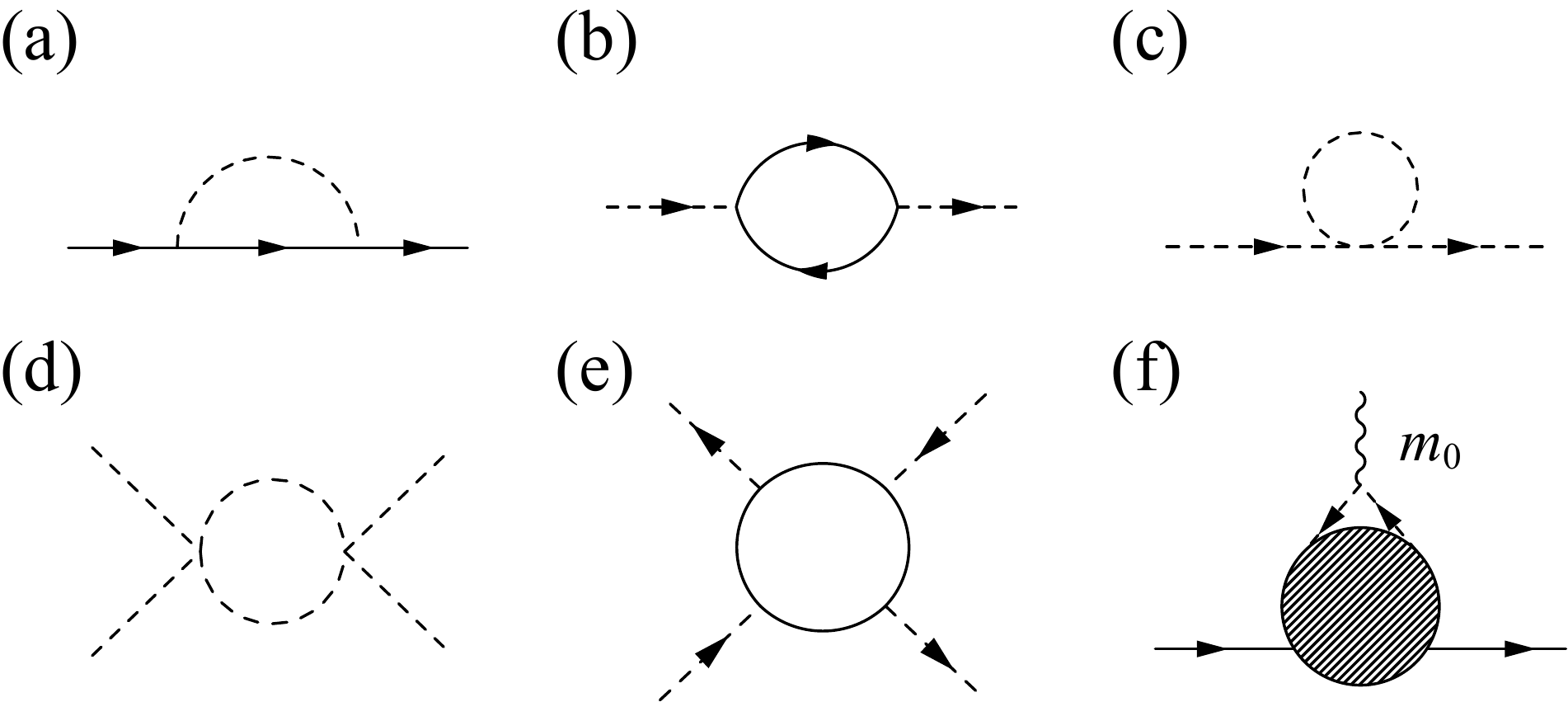}
	\caption{(a-e) One-loop Feynman diagrams included in the renormalized perturbation analysis. 
(f) $\Gamma^{(2,2)}(k,k;q,q)$. The solid/broken line is fermion/boson line respectively. A wavy 
line represents a boson mass $m_0$, that comes with a pair of boson lines.}
	\label{sfig:1}
\end{figure}

A 1-loop proper part of the fermion self energy has a weak infrared (IR) singularity 
(Fig.~\ref{sfig:1}(a); $\omega \log \omega$ in small external frequency $\omega$ or $l$; see Sec.~IVC). 
Thus, the condition on its derivatives is imposed at finite $\omega$ 
or $v_{\pm}l=K$. A 1-loop proper part of the boson self-energy has an IR logarithmic 
divergence, that comes from the fermion's polarization function (Fig.~\ref{sfig:1}(b); see Sec.~IVD). 
The polarization function has a closed fermion loop, so that it is proportional to $k^2_F$. 
When being taken a derivative of with respect to $q$, the IR 
logarithmic divergence leads to a IR linear divergence. To circumvent these IR divergences, we set 
the external boson frequency and momentum at finite value,  $\varepsilon$ or $c|{\bm q}|=Q$. 
Besides, we expand in the power of $k^2_F$ the right hand 
side of the condition for the derivatives of the boson self-energy. The zero-th order 
in the expansion would take care of the usual logarithmic UV 
divergence in the quadratic terms of the external boson momenta, that come from 2-loop corrections 
with ${\cal O}(\lambda^2)$ order. The first order in the expansion 
is for canceling the IR linear divergence from the fermion's polarization function. See 
Eq.~(\ref{ex-g}) and text around this equation for an explanation of the $k^2_F$-expansion.

A 1-loop four-point boson vertex function can have an IR quadratic divergence, depending on its four 
external boson momenta (Fig.~\ref{sfig:1}(e)). The IR quadratic divergence comes from an integral 
over an internal fermion momentum and frequency. We choose the external boson frequencies, 
in such a way that the one-loop integral over the fermion momentum becomes zero (See Sec.~IVE);
\begin{align}
&q_1 \equiv (\varepsilon_1,{\bm q}_1)= (P,{\bm 0}), \!\ q_2 = (P/3,{\bm 0}), \nonumber \\
&\!\ q_3 = (2P/3,{\bm 0}), \!\ q_4 = (2P/3,{\bm 0}).  
\label{four-point-boson-vertex}
\end{align}
For the other choice of the external boson frequencies, the one-loop integral has an IR 
quadratic divergence and in that case, we could modify the conditions. Namely, we could expand 
in the power of $k^2_F$ the right hand side of the condition for the four-point boson vertex function, 
and add the first order term in $k^2_F$, that cancels the 
IR quadratic divergent term.         

$Q$, $K$ and $P$ in Eq.~(\ref{boundary}) are external frequencies or 
momenta, at which the renormalization conditions are imposed on the 2-point fermion, 2-point boson and 
4-point boson vertex functions respectively. 
When these three energy scales were treated as three independent renormalization group (RG) 
scale variables, a $\beta$ function 
of the same physical quantity would depend on the type of the vertex function, for 
which a Callan-Symanzik (CS) equation with the $\beta$ function would be solved. Such is 
against our intuition of universality in critical phenomena. We thus set $Q$, $K$ and $P$ as the {\it same} RG 
scale variable $\kappa$, 
\begin{eqnarray}
Q = K = P \equiv \kappa.  \label{Q=K=P}
\end{eqnarray}   
The renormalization conditions in Eq.~(\ref{boundary}) eliminate all the UV divergences in the renormalized 
vertex functions, while the UV dependences are put into the renormalizations of the field 
operator amplitudes ($Z_{\pm}$, $Z_{\phi}$), and bare quantities 
($m^2_0,v_{+0},v_{-0}, c^2_0, g_0, \lambda_0$).  

In terms of the renormalized 
perturbation theory, the renormalization conditions shall be satisfied perturbatively in 
the Yukawa coupling $g$ and the $\phi^4$ coupling $\lambda$. To this end, we treat as 
perturbations $g$ and $\lambda$ in $S_E$ as well as the counterterms $S_{c}$ and treat the quadratic part 
in $S_E$ as the free part. Eq.~(\ref{Lexp}) shows that for fixed $N_F$ and $N_B$, the number of the internal 
integral variables ($L$) increases with a unit of $g^2$ or $\lambda$. We thus treat $g^2$ and $\lambda$ 
as the same order of the smallness in the perturbation analysis. The standard Dyson-Feynman 
perturbation theory leads to following one-loop terms ($L=1$)
for the amputated 1PI connected Green's functions (Fig.~\ref{sfig:1}(a-e)),  
\begin{align}
&\overline{G}^{-1}_{\pm}(k) = \overline{G}^{-1}_{\pm,0}(k) 
+ (\delta_{\pm} i\omega \mp \delta v_{\pm} \!\ l) \nonumber \\
&\hspace{0.5cm} - g^2 \int_{q} \overline{G}_{\mp,0}(k+q) 
\overline{G}_{\phi,0}(q)  + {\cal O}(g^4,g^2\lambda,\lambda^2),\label{1-loop-Gab} \\  
&\overline{G}^{-1}_{\phi}(q) = \overline{G}^{-1}_{\phi,0}(q) 
+ (\delta m^2 + \delta c^2 \!\ {\bm q}^2 + \delta_{\phi} \!\ \varepsilon^2) \nonumber \\ 
& \hspace{0.2cm} + g^2 \int_{k} \overline{G}_{+,0}(k+q) \overline{G}_{-,0}(k) + \lambda 
\int_{q^{\prime}} \overline{G}_{\phi,0}(q^{\prime}) \nonumber \\
&\hspace{1cm} + {\cal O}(g^4, g^2\lambda,\lambda^2), \label{1-loop-Gphi} \\
&\overline{\Gamma}_g(k,k+q;q) = - g - \delta g + {\cal O}(g^5,g^3\lambda,g \lambda^2), \label{1-loop-Gg} \\
&\overline{\Gamma}_{\lambda}(q_1,q_2,q_3) = - \lambda - \delta \lambda + \frac{\lambda^2}{2} 
\int_q \overline{G}_{\phi,0}(q) \overline{G}_{\phi,0}(q_1+q_2-q) \nonumber \\
& + \lambda^2 
\int_q \overline{G}_{\phi,0}(q) \Big(\overline{G}_{\phi,0}(q+q_2-q_3)  + \overline{G}_{\phi,0}(q+q_1-q_3)\Big) \nonumber \\
& - g^4 \int_{k}  \overline{G}_{+,0}(k) \overline{G}_{-,0}(k+q_1) \Big( 
\overline{G}_{+,0}(k+q_1-q_3) \nonumber \\
& \hspace{0.5cm} \times \overline{G}_{-,0}(k+q_1+q_2-q_3)  
+ \overline{G}_{+,0}(k-q_2+q_3) \nonumber \\
& \hspace{0.7cm} \times \overline{G}_{-,0}(k+q_3)\Big) + {\cal O}(\lambda^3,g^2 \lambda^2, g^4 \lambda,g^6),  
\label{1-loop-Glam}
\end{align}
where renormalized free Green's functions for fermions and boson are given by 
\begin{align}
\overline{G}^{-1}_{\pm,0}(k) = i \omega \mp v_{\pm} l, \ \ 
\overline{G}^{-1}_{\phi,0}(q) = m^2 + c^2 {\bm q}^2 + \varepsilon^2.    \label{free-re}
\end{align}
%with $q \equiv (\varepsilon,{\bm q})$, $k \equiv (\omega,{\bm k})$ and ${\bm k}=(k_F+l)\hat{\Omega}$.  
The integrals over $q$ or $k$ are defined in Eqs.~(\ref{k-int},\ref{q-int},\ref{kq-int}). All the one-loop 
terms except for the last term in Eq.~(\ref{1-loop-Glam}) have ultraviolet divergences.  The 
last term in Eq.~(\ref{1-loop-Glam}) has no influence on the $\beta$ function for $\lambda$; 
$\beta_{\lambda}(\cdots)$. $\beta_{\lambda}$ within the one-loop level is identical to 
that in the pure  $\phi^4$ theory. Note also that $\overline{\Gamma}_{g}(k,k+q;q)$ 
has no one-loop term beacuse of the Yukawa coupling; $\sigma_{\pm}\sigma_{\pm}\sigma_{\mp} = 0$. 

\subsection{Calculation of $\int_{q} \overline{G}_{\mp,0}(k+q) 
\overline{G}_{\phi,0}(q)$ at $d=3$} 
The 1-loop term in the fermion self-energy (Fig.~\ref{sfig:1}(a)) has no linear divergence in 
$\Lambda$, while it has the logarithmic divergence in large $\Lambda$. 
To see this, we take $\overline{\Lambda}_B$ to the infinite in Eq.~(\ref{q-int}) 
and see how the integral depends 
on the UV cutoff for the boson momentum, $\Lambda_B$. Carry out first an integral over an angle 
between the internal boson momentum and external fermion momentum, 
\begin{align}
&\int_{q} \overline{G}_{+,0}(k+q) 
\overline{G}_{\phi,0}(q) =  \frac{1}{8\pi^3} \int^{\infty}_{-\infty} d\varepsilon \int^{\Lambda_B}_{0} 
Q^2 dQ  \nonumber \\
& \hspace{0.8cm}  \times \int^{1}_{-1} dt \!\ \frac{1}{\varepsilon^2 + c^2 Q^2 + m^2}
 \frac{1}{i\omega-v_{+}l+i\varepsilon-v_{+}Qt} \nonumber \\
&  \ \ \  = - \frac{1}{8\pi^3 v_{+}} \int^{\Lambda_B}_{0} Q dQ  \nonumber \\
& \int^{\infty}_{-\infty}  
\frac{d\varepsilon}{(\varepsilon+iM)(\varepsilon-iM)} \!\ {\rm Log}\!\ 
\Big[ \frac{i\omega+i\varepsilon-v_{+}l-v_{+}Q}{i\omega+i\varepsilon-v_{+}l+v_{+}Q}\Big],  
\label{GaGp} 
\end{align}
with $q\equiv (i\varepsilon,{\bm q})$, $k\equiv (i\omega,{\bm k})$, ${\bm k}=(k_F+l)\hat{\Omega}$, 
$|{\bm q}| \equiv Q$, $Qt \equiv {\bm q}\cdot \hat{\Omega}$ and $M^2 \equiv c^2 Q^2 + m^2$. The 
integral is odd under $(\omega,l) \rightarrow -(\omega,l)$. Thus, without loss of generality, we can assume 
that $l$ is positive. In the complex variable plane of $\varepsilon$, the integrand in the right hand side 
has a branch cut of the logarithmic function, which 
runs from $\varepsilon=-\omega-i \!\ v_{+}(l+Q)$ to $\varepsilon=-\omega-i v_{+}(l-Q)$. For $(0<) Q<l$, the integral can 
be given by a pole contribution at $\varepsilon=\pm iM$. For $l<Q$, The real axis of $\varepsilon$ crosses the 
branch cut, so that the integral is composed of the pole contribution and an integral along the branch cut 
from $\varepsilon = - \omega$ to $\varepsilon = - \omega-i \!\ v_{+}(l \pm Q)$ respectively. This leads to  
\begin{align}
& \int_{q} \overline{G}_{+,0}(k+q) 
\overline{G}_{\phi,0}(q) = - \frac{1}{8 \pi^2 v_{+}} \nonumber \\
&  \Bigg( \int^{l}_{0} dQ \!\ \frac{Q}{M} \!\ 
{\rm Log}\Big[\frac{i\omega - M - v_{+}l - v_{+}Q}{i\omega - M-v_{+}l + v_{+}Q} \Big] + 
\int^{\Lambda_B}_{l} dQ \!\ \frac{Q}{M}\nonumber \\ 
& \ \ \bigg\{ {\rm Log}\Big[\frac{M+ v{+}l + v_{+}Q - i\omega}{M-v_{+}l+v_{+}Q+i\omega}\Big]  
+  {\rm Log}\Big[\frac{M + i\omega}{M-i\omega}\Big] \bigg\} \Bigg), \label{massless-aa} 
\end{align} 
for $l>0$. At the massless point $(m=0)$, the  integral over $Q$ can be further carried out 
explicitly. Besides, replacing $v_{+}$ by $-v_{-}$ in Eq.~(\ref{GaGp}) is equivalent to changing 
$\omega$ by $-\omega$ and adding the overall minus sign in Eq.~(\ref{GaGp}). This leads to 
\begin{widetext}
\begin{align}
\int_{q} \overline{G}_{\sigma,0}(k+q) 
\overline{G}_{\phi,0}(q) \Big|_{m=0} =& - \frac{\sigma}{8\pi^2 c v_{\sigma}} 
\bigg\{ - \Big(\frac{1}{c+v_{\sigma}} - \frac{1}{c-v_{\sigma}} \Big) (v_{\sigma}l-i \sigma \omega) \!\ 
{\rm Log}(v_{\sigma}l -i\sigma \omega)   \nonumber \\ 
& \hspace{-1.2cm} + \Big(\frac{1}{c+v_{\sigma}} - \frac{1}{c} \Big) (cl + i\sigma \omega) \!\ {\rm Log}(cl + i\sigma \omega) 
 - \Big(\frac{1}{c-v_{\sigma}} - \frac{1}{c} \Big) (cl - i\sigma \omega) \!\ {\rm Log}(cl - i\sigma \omega) \nonumber \\ 
& \hspace{-2.4cm} - 2 \!\ \frac{i\sigma \omega - v_{\sigma}l}{c+v_{\sigma}} \!\ \log\big((c+v_{\sigma}) \Lambda_B\big) 
+ 2 \!\ \frac{i\sigma \omega}{c} \!\ \log\big(c\Lambda_B\big) - 2 \!\ \frac{i\sigma \omega - v_{\sigma}l}{c+v_{\sigma}} 
+ 2 \!\ \frac{i\sigma \omega}{c} 
+ {\cal O}(\omega/\Lambda_{B},l/\Lambda_B) \bigg\}, \label{massless-a}
\end{align}
\end{widetext}
with $\sigma=\pm$. When substituted into Eq.~(\ref{1-loop-Gab}) in favor 
for $\overline{G}^{-1}_{\pm}(k)$, Eq.~(\ref{massless-a}) leads to the UV logarithmic divergence 
in the linear coefficients of the frequency 
$\omega$ and the one-dimensional momentum $l$;
\begin{align}
\overline{G}^{-1}_{\pm}(k) = &\overline{G}^{-1}_{\pm,0}(k) + 
(\delta_{\pm} i\omega \mp \delta v_{\pm} l) \nonumber \\
& \ \  + (\ldots) \!\ \ln\Lambda \!\ i\omega 
+ (\ldots) \!\ \ln \Lambda \!\ l + \cdots.  \label{massless-a-dash}
\end{align}
Eq.~(\ref{massless-a}) also has the weak IR singularity ($\omega \ln \omega$ type; 
omitted as `$\cdots$' in Eq.~(\ref{massless-a-dash})).  When being taken derivatives of 
with respect to $\omega$ or $l$, the weak IR singularity leads to the IR logarithmic 
singularity. The IR logarithmic singularity is controlled by the finite RG scale 
$\kappa$ in the RG conditions;  
Eq.~(\ref{boundary}) with $\omega$ or $v_{\pm}l=\kappa$. Thereby, 
$\ln \kappa$ comes in pairs with $\ln \Lambda$, 
\begin{align}
&\overline{G}^{-1}_{\pm}(\omega=l=0) = \overline{G}^{-1}_{\pm,0}(\omega=l=0) = 0, \nonumber \\
&\frac{\partial \overline{G}^{-1}_{\pm}(\omega,l)}{\partial (i\omega)} \Big|_{v_{\pm} l = 0, \omega=\kappa} 
= 1 + \delta_{\pm} + (\cdots) \ln \bigg(\frac{\Lambda}{\kappa}\bigg) + \cdots, \nonumber \\ 
&\frac{\partial \overline{G}^{-1}_{\pm}(\omega,l)}{\partial (\mp v_{\pm} l )}\Big|_{v_{\pm} l = \kappa, \omega=0} 
= 1 + \frac{\delta v_{\pm}}{v_{\pm}} + (\cdots) \ln \bigg(\frac{\Lambda}{\kappa}\bigg) + \cdots. \nonumber 
\end{align}
 
\subsection{Calculation of $\int_{k} \overline{G}_{+,0}(k+q) \overline{G}_{-,0}(k)$ and $\int_{q} \overline{G}_{\phi,0}(q)$ 
at $d=3$}
The 1-loop terms in the boson self-energy have two contributions (Fig.~\ref{sfig:1}(b,c)). 
One with an integral over internal boson loop (Fig.~\ref{sfig:1}(c)), and the other 
with an integral over internal fermion loop (Fig.~\ref{sfig:1}(b)). The integral over the boson 
loop does not depend on external momentum and frequency,
\begin{align}
\int_{q} \overline{G}_{\phi,0}(q) &= \frac{1}{c^3} \int^{+\infty}_{-\infty} \frac{d\varepsilon}{2\pi} \int_{|{\bm q}|<\Lambda_B} 
\frac{d {\bm q}}{(2\pi)^3} 
\frac{1}{\varepsilon^2 + c^2 {\bm q}^2 + m^2} \nonumber \\
%&\hspace{-2cm} 
%= \frac{1}{16 \pi^2 c^3} \bigg\{\Lambda_B \sqrt{\Lambda^2_B+m^2}  - m^2 \log\Big(\frac{\Lambda_B
%+\sqrt{\Lambda^2_B+m^2}}{m}\Big)\bigg\} 
&\rightarrow \frac{\Lambda^2_B }{16 \pi^2 c^3} \ \ \!\ (m\rightarrow 0). \label{gphi0}
\end{align} 
The integral over the fermion loop has the logarithmic divergence in the ultraviolet cutoff. 
To see this UV divergence, integrate first an angle between the internal fermion momentum and external 
boson momentum, $\theta$, 
%$\hat\Omega$ first with ${\bm k} \equiv (k_F+l)\hat{\Omega}$, 
\begin{align}
&\int_{k} \overline{G}_{+,0}(k+q) \overline{G}_{-,0}(k) = k^2_F \int \frac{d^2\hat{\Omega}}{(2\pi)^2}  \nonumber \\
& \!\ \int^{+\Lambda_F}_{-\Lambda_F} 
\frac{dl}{2\pi} \int^{\overline{\Lambda}_F}_{-\overline{\Lambda}_F} \frac{d \omega}{2\pi} 
\frac{1}{i\omega - v_+ l} \frac{1}{i(\omega + \varepsilon) + v_{-} (l+m)}  \nonumber \\
& = \frac{1}{2\pi} \frac{k^2_F}{v_{-} q}
\int^{+\Lambda_F}_{-\Lambda_F} 
\frac{dl}{2\pi} \int^{\overline{\Lambda}_F}_{-\overline{\Lambda}_F} \frac{d \omega}{2\pi} \nonumber \\
&\hspace{1.6cm} \frac{1}{i\omega - v_{+} l} \!\ 
{\rm Log}\bigg[\frac{i(\omega + \varepsilon) + v_{-} l + v_{-} q}{i(\omega + \varepsilon) + v_{-} l - v_{-} q}\bigg], \label{int0}
\end{align}
where $k \equiv (\omega,{\bm k})$, ${\bm k} \equiv (k_F + l )\hat{\Omega}$,  $q \equiv (\varepsilon,{\bm q})$, 
and ${\bm q}\cdot \hat{\Omega}  = q \cos\theta = m$ (Henceforth we will often write 
$|{\bm q}|$ as $q$, as far as the scalar $|{\bm q}|$ can be obviously distinguishable from 
the four-dimensional $q=(\varepsilon,{\bm q})$). 
For simplicity, we consider the case with $v_{+}=v_{-}=v$, while putting only 
the result for the general case ($v_{+} \ne v_{-}$) in Eq.~(\ref{int3}).  By choosing 
$v \Lambda_{F} = \overline{\Lambda}_F $, we introduce 
$Q/z \equiv \omega + i v l$ with $Q^2 \equiv \omega^2 + (vl)^2$;
\begin{align}
&\int_{k} \overline{G}_{+,0}(k+q) \overline{G}_{-,0}(k) \Big|_{v_+=v_{-}=v} = - \frac{1}{2\pi} 
\frac{k^2_F}{v^2 q} \nonumber \\
&  \ \ \times \int^{\overline{\Lambda}_F}_{0} \frac{dQ}{2\pi}\oint_{|z|=1} \frac{dz}{2\pi}  \!\ 
{\rm Log}\bigg[\frac{i z + (i \varepsilon + v q)/Q }{i z+ (i \varepsilon - v q)/Q}\bigg].  \label{int1}
\end{align}
As a function of the complex variable $z$, the integrand in the right hand side has a branch 
cut from $z=- \xi \equiv - (\varepsilon + i vq)/Q$ to $z = -\xi^{*} \equiv - (\varepsilon - ivq)/Q$. 
For $Q < \varepsilon$, the $z$-integral along the unit circle is contractible.  For 
$\varepsilon < Q < R \equiv \sqrt{\varepsilon^2 + (vq)^2}$,  the $z$-integral along the unit circle is 
contracted into an integral along a part of 
the branch cut. Thereby, the integral runs along one side and the other side of a line, that runs 
from $z = z_0  \equiv - \varepsilon/Q + i \sqrt{1-(\varepsilon/Q)^2}$ 
to $z=z^{*}_0$ with $|z_0|=1$. For $R<Q$, the $z$-integral along the unit circle can be contracted 
into an integral along a loop, that goes around the whole branch cut. That says, the integral reduces to 
\begin{align} 
&\int_{k} \overline{G}_{+,0}(k+q) \overline{G}_{-,0}(k) \Big|_{v_+=v_{-}=v}  \nonumber \\
& = \frac{i}{4\pi^2} \frac{k^2_F}{v^2 q} \Bigg( \int^{\overline{\Lambda}_F}_{R} dQ  (-\xi^{*}+\xi) 
+  \int^R_{\varepsilon} dQ  (z_0-z^{*}_0) \Bigg) \nonumber \\
& = -\frac{k^2_F}{2\pi^2 v} \bigg\{\log\Big(\frac{\overline{\Lambda}_F}{R}\Big) + 
1 - \frac{\varepsilon}{vq} {\rm ArcTan}\big(\frac{vq}{\varepsilon}\Big) \bigg\}.  \label{int2}
\end{align} 
In the general case of $v_{+}\ne v_{-}$, Eq.~(\ref{int0}) can be similarly evaluated. The result is  
\begin{widetext}
\begin{align}
& \int_{k} \overline{G}_{+,0}(k+q) \overline{G}_{-,0}(k) =\nonumber \\
&\ \ \  - \frac{k^2_F}{\pi^2 (v_{+}+v _{-})} 
\Bigg\{\log\bigg(\frac{\overline{\Lambda}_F}{R}\bigg) + 1 - 
\frac{\varepsilon}{v_{-} q} \frac{(v_{+}+v_{-})}{2v_{+}} {\rm ArcTan}\Big[\frac{v_{-} q}{\varepsilon}\Big] 
- \frac{\varepsilon}{2v_{+}q} 
{\rm ArcTan}\bigg[\frac{(v_{+}-v_{-})\sin 2\theta}{(v_{+}+v_{-}) - (v_{+}-v_{-}) \cos 2\theta}\bigg] \nonumber \\
& \ \ \ \ - \frac{1}{4} \log\bigg[\frac{(v_{+}+v_{-})^2+(v_{+}-v_{-})^2-2(v_{+}+v_{-})(v_{+}-v_{-}) 
\cos2\theta}{(v_{+}+v_{-})^2}\bigg] \Bigg\} 
+ {\cal O}(q/\Lambda_F), 
\label{int3}
\end{align}
\end{widetext}
with $\overline{\Lambda}_F \equiv v_{-} \Lambda_F$, $R^2 \equiv \varepsilon^2+v^2_{-} q^2$ and 
$(\cos\theta,\sin \theta) \equiv (\varepsilon,v_{-} q)/R$. In Eq.~(\ref{1-loop-Gphi}) in favor 
for $\overline{G}^{-1}_{\phi}(q)$, Eqs.~(\ref{int2},\ref{int3}) bring the UV logarithmic 
divergence into the boson mass. The UV divergence comes with the IR logarithmic divergence;
\begin{align}
\overline{G}^{-1}_{\phi}(q) = &\overline{G}^{-1}_{\phi,0}(q) + (\delta m^2 + \delta c^2 q^2 
+ \delta_{\phi} \varepsilon^2) \nonumber \\
& \ \ + (\ldots) \!\ k^2_F \!\ \log \Big(\frac{\Lambda}{R}\Big) + \cdots, \nonumber  
\end{align}  
with $R^2 \equiv \varepsilon^2+(v{\bm q})^2$. The IR divergence is controlled by the 
RG scale variable $\kappa$ in Eq.~(\ref{boundary}), where $R$ is replaced by $\kappa$.  

\subsection{Calculation of the boson and fermion integrals in Eq.~(\ref{1-loop-Glam}) at $d=3$} 
The 1-loop term in the four-point boson vertex function has two contributions (Fig.~\ref{sfig:1}(d,e)). 
The 1-loop term with the internal fermion integral has the IR quadratic divergence (Fig.~\ref{sfig:1}(e)). 
To see this, set only the external boson frequencies to be finite, $q_i=(p_i,{\bm 0})$, and 
integrate over the fermion frequency $\omega$, 
\begin{align}
&\int_{k} \overline{G}_{+,0}(k) \overline{G}_{-,0}(k+q_1) \nonumber \\
& \hspace{1cm} \times \overline{G}_{+,0}(k+q_1-q_3) \overline{G}_{-,0}(k+q_1+q_2-q_3) \nonumber \\
& \hspace{-0.1cm} = - i\frac{k^2_F}{\pi(v_{+}+v_{-})} \int^{+\infty}_{-\infty} 
\frac{dX}{2\pi} \Big\{\frac{1}{X+ip_1}\frac{1}{X+ip_3}\frac{1}{p_2-p_3} \nonumber \\
&\hspace{2.5cm} + \frac{1}{X+ip_2}\frac{1}{X+ip_4} \frac{1}{-p_2+p_3}\Big\} \label{int4a}
\end{align} 
with $k \equiv (\omega,{\bm k})$, ${\bm k} \equiv (k_F+l)\hat{\Omega}$ and $X \equiv (v_{+}+v_{-})l$. 
Here $p_1$, $p_2$, $p_3$, $p_4=p_1+p_2-p_3$ are external boson frequencies of $q_1$, $q_2$, $q_3$ and $q_4$ respectively. 
When these frequencies are chosen with either ${\rm sgn}(p_1p_3)<0$ or ${\rm sgn}(p_2 p_4)<0$, the integral over the 
one-dimensional fermion momentum $l \equiv X/(v_{+}+v_{-})$ 
gives a quadratic divergent term in the small external frequencies. 
When these frequencies are chosen with 
${\rm sgn}(p_1p_3)>0$ and ${\rm sgn}(p_2 p_4)>0$ as in 
Eq.~(\ref{four-point-boson-vertex}), the integral over $X$ leads to zero. 

The 1-loop term with the internal boson integral has the UV logarithmic divergence 
(Fig.~\ref{sfig:1}(d)). In the massless case ($m=0$), the UV divergence comes with an 
IR logarithmic divergence. The IR divergence can be controlled by choosing 
the external boson frequencies as in Eqs.~(\ref{four-point-boson-vertex},\ref{Q=K=P}). 
Thereby, the external boson frequencies play role of the RG scale variable $\kappa$,
\begin{align}
& \int_q \overline{G}_{\phi,0}(q) \overline{G}_{\phi,0}(q_1+q_2-q) \nonumber \\  
& = \int^{+\infty}_{-\infty} \frac{d\varepsilon}{2\pi} \int_{|{\bm q}|<\Lambda_B} \frac{d^3 {\bm q}}{(2\pi)^3} 
\frac{1}{\varepsilon^2+c^2 {\bm q}^2} \frac{1}{(\varepsilon-y)^2 + c^2{\bm q}^2}  \nonumber \\
& = \frac{\log\big(\frac{2c\Lambda_B}{y}\big) }{8\pi^2 c^3}
+ {\cal O}(\Lambda^{-2}_B). \label{int4b}
\end{align} 
Here $y=4\kappa/3$ for the first term in Eq.~(\ref{1-loop-Glam}) 
(particle-particle channel) and $y=\kappa/3$ for the other two in Eq.~(\ref{1-loop-Glam}) (particle-hole channel). 
When substituted into Eq.~(\ref{1-loop-Glam}) in favor for $\overline{\Gamma}_{\lambda}$, 
Eq.~(\ref{int4b}) leads to the UV logarithmic divergence in the $\phi^4$ coupling, 
\begin{align}
&\overline{\Gamma}_{\lambda}(q_1,q_2,q_3)\Big|_{q_1=(\kappa,0),q_2=(\kappa/3,0),q_3=(2\kappa/3,0)} 
\nonumber \\
& \ \ \  = -\lambda - \delta \lambda + 
(\ldots) \!\ \log \Big(\frac{\Lambda}{\kappa}\Big) + \cdots. \nonumber 
\end{align}  

\subsection{determination of the counterterms and $\beta$ functions} 
By substituting  Eqs.~(\ref{massless-a},\ref{gphi0},\ref{int3},\ref{int4b}) 
into Eqs.~(\ref{1-loop-Gab},\ref{1-loop-Gphi},\ref{1-loop-Glam}) and 
imposing the conditions Eq.~(\ref{boundary}) on the renormalized vertex functions, 
we determine the counterterms in $S_c$; $\delta m^2$, $\delta c^2$, 
$\delta_{\phi}$, $\delta_{\pm}$, $\delta v_{\pm}$, $\delta g$ and $\delta \lambda$. 
The counterterms are given as functions of the renormalized physical quantities, 
$m^2$, $c^2$, $v_{\pm}$, $g$, $\lambda$, the RG scale $\kappa$, the Fermi wavelength $k_F$ 
and the UV cutoff $\Lambda$. So are the bare 
quantities, $m^2_0$, $c^2_0$, $v_{\pm 0}$, $g_0$, $\lambda_0$, 
and the renormalization of the field operator amplitudes, $Z_{\phi}$ and $Z_{\pm}$;
\begin{eqnarray}
\left\{\begin{array}{c}
m^2_{0} = M^2_0(m^2,c^2,v_{+},v_{-},g,\lambda,\kappa,k_F,\Lambda), \\
c^2_0 = C^2_0(m^2,c^2,\cdots,g,\lambda,\kappa,k_F,\Lambda),  \\
v_{\sigma0}  = V_{\sigma0}(m^2,c^2,\cdots,g,\lambda,\kappa,k_F,\Lambda),  \\
g_{0} = G_{0}(m^2,c^2,\cdots,g,\lambda,\kappa,k_F,\Lambda),  \\
\lambda_{0}  = \Lambda_{0}(m^2,c^2,\cdots,g,\lambda,\kappa,k_F,\Lambda), \\ 
\end{array}\right. \label{bare}
\end{eqnarray}
and 
\begin{eqnarray}
\left\{\begin{array}{c}
Z_{\phi} = Z_{\phi}(m^2,c^2,v_{+},v_{-},g,\lambda,\kappa,k_F,\Lambda), \\
Z_{\sigma} = Z_{\sigma}(m^2,c^2,\cdots,g,\lambda,\kappa,k_F,\Lambda), \\
\end{array}\right. \label{Z} 
\end{eqnarray} 
with $\sigma=\pm$.
Eqs.~(\ref{bare},\ref{Z}) define a relation between bare quantities and renormalized physical quantities, in which all the 
universal information of the quantum criticality are encoded. To decode the information of 
the criticality, Eq.~(\ref{relation}) 
and their derivative with respect to the RG scale variable $\kappa$ are used 
in combination with Eqs.~(\ref{bare},\ref{Z}) (see Sec.~VA). 

The renormalized perturbation theory determines 
Eqs.~(\ref{bare},\ref{Z}) perturbatively in $g^2$ and $\lambda$.  In this work, we shall focus only on the relation 
at the massless point, and set the renormalized physical mass $m$ to be zero in Eqs.~(\ref{bare},\ref{Z}). 
Thereby, the first line of Eq.~(\ref{bare}) defines a subspace of the quantum critical `point' in a 
multiple-dimensional parameter space subtended by the bare quantities, 
$m_0$, $c_0$, $v_{\sigma0}$, $g_0$, $\lambda_0$, and the RG scale 
variable $\kappa$;
\begin{align}
m^2_{0} = M^2_0(m^2=0,c^2,v_{+},v_{-},g,\lambda,\kappa,k_F,\Lambda). \label{QCP}
\end{align}  
Meanwhile, the other relations in Eqs.~(\ref{bare},\ref{Z}) with $m^2=0$ 
define the critical properties at the quantum critical point;
\begin{eqnarray}
\left\{\begin{array}{c}
c^2_0 = C^2_0(m^2=0,c^2,v_{+},v_{-},g,\lambda,\kappa,k_F,\Lambda),  \\
v_{\sigma0}  = V_{\sigma0}(m^2=0,c^2,\cdots,\lambda,\kappa,k_F,\Lambda),  \\
g_{0} = G_{0}(m^2=0,c^2,\cdots,\lambda,\kappa,k_F,\Lambda),  \\
\lambda_{0}  = \Lambda_{0}(m^2=0,c^2,\cdots,\lambda,\kappa,k_F,\Lambda), \\ 
Z_{\phi} = Z_{\phi}(m^2=0,c^2,\cdots,\lambda,\kappa,k_F,\Lambda), \\
Z_{\sigma} = Z_{\sigma}(m^2=0,c^2,\cdots,\lambda,\kappa,k_F,\Lambda). \\
\end{array}\right. \label{bare-Z-prime} 
\end{eqnarray} 

To be more specific, at the critical point ($m=0$), the counterterms in $S_c$ are given 
as functions of renormalized physical quantities, $\kappa$, $k_F$, and $\Lambda$ perturbatively in 
$g^2$ and $\lambda$;
\begin{widetext}
\begin{eqnarray}
&& \ \ \ \ \delta m^2  + \delta_{\phi} \kappa^2 - \frac{g^2 k^2_F}{\pi^2 (v_{+}+v_{-})} 
\bigg\{\log\Big(\frac{\overline{\Lambda}_F}{\kappa}\Big) - \frac{1}{2} \log \Big(\frac{2v_{-}}{v_{+}+v_{-}}\Big)\bigg\} 
+ \frac{\lambda \Lambda^2_B}{16 \pi^2 c^3} = 0,  \label{36-}  \\ 
&& \left\{\begin{array}{l}
\delta_{\sigma} = - \frac{g^2}{8\pi^2 c v_{\overline{\sigma}}} \bigg\{
\frac{2v_{\overline{\sigma}}}{(c+v_{\overline{\sigma}}) c} 
\log\Big(\frac{\Lambda_B}{\kappa}\Big) - \frac{2}{c+v_{\overline{\sigma}}}\log(c+v_{\overline{\sigma}}) 
+ \frac{2}{c} \log c \bigg\} \\
\delta v_{\sigma}  = - \frac{g^2}{8\pi^2 c v_{\overline{\sigma}}} \bigg\{
\frac{2v_{\overline{\sigma}}}{c+v_{\overline{\sigma}}} 
\log\Big(\frac{v_{\sigma}\Lambda_B}{\kappa}\Big) 
+ \frac{2v_{\overline{\sigma}}}{c+v_{\overline{\sigma}}}\log(c+v_{\overline{\sigma}}) 
- \frac{2v_{\overline{\sigma}}}{c+v_{\overline{\sigma}}} \frac{1}{c-v_{\overline{\sigma}}} (c \log c - 
v_{\overline{\sigma}} \log v_{\overline{\sigma}})  \bigg\},  \\
\delta \lambda = \frac{\lambda^2}{16\pi^2 c^3} \bigg\{5 \log\Big(\frac{3c\Lambda_B}{\kappa}\Big) + 3\log 2 \bigg\}, 
\ \  \delta_{\phi} = 0,  \  \  \delta c^2 = 0, \ \ \delta g = 0.  \\ 
\end{array}\right.  \label{36} 
\end{eqnarray}  
with $\overline{\sigma}=\mp$ for $\sigma=\pm$. Here the followings were used from Eq.~(\ref{int3}), 
\begin{align} 
& \frac{\partial}{\partial \varepsilon} \bigg( \int_k \overline{G}_{+,0}(k+q) \overline{G}_{-,0}(k) 
\bigg)\Big|_{\varepsilon= \kappa, {\bm q}=0} 
= \frac{\partial}{\partial (c|{\bm q}|) } \bigg( \int_k \overline{G}_{+,0}(k+q) \overline{G}_{-,0}(k) 
\bigg)\Big|_{\varepsilon= 0, c|{\bm q}|=\kappa} 
= \frac{k^2_F}{\pi^2 (v_{+}+v_{-}) \kappa}.     \label{int5}
\end{align}
\end{widetext}
When substituted in Eqs.~(\ref{1-loop-Gphi},\ref{boundary}), Eq.~(\ref{int5}) 
cancels the first order term in $k^2_F$ in the right hand side of the conditions for the 
boson self energy in Eq.~(\ref{boundary}).    

In terms of Eqs.~(\ref{36},\ref{36-}) and $m=0$, a set of the bare physical quantities, 
$m_0$, $c_0$, $v_{\sigma0}$, $g_{0}$, $\lambda_0$, are given as 
functions of renormalized physical quantities, 
the RG scale variable $\kappa$, the Fermi wavelength $k_F$ and the UV cutoff $\Lambda$,
\begin{align}
m^2_0 & = Z^{-1}_{\phi} (m^2 + \delta m^2) = \delta m^2 = - \frac{\lambda \Lambda^2_B}{16 \pi^2 c^3} \nonumber \\
& \hspace{-0.5cm}  + \frac{g^2 k^2_F}{\pi^2 (v_{+}+v_{-})} 
\bigg\{\log\Big(\frac{\overline{\Lambda}_F}{\kappa}\Big) - \frac{1}{2} \log \Big(\frac{2v_{-}}{v_{+}+v_{-}}\Big)\bigg\} 
 + \cdots \label{m0}  \\ 
v_{\sigma0} & = Z^{-1}_{\sigma} (v_{\sigma}+\delta v_{\sigma}) = v_{\sigma} - \delta_{\sigma} 
v_{\sigma}  + \delta v_{\sigma} + \cdots , \label{va0} \\
 c^2_0 &= c^2 + \cdots, \ \ \ \  \lambda_0  = \lambda + \delta \lambda,  \label{vb0c0} \\
g_0 &=(1 -\frac{1}{2}\delta_{\phi}) (1-\frac{1}{2} \delta_{+}) (1-\frac{1}{2} \delta_{-}) g \nonumber \\
&= (1-\frac{1}{2} \delta_{+} -\frac{1}{2} \delta_{-}) g + \cdots, \label{g0} 
\end{align}
with $\sigma=\pm$. Here $\delta_{\sigma}$, $\delta v_{\sigma}$, $\delta \lambda$ in the 
right hand sides are given in Eq.~(\ref{36}).`$\cdots$' in the right 
hand sides stands for the higher-order 
contributions in $g^2$ and $\lambda$. By inverse solutions of Eqs.~(\ref{va0},\ref{vb0c0},\ref{g0},\ref{36}), 
$c$, $v_{\pm}$, $g$, $\lambda$ are given as functions of $c_0$, $v_{\pm0}$, 
$g_0$, $\lambda_0$, $\kappa$, $k_F$ and $\Lambda$. On substitution of such solutions into Eq.~(\ref{m0}), 
$m_0$ is given as a function of  $c_0$, $v_{\pm0}$, $g_0$, $\lambda_0$, $\kappa$, $k_F$ 
and $\Lambda$. Such a function for $m_0$ defines a subspace of 
$m=0$ in a multiple-dimensional parameter space subtended by the bare physical 
quantities and the RG scale variable $\kappa$. The subspace of $m=0$ defines 
the quantum critical `point' in the parameter space of the bare quantities and $\kappa$.

Let us introduce a derivative with respect to the RG 
scale $\kappa$ within the subspace of $m=0$ and with fixed values of 
the other bare quantities, $k_F$ and the UV cutoff;
\begin{align}
\frac{\partial}{\partial \ln \kappa}  \Big|_{c_0,v_{\pm 0},g_0,\lambda_0,k_F,\Lambda: {\rm fixed}}  
\equiv \frac{\partial}{\partial  \ln \kappa} \Big|_{\cdots}.  
\label{kappa-deri}
\end{align} 
The following identities hold true trivially;
\begin{align}
\frac{\partial c_0}{\partial \ln \kappa}\Big|_{\cdots} 
= \frac{\partial v_{\sigma0}}{\partial \ln \kappa}\Big|_{\cdots}  
= \frac{\partial g_{0}}{\partial \ln \kappa}\Big|_{\cdots} 
= \frac{\partial \lambda_{0}}{\partial \ln \kappa}\Big|_{\cdots} = 0, \label{trivial}
\end{align} 
with $\sigma=\pm$. On application of such $\ln \kappa$-derivative onto Eqs.~(\ref{va0},\ref{vb0c0},\ref{g0}) with 
Eq.~(\ref{trivial}) in their left hand sides, $\beta$ functions of the physical quantities are obtained  
as their $\ln \kappa$-derivatives. The $\beta$ functions are calculated perturbatively in the Yukawa coupling 
$g^2$ and the $\phi^4$ coupling $\lambda$,
\begin{align}
&\frac{\partial v_{\sigma}}{\partial \ln \kappa}\Big|_{\cdots}  
 = \frac{\partial \delta_{\sigma}}{\partial \ln \kappa} v_{\sigma} - \frac{\partial \delta v_{\sigma}}{\partial \ln \kappa} 
+ {\cal O}(g^4, g^2\lambda, \lambda^2) \nonumber \\
& = \frac{g^2}{4\pi^2 c(c+v_{\overline{\sigma}})} \Big(\frac{v_{\sigma}}{c} - 1\Big) + 
\cdots \equiv \beta_{v_{\sigma}}(c,\cdots), \label{beta-va} \\  
&\frac{\partial c}{\partial \ln \kappa}\Big|_{\cdots} 
= 0 +\cdots \equiv \beta_c(c,\cdots), \label{beta-c} \\
& \frac{\partial g}{\partial \ln \kappa} \Big|_{\cdots}  
= \frac{g}{2} \Big(\frac{\partial \delta_{+}}{\partial \ln \kappa} 
+ \frac{\partial \delta_{-}}{\partial \ln \kappa}\Big) + {\cal O}(g^5, g^3\lambda, g\lambda^2) 
\nonumber \\
& =  \frac{g^3}{8 \pi^2 c^2} \Big(\frac{1}{c+v_{+}} + \frac{1}{c+v_{-}} \Big) + \cdots 
\equiv \beta_{g}(c,\cdots), \label{beta-g} \\
& \frac{\partial \lambda}{\partial \ln \kappa} \Big|_{\cdots}   
= \frac{5\lambda^2}{16\pi^2 c^3} + {\cal O}(\lambda^3,\lambda^2 g^2, \lambda g^4, g^6)
\equiv \beta_{\lambda}(c,\cdots),   \label{beta-lambda}
\end{align}
with $\sigma=\pm$ and $\overline{\sigma}=\mp$. 
Here `$\cdots$' in the right hand side refers to the higher-order terms in $g^2$ and $\lambda$.  Importantly, 
the $\beta$ functions are given only by the renormalized physical quantities themselves.

The leading-order $\beta$ functions of $g$ as well as $\lambda$ dictate that in the low-energy limit 
($\ln \kappa \rightarrow -\infty$), the Yukawa coupling $g$ and $\phi^4$ coupling $\lambda$ are   
renormalized into the zero. The marginal irrelevance of $g$ and $\lambda$ in the IR limit 
justifies a posteriori the perturbative calculation in small $g^2$ and $\lambda$. 
The $\beta$ functions of the two Fermi velocities shows that in the IR limit, 
both $v_{+}$ and $v_{-}$ are renormalized into the same critical velocity as 
the boson velocity $c$. Within the one-loop level, the boson velocity has no renormalization; $\beta_c=0$.

\section{Callan-Symanzik equation and its approximate solution}
The $\beta$ functions appear in the Callan-Symanzik (CS) equations for the 
renormalized Green's functions. The role of the $\beta$ functions become clear, 
when the CS equations are solved in favor for the Green's 
functions~\cite{peskin-schroeder,amit}. In this section, the CS equation for 
the two-point fermion Green's functions will be derived first. In the latter part of this 
section, the CS equations are solved in favor for the renormalized fermion Green's functions. 

\subsection{derivation of Callan-Symanzik equation at QCP}
Let us begin with the relation between the renormalized two-point 
fermion Green's function and the bare one. From Eq.~(\ref{relation}), the 
relation is,
\begin{align}
&Z_{\sigma}(m^2=0,c^2,v_{+},v_{-},g,\lambda,\kappa,\Lambda)  \nonumber \\
& \ \ \ \ \ \times \overline{G}_{\sigma}(k;m^2=0, c^2,v_{+},v_{-},g,\lambda,\kappa) \nonumber \\ 
& \ \ \ \hspace{1.5cm} = G_{\sigma}(k;m^2_0,c^2_0,v_{+0},v_{-0},g_0,\lambda_0,\Lambda).  \label{source}
\end{align}
Here the $k_F$-dependence was trivially 
omitted, and will be omitted henceforth unless dictated otherwise. In Eq.~(\ref{source}), 
$Z_{\sigma}$ and $\overline{G}_{\sigma}$ in the left hand sides are given as functions of the renormalized 
physical quantities, $c$, $v_{\pm}$, $g$, $\lambda$, the RG scale variable $\kappa$ and the UV 
cutoff $\Lambda$. By Eqs.~(\ref{va0},\ref{vb0c0},\ref{g0},\ref{36}), such renormalized physical quantities 
are given as functions of $c_0$, $v_{\pm0}$, $g_0$, $\lambda_0$, the 
RG scale variable $\kappa$, and $\Lambda$. $G_{\sigma}$ in the right hand side is originally 
given as a function of $m_0$, $c_0$, $v_{\pm0}$, $g_0$, $\lambda_0$ 
and the UV cutoff $\Lambda$. By Eq.~(\ref{m0}) with Eqs.~(\ref{va0},\ref{vb0c0},\ref{g0},\ref{36}), 
$m_0$ is given by $c_0$, $v_{\pm0}$, $g_0$, $\lambda_0$, the RG scale variable $\kappa$, and $\Lambda$. 
With this in mind, take the $\kappa$-derivative of Eq.~(\ref{source}) with respect to  
Eq.~(\ref{kappa-deri}). The derivative 
gives the following inhomogeneous CS equation 
\begin{align} 
&\Big\{\frac{\partial}{\partial \ln \kappa} 
+ \beta_c \frac{\partial} {\partial c} + \sum_{\sigma^{\prime}=\pm } 
\beta_{v_{\sigma^{\prime}}} \frac{\partial}{\partial v_{\sigma^{\prime}}} 
+ \beta_{g} \frac{\partial}{\partial g} + \beta_{\lambda} 
\frac{\partial}{\partial \lambda} \nonumber \\
& \ \  + \gamma_{\sigma} \Big\} \overline{G}_{\sigma}(k;m^2=0,c^2,\cdots,\lambda,\kappa)  
= \frac{\partial m^2_0}{\partial \ln \kappa} \frac{\partial G_{\sigma}} {\partial m^2_0}. \label{CS-1}
\end{align}
Here $\gamma_{\sigma}$ ($\sigma=\pm$) is called as $\gamma$ function and it is 
given by the $\kappa$-derivative of the renormalizations of the field operator amplitudes, 
\begin{align}
\frac{\partial  \ln Z_{\sigma}}{\partial \ln \kappa}\Big|_{\cdots} 
&= \frac{\partial \ln (1+\delta_{\sigma})}{\partial \ln \kappa}\Big|_{\cdots}  \nonumber \\ 
&= \frac{g^2}{4\pi^2 c^2 (c+v_{\overline{\sigma}})} + {\cal O}(g^4,g^2\lambda,\lambda^2) 
\equiv \gamma_{\sigma}. \label{gamma}
\end{align}
with $\sigma=\pm$ and $\overline{\sigma}=\mp$. 

Note that the right hand side of Eq.~(\ref{CS-1}) is at most on the order of ${\cal O}(g^4)$ and 
negligible within the one-loop level calculation. Namely, Eq.~(\ref{m0}) leads to  
\begin{align}
\frac{\partial m^2_0}{\partial \ln \kappa}\Big|_{\cdots} 
= -\frac{g^2 k^2_F}{\pi^2 (v_{+}+v_{-})} + {\cal O}(g^4,g^2\lambda,\lambda^2). 
\label{dmdlnk}  
\end{align} 
The partial derivative of the bare fermion Green's function with respect to the bare boson mass results in an  
amputated 1PI connected Green's function with two boson lines and two fermion lines. 
Such is at most on the order of ${\cal O}(g^2)$ too;  
\begin{align}
& \frac{\partial G_{\sigma}(k)} {\partial m^2_0} = G_{\sigma}(k) 
\frac{\partial G^{-1}_{\sigma}(k)} {\partial m^2_0} G_{\sigma}(k) \nonumber \\
& \ \ \ \equiv \int_q G_{\sigma}(k) \Gamma^{(2,2)}_{\sigma}(k,k;q,q) G_{\sigma}(k) = {\cal O}(g^2).  \label{dgdm0} 
\end{align}
Here $\Gamma^{(2,2)}_{\sigma}(k,k;q,q)$ is the amputated 1PI connected Green's function with two external 
bosons and two external fermions lines (Fig.~\ref{sfig:1}(f)). 
From Eqs.~(\ref{dmdlnk},\ref{dgdm0}), the right hand side of Eq.~(\ref{CS-1}) can be neglected 
within the one-loop order;
\begin{align}
& \Big\{\frac{\partial}{\partial \ln \kappa} + \sum_{\sigma^{\prime}=\pm}\beta_{v_{\sigma^{\prime}}} 
\frac{\partial}{\partial v_{\sigma^{\prime}}} 
+ \beta_{g} \frac{\partial}{\partial g}  \nonumber \\
&\hspace{0.3cm} + \gamma_{\sigma} \Big\} \overline{G}_{\sigma}(l,\omega;m^2=0,c^2,\cdots,\kappa)  
= {\cal O}(g^4, g^2 \lambda, \lambda^2),  \label{CS-2}
\end{align} 
with $k \equiv ({\bm k},\omega)$ and ${\bm k} \equiv (k_F+l)\Omega$. 
Here those terms with $\partial_c$ and $\partial_{\lambda}$ are omitted from the left hand side in Eq.~(\ref{CS-1}),
because $\beta_c=0$ at the one-loop level, and the $\lambda$-dependence of 
$\overline{G}_{\sigma}$ starts from ${\cal O}(g^2 \lambda)$, which gives out  
${\cal O}(g^2 \lambda^2)$ in the right hand side of Eq.~(\ref{CS-2}).  
In the next section, we shall solve this homogeneous CS equation in favor for 
the renormalized fermion Green's functions.  

\subsection{approximate solutions of Callan-Symanzik equation}

The $\beta$ functions for the fermion velocities suggest that the two fermion velocities are 
renormalized into the boson velocities $c$ in the infrared (IR) limit ($\ln \kappa \rightarrow -\infty$);
\begin{align}
\frac{dv_{\pm}}{d\ln \kappa} = \frac{g^2}{4\pi^2 c(c+v_{\mp})} \Big(\frac{v_{\pm}}{c} - 1\Big). \nonumber
%\frac{dv_{-}}{d\ln \kappa} = \frac{g^2}{4\pi^2 c(c+v_{+})} \Big(\frac{v_{-}}{c} - 1\Big). \nonumber 
\end{align}    
That says, at the quantum critical point ($m=0$), the two fermion bands proximate to the Fermi 
surface have asymptotically the  same critical velocity as the boson velocity. 
We thus assume that $v_{\pm}=c$ in the following analysis.  With 
$v_{\pm}=c$, the CS equation for the fermion Green's function is given by the homogeneous 
equation with the one-parameter scaling
\begin{align}
\Big\{\frac{\partial}{\partial \ln \kappa} +
 \beta_{g}(g) \frac{\partial}{\partial g} + \gamma_{\sigma}(g) \Big\} \overline{G}_{\sigma}(l,\omega;g,\kappa)  
= 0.  \label{CS-3}
\end{align}
The $c$ and $v_{\pm}$-dependences of $\overline{G}_{\sigma}$, $\beta_g$ and $\gamma_{\sigma}$ 
can be trivially omitted; $\beta_{v_{\pm}}=\beta_c=0$ at $v_{\pm}=c$. From 
a dimensional analysis, the fermion Green's function has the following scaling form,
\begin{align}
\overline{G}_{\sigma}(l,\omega;g,\kappa) = \frac{1}{i\omega - \sigma cl} f_{\sigma}(l/\kappa,\omega/\kappa:g),  
\end{align}
with $\sigma=\pm$. Thus, with $\omega=p\sin\theta$ and 
$cl=p\cos\theta$, the $\kappa$-derivative is replaced by the $p$-derivative with fixed $\theta$;
\begin{align}
\Big\{p \frac{\partial}{\partial p} 
- \beta_{g}(g) \frac{\partial}{\partial g} + \big(1 - \gamma_{\sigma}(g)\big) \Big\} \overline{G}_{\sigma}(p,\theta;g,\kappa)  
= 0. \label{CS-4}
\end{align} 
From Eqs.~(\ref{beta-g},\ref{gamma}), $\beta_g$ and $\gamma_{\sigma}$ are given at $v_{\pm}=c$ as;
\begin{align}
\beta_g(g) = \frac{g^3}{8\pi^2 c^3}, \  \ \ \gamma_{\sigma}(g) = \frac{g^2}{8\pi^2 c^3}. 
\end{align}
The solution of Eq.~(\ref{CS-4}) is given by
\begin{align}
& \overline{G}_{\sigma}(p,\theta;g,\kappa) = \overline{\cal G}_{\sigma}(\theta;\overline{g}(t;g),\kappa) \nonumber \\
& \hspace{2cm} \times \exp \bigg[\int^g_{\overline{g}(t;g)} \frac{1-\gamma_{\sigma}(g^{\prime})}{\beta_g(g^{\prime})} dg^{\prime}\bigg], \label{sol1}
\end{align}
with $t\equiv \ln (p/\kappa)$. $\overline{\cal G}_{\sigma}(\theta;g,\kappa)$ is the 
Green's function at $p=\kappa$ ($t=0$);
\begin{align}
\overline{G}_{\sigma}(p=\kappa,\theta;g,\kappa) \equiv \overline{\cal G}_{\sigma}(\theta;g,\kappa).  \label{sol1a}
\end{align} 
$\overline{g}(t;g)$ is a running coupling constant determined by a one-parameter scaling 
equation with $\beta_g$;
\begin{align}
\frac{\partial \overline{g}(t;g)}{\partial t} = \beta_{g}(\overline{g}) = \frac{\overline{g}^3}{8\pi^2 c^3},  \ \ \overline{g}(t=0;g) = g. 
\end{align}
The running coupling constant becomes renormalized into the smaller value in the IR limit, 
\begin{align}
\overline{g}(t;g) = g \sqrt{\frac{1}{1-g^2 \frac{t}{4\pi^2 c^3}}}.  \label{gbar}
\end{align} 
Namely, for $p\ll \kappa$, $t = \ln (p/\kappa)$ is negative and logarithmically large. 
$\overline{g}/g$ goes to the zero in the low-energy limit ($p \ll \kappa$). Thus, for 
smaller $p$ in Eq.~(\ref{sol1}), $\overline{\cal G}_{\sigma}(\theta;\overline{g},\kappa)$ 
in its right hand side might be evaluated perturbatively in the small 
$\overline{g}$. Eq.~(\ref{1-loop-Gab}) with $v_{\pm}=c$ gives  
the second order expression for $\overline{G}_{\sigma}(l,\omega;g,\kappa)$ for 
small $g$,  
\begin{align}
\overline{G}^{-1}_{\sigma}(l,\omega;g,\kappa) &= (i\omega- \sigma cl)  \nonumber \\
& \ \hspace{-2cm} - 
\frac{g^2}{8\pi^2 c^3} (i\omega- \sigma cl) 
\bigg\{\log \Big[\frac{\sqrt{(cl)^2+\omega^2}}{\kappa}\Big] - 1 \bigg\}, \label{sol2}
\end{align}
at $v_{\pm}=c$. Here the following relations are used from Eqs.~(\ref{massless-a},\ref{36}), 
\begin{align}
&\lim_{v_{\pm} \rightarrow c} \int_q \overline{G}_{\sigma}(k+q) \overline{G}_{\phi}(q) \Big|_{m=0} \nonumber \\
& =  -\frac{\sigma}{8\pi^2 c^2} \Bigg\{-\frac{1}{2c} (cl + i\sigma \omega) 
\log \big(cl+i\sigma \omega\big) \nonumber \\
&\hspace{1.6cm} - \frac{1}{2c} (cl+i\sigma \omega) \log\big(cl-i\sigma \omega\big) + 
\frac{i\sigma \omega}{c} \nonumber \\
& \hspace{2.0cm}  - \frac{i\sigma \omega-cl}{c} \log 2 
+ \frac{i\sigma \omega + c l}{c} \log \big(c \Lambda_B\big) \Bigg\}, \nonumber \\
&\lim_{v_{\pm} \rightarrow c}\delta_{\sigma} = - \frac{g^2}{8\pi^2 c^2} \bigg\{\frac{1}{c} 
\log \Big(\frac{\Lambda_B}{\kappa}\Big) 
+ \frac{1}{c} \log\Big(\frac{c}{2}\Big) \bigg\}, \nonumber \\ 
& \lim_{v_{\pm} \rightarrow c} \delta v_{\sigma} =  - \frac{g^2}{8\pi^2 c^2} \bigg\{\log\Big(\frac{c\Lambda_B}{\kappa}\Big) 
+ \log 2 - 1\bigg\}. 
\end{align}
A comparison between Eqs.~(\ref{sol1a},\ref{sol2}) gives 
\begin{align}
\overline{\cal G}^{-1}_{\sigma}(\theta;\overline{g},\kappa) 
= - \sigma \kappa e^{-\sigma i\theta} \Big(1 + \frac{\overline{g}^2}{8\pi^2 c^3}\Big).
\label{sol3} 
\end{align}
Meanwhile, the exponential factor in Eq.~(\ref{sol1}) is calculated as
\begin{align}
& \exp \bigg[\int^g_{\overline{g}(t;g)} 
\frac{1-\gamma_{\sigma}(g^{\prime})}{\beta_g(g^{\prime})} dg^{\prime}\bigg] \nonumber \\
&= \exp \bigg[\int^{0}_{t} dt^{\prime}\bigg] \exp \bigg[-\int^g_{\overline{g}} \frac{1}{g^{\prime}} 
dg^{\prime}\bigg] = \frac{\kappa}{p} \frac{\overline{g}}{g}. \label{sol4} 
\end{align}
Thus, the renormalized fermion Green's function near the Fermi surface is obtained as 
\begin{align}
\overline{G}_{\sigma}(l,\omega;g,\kappa) 
&= \frac{1}{i\omega-\sigma cl} \Big(1 + \frac{\overline{g}^2}{8\pi^2 c^3}\Big)^{-1}
 \frac{\overline{g}(t)}{g} \nonumber \\
&= \frac{1}{i\omega-\sigma cl} \Big(1 + \frac{\overline{g}^2}{8\pi^2 c^3}\Big)^{-1} 
\sqrt{\frac{1}{1-g^2 \frac{t}{4\pi^2 c^3}}}, \label{final-G} 
\end{align}
with $t \equiv \ln (\sqrt{(cl)^2+\omega^2}/\kappa)$. $\overline{g}$ in the right hand side is given by $t$ 
and $g$ by Eq.~(\ref{gbar}). 

\section{Fermionic spectral function and density of states in the quantum critical regime}
Previously, the imaginary-time time-ordered fermion Green's function is 
obtained as a solution of the Callan-Symanzik equation at the QCP. In this section, we 
use an analytic continuation ($i\omega \rightarrow \omega + i\delta $) and obtain the real-time 
retarded fermion Green's function, whose imaginary part gives the fermion's spectral 
function at the QCP. The spectral function thus 
obtained has no quasi-particle spectral weight, indicating the breakdown of the `quasi-particle Fermi-liquid' 
picture at the QCP. The result also suggests a non-Fermi liquid property in a finite-$T$ 
side of the QCP (quantum critical regime; QCR). To this end, we next integrate the spectral function with 
respect to the momentum, and obtain the fermion's density of states (DOS) near the Fermi level at 
the QCP as well as the DOS at the Fermi level in the finite-$T$ side of the QCP. Based on them, we will 
discuss in the next section the temperature-dependence of the specific heat, Pauli paramagnetic 
susceptibility and NMR relaxation time in the QCR (summarized in Table ~\ref{table:1}). 

\subsection{fermion's spectral function around the QCP}
\begin{figure}[t]
	\centering
	\includegraphics[width=0.95\linewidth]{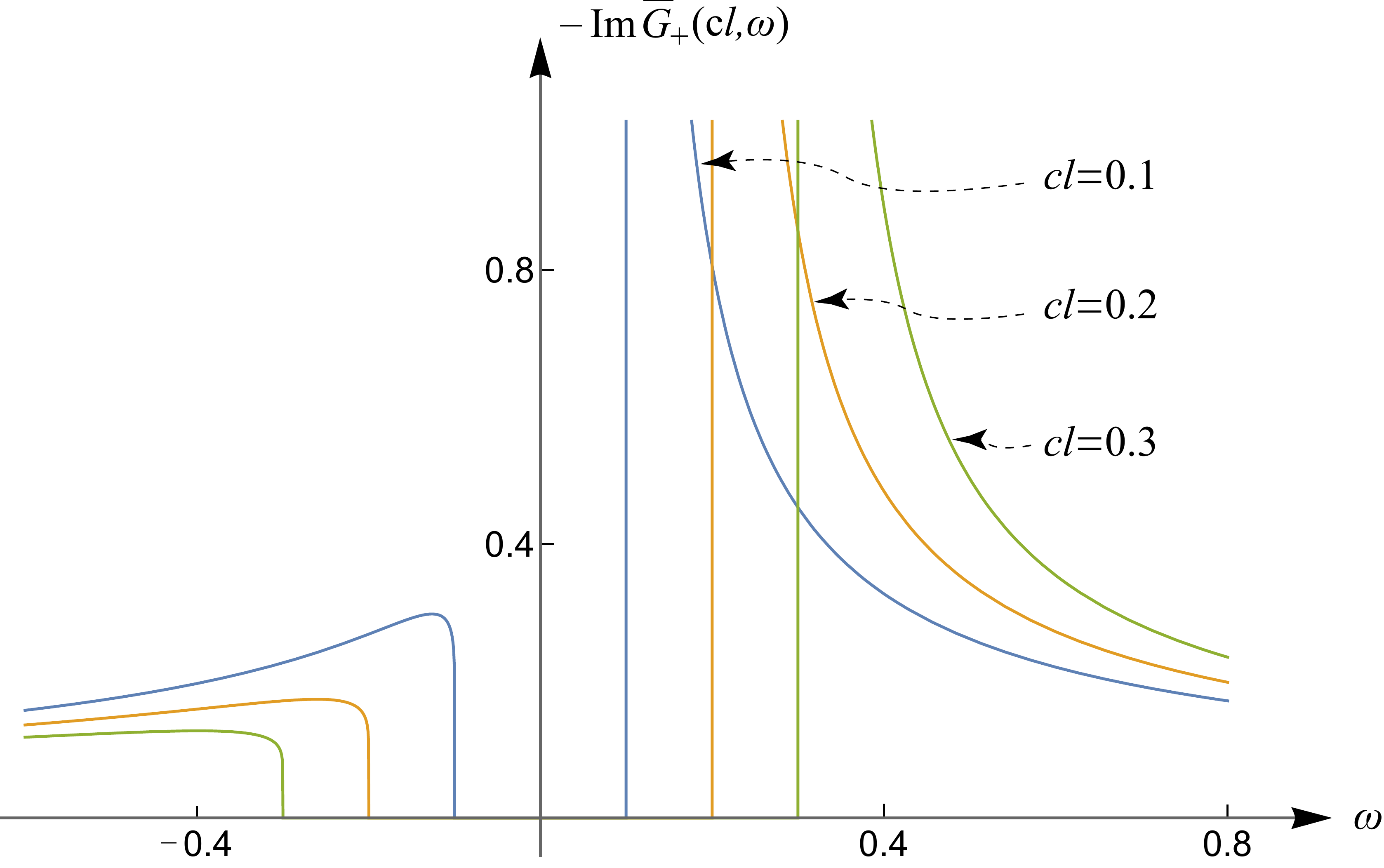}
	\caption{Fermionic spectral function $-{\rm Im}\overline{G}_{+}(l,\omega)$ as a function of the frequency $\omega$ for 
several $cl$. Eq.~(\ref{spectral}) is plotted for $cl=0.1$ (blue), $cl=0.2$ (purple), $cl=0.3$ (ocher). $\kappa=1$, 
$\alpha=0.1$, and the horizontal axis is $\omega$.}
	\label{sfig:2}
\end{figure}

By the analytic continuation $i\omega \rightarrow \omega+i\delta$, the imaginary part of 
the retarded Green's function (fermion spectral function) is obtained;
\begin{align}
\overline{G}^R_{\sigma}(l,\omega)  &\equiv 
\overline{G}_{\sigma}(l,\omega) \big|_{i\omega \rightarrow \omega+i\delta} \nonumber \\ 
{\rm Im}\!\ \overline{G}^{R}_{\sigma}(l,\omega) &= 
\frac{{\cal P}}{\omega-\sigma cl} (\overline{A}\!\ \overline{D}-\overline{B}\!\ 
\overline{C})  \nonumber \\
& \ \ \ - \pi \delta(\omega-\sigma cl) 
(\overline{A} \!\ \overline{C}+\overline{B} \!\ \overline{D}), \label{spectral}
\end{align} 
where 
\begin{eqnarray}
\left\{\begin{array}{l}
\overline{A} \equiv \frac{(1-2\alpha A)^2 + \alpha(1-2\alpha A) + (2\alpha B)^2}{(1+\alpha-
2\alpha A)^2+(2\alpha B)^2}, \\ 
\overline{B} \equiv \frac{2\alpha^2 B}{(1+\alpha-2\alpha A)^2+(2\alpha B)^2}, \\
\overline{C}  \equiv \frac{{\rm Cos}\big[\frac{1}{2}{\rm Arg}(1-2\alpha A- i 2\alpha B)\big]}{[(1-2\alpha A)^2 + (2\alpha B)^2]^{\frac{1}{4}}}, \\ 
\overline{D}  \equiv - \frac{{\rm Sin}\big[\frac{1}{2}{\rm Arg} (1-2\alpha A- 
i 2\alpha B)\big]}{[(1-2\alpha A)^2 + (2\alpha B)^2]^{\frac{1}{4}}}, \\
 \alpha \equiv \frac{g^2}{8\pi^2 c^3}, \ \ A \equiv \log \big[\frac{\sqrt{|(cl)^2 - \omega^2|}}{\kappa}\big], \\
B \equiv \frac{\pi}{2} \big[\theta(-\omega-|cl|) - \theta(\omega-|cl|)\big]. \\
\end{array}\right. \label{ABCD}
\end{eqnarray}
Note that for low-energy fermions, $|(cl)^2 -\omega^2| \ll \kappa^2$ and $A$ is 
negative definite. Thus, $\overline{A}$ and $\overline{C}$ are positive definite. $B$, $\overline{B}$ and 
$\overline{D}$ is positive (negative) definite for $\omega<0$ ($\omega>0$).  
Especially, when $\omega \rightarrow \pm cl\pm \eta$ with positive $\eta \ll \kappa$, 
$A$ becomes negatively large and show a logarithmic divergence,  
\begin{align}
\lim_{\omega \rightarrow |cl| \pm \eta}  A = \frac{1}{2} \log \eta + \cdots.  
\end{align}  
Because of the logarithmic divergence, the quasi-particle weight in the spectral function is zero. 
Namely, in the limit of $\eta=0$ with $\omega^2 = (cl)^2\pm 2(cl)\eta$, 
\begin{align}
& \lim_{\eta \rightarrow 0}\overline{A} = 1, \ \  \lim_{\eta \rightarrow 0}\overline{B} = \frac{B}{2|A|^2}, \nonumber \\
& \lim_{\eta \rightarrow 0}\overline{C} = \frac{1}{\sqrt{2\alpha |A|}}, \ \  \lim_{\eta \rightarrow 0}\overline{D} 
= \frac{1}{\sqrt{2\alpha |A|}}\frac{B}{2|A|}, \nonumber \\
&\lim_{\eta \rightarrow 0} (\overline{A} \!\ \overline{C}+ \overline{B} \!\ \overline{D})  = 
\frac{1}{\sqrt{2\alpha} |A|^{\frac{1}{2}}} 
+ \cdots \rightarrow 0, \label{limit1} \\
&\lim_{\eta \rightarrow 0} (\overline{A} \!\ \overline{D} - \overline{B} \!\ \overline{C})  
= \frac{B}{2\sqrt{2\alpha} |A|^{\frac{3}{2}}} + \cdots \rightarrow 0. \label{limit2}
\end{align}
The spectral function is comprised only of continuum spectrum, that appears in $\omega>|cl|$ 
(Fig.~\ref{sfig:2}, Fig.~\ref{fig:3}). From Eqs.~(\ref{spectral},\ref{limit2}), the 
continuum spectrum of ${\rm Im}\overline{G}^R_{\sigma}$ 
show a weak divergence at $\omega = \sigma cl$;
\begin{eqnarray}
- {\rm Im}\overline{G}^{R}_{\sigma}(l,\omega) = 
\left\{\begin{array}{cc} \sqrt{\frac{1}{\alpha}} \frac{\pi}{2} 
\frac{1}{\eta} \frac{1}{[|\log \eta|]^{\frac{3}{2}}} & (\omega = \sigma cl + \eta)  \\
0 & (-\sigma cl < \omega < \sigma cl) \\
\sqrt{\frac{1}{\alpha}} \frac{\pi}{4\sigma cl} \frac{1}{[|\log \eta|]^{\frac{3}{2}}} &  (\omega = - \sigma cl - \eta) \\ 
\end{array}\right. \label{asy-a1}
\end{eqnarray}
for $\sigma cl>0$ 
and 
\begin{eqnarray}
- {\rm Im}\overline{G}^{R}_{\sigma }(l,\omega) = \left\{\begin{array}{cc} 
\sqrt{\frac{1}{\alpha}} \frac{\pi}{(-4\sigma cl)} \frac{1}{[|\log \eta|]^{\frac{3}{2}}} 
 & (\omega = - \sigma cl + \eta)  \\
0 & (\sigma cl < \omega < -\sigma cl) \\
\sqrt{\frac{1}{\alpha}} \frac{\pi}{2} 
\frac{1}{\eta} \frac{1}{[|\log \eta|]^{\frac{3}{2}}} &  (\omega = \sigma cl - \eta) \\ 
\end{array}\right. \label{asy-a2}
\end{eqnarray}
for $\sigma cl<0$ with $\sigma =\pm$. Note that the asymptotic behaviours of 
the fermion spectral function near the Fermi edge come from a factor of 
$\overline{g}/g$ in Eq.~(\ref{sol4}); 
Eqs.~(\ref{asy-a1},\ref{asy-a2}) hold true even if 
$\overline{\cal G}^{-1}_{\sigma}(\theta;\overline{g},\kappa)$ in Eq.~(\ref{sol3}) is replaced 
by the free Green's function; $-\sigma \kappa e^{-i\sigma \theta}$. 

When the temperature is finite, the quasi-particle spectral weight at $\omega=\sigma cl$ 
becomes finite in ${\rm Im}\!\ \overline{G}^{R}$ and it is scaled by $1/\sqrt{|\log T|}$;
\begin{align}
- {\rm Im} \overline{G}^{R}_{\sigma}(l,\omega,T\ne 0) = 
\frac{\pi}{\sqrt{2\alpha}} \frac{1}{\sqrt{|\log T|}} \!\ \delta(\omega - \sigma cl) + \cdots. 
\label{finite-T-spectral-weight-0}
\end{align}
To see this, %, one can go back to Eqs.~(\ref{sol1},\ref{sol4},\ref{final-G}) and 
note first that the vanishing 
quasi-particle spectral weight at $T=0$ comes from the vanishing 
$\overline{g}/g$ in Eq.~(\ref{final-G}) in the limit of 
small $p^2  \equiv (cl)^2-\omega^2 -2i\omega \delta$. Namely at $T=0$, one can 
renormalize infinitely many times, until infinitely small $p$ could be scaled up to the RG scale $\kappa$ 
on the renormalization. The infinitely many-times renormalization makes $\overline{g}/g$ 
to be zero; $t \rightarrow -\infty$ in Eq.~(\ref{gbar}). 
At the finite temperature, however, the temperature as well as the small $p$ 
is also scaled up to a larger value upon the renormalization. 
Thus, the renormalization must be terminated, either when the renormalized temperature reaches a certain 
high temperature scale $T_0$ or when the renormalzed $p$ could reach the RG scale. This consideration 
naturally lets us replace $\overline{g}/g$ in Eq.~(\ref{final-G}) 
by the following expression at $p=0$ and at $T \ne 0$, 
\begin{align}
\bigg(\frac{\overline{g}}{g}\bigg)\Big|_{p=0,T\ne 0} &= \sqrt{\frac{1}{1-2\alpha t}}\!\ \Big|_{t=\log (\frac{T}{T_0})} \nonumber \\ 
&\simeq \frac{1}{\sqrt{2\alpha}} \frac{1}{\sqrt{|\log T|}} + \cdots.  
\label{finite-T-spectral-weight}
\end{align}
Putting this back to Eq.~(\ref{final-G}) gives Eq.~(\ref{finite-T-spectral-weight-0}).

\subsection{density of states around the QCP}
\begin{figure}[t]
	\centering
	\includegraphics[width=0.95\linewidth]{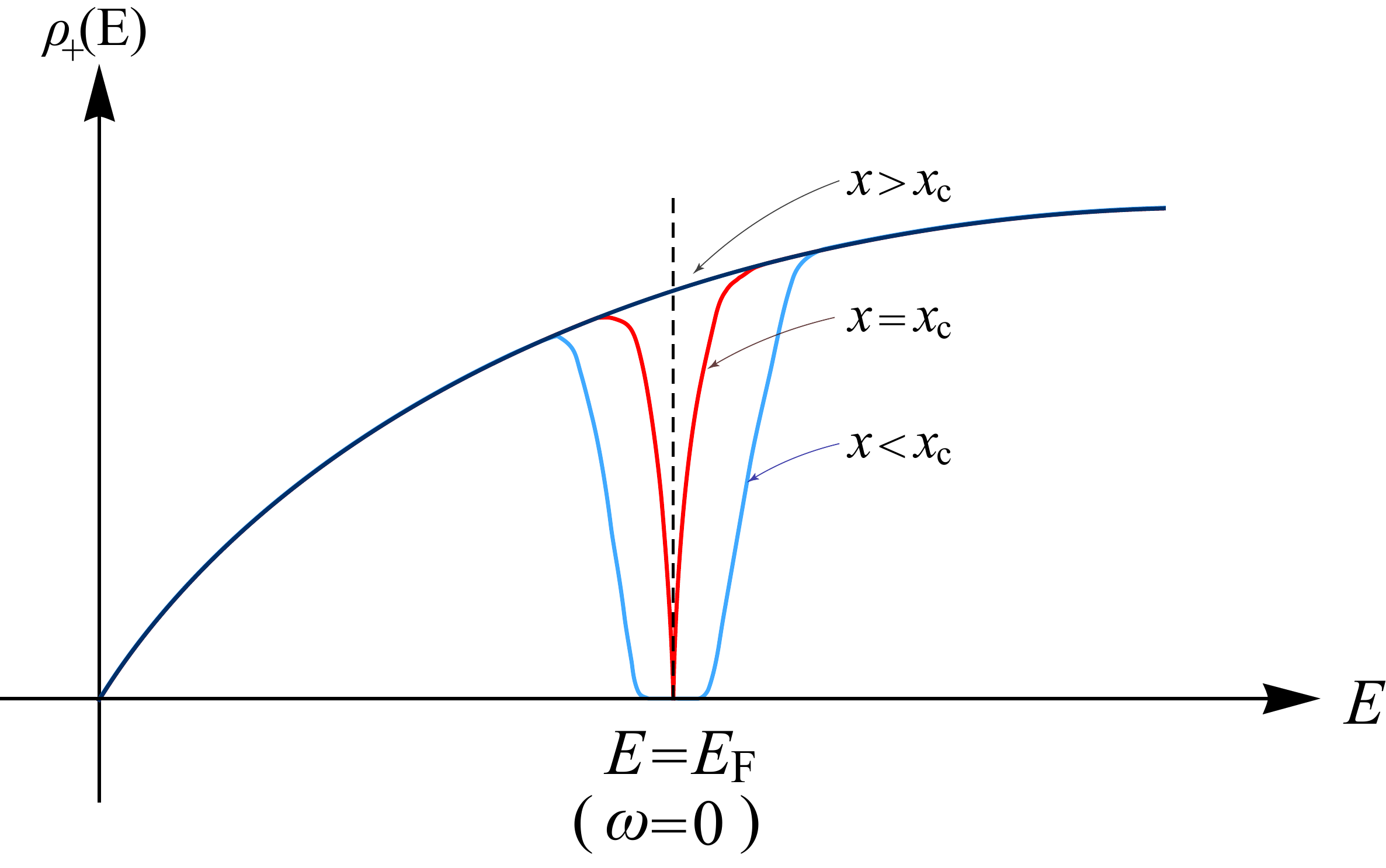}
	\caption{Schematic picture of the density of states (DOS) for electron-type energy band near the 
quantum critical point. For $x>x_c$, the DOS is finite at $E=E_F$ (FL phase), while for $x<x_c$, 
the DOS has a finite band gap (band insulator phase). At the QCP ($x=x_c$), the DOS shows 
a pseudo-gap behaviour; $\rho(E)\propto 1/|\log|[E-E_F]||^{\frac{1}{2}}$.}
	\label{sfig:4}
\end{figure}

The density of states (DOS) is given by the integral of the fermion spectral function 
over the momentum ${\bm k}$;
\begin{align}
- \rho_+(\omega) &= \frac{1}{\pi}  
\int \frac{d^3{\bm k}}{(2\pi)^3}  {\rm  Im} \overline{G}^R_{+} (l,\omega) \nonumber \\
&= \frac{k^2_F}{c\pi^2} 
\int^{\omega}_{-\omega} {\rm Im} \overline{G}^R_{+} (l,\omega) \frac{d(cl)}{2\pi}. \label{def-DOS}   
\end{align}
The subscript `$+$' refers to the DOS for the electron-type band. The following argument 
holds true in the same way for the hole-type band. Thus, we focus on the DOS of only 
the electron-type band. For small $\omega >0$ 
($\omega=0$ corresponds to the Fermi level), the integral over the one-dimensional momentum $l$ 
is dominated by the singular spectral weight near $cl=\omega$. 
One can see this by dividing the integral into two regions;
\begin{align}
&\int^{\omega}_{-\omega} {\rm Im} \overline{G}^R_{+} (l,\omega) \frac{d(cl)}{2\pi} \nonumber \\
& = \int^{\omega}_{s\omega} {\rm Im}\overline{G}^R_{+} (l,\omega) \frac{d(cl)}{2\pi} + 
\int^{s\omega}_{-\omega} {\rm Im} \overline{G}^R_{+} (l,\omega) \frac{d(cl)}{2\pi}.  \label{divide0}
\end{align}
Here `$s$' is a positive constant smaller than 1. Since the integrand increases monotonically in $cl$ for small 
positive $\omega$ (Fig.~\ref{fig:3}), the second term can be bounded from above;
\begin{align}
&- \int^{s\omega}_{-\omega} {\rm Im} \overline{G}^R_{+} (l,\omega) \frac{d(cl)}{2\pi} \nonumber \\
& \ \ < - \frac{(1+s)\omega}{2\pi} 
{\rm Im} \overline{G}^R_{+} (cl=s\omega,\omega) = {\cal O}\bigg(\frac{1}{|\log \omega|^{\frac{3}{2}}}\bigg).  
\label{divide}
\end{align} 
Here we set $cl=s\omega$ in Eqs.~(\ref{spectral},\ref{ABCD}) and assume that $\omega$ is sufficiently small;
\begin{align}
&\lim_{\omega \rightarrow 0} |A| = |\log \omega| + \cdots, \  
\lim_{\omega \rightarrow 0}\overline{A} = 1, \ \ \lim_{\omega \rightarrow 0}\overline{B} = -\frac{\pi}{4|A|^2}, \ \  \nonumber \\
&
 \lim_{\omega \rightarrow 0}\overline{C} = \frac{1}{\sqrt{2\alpha |A|}}, \ \ 
 \lim_{\omega \rightarrow 0}\overline{D} 
 = - \frac{1}{\sqrt{2\alpha |A|}}\frac{\pi}{4|A|}, \nonumber \\ 
& \lim_{\omega \rightarrow 0} (\overline{A} \!\ \overline{D} - \overline{B} \!\ \overline{C})  
= -\frac{\pi}{4\sqrt{2\alpha} |\log \omega |^{\frac{3}{2}}} + \cdots. \label{limit2a}
\end{align}
Meanwhile, for those $s$ close to 1 ($1-s$ is small but finite), the integrand of the first term in 
Eq.~(\ref{divide0}) can be approximated by the asymptotic form 
given in Eq.~(\ref{asy-a1});
\begin{align}
&\int^{\omega}_{s\omega} (-){\rm Im} \overline{G}^R_{+} (l,\omega) \frac{d(cl)}{2\pi} \nonumber \\
&\ \ \simeq \sqrt{\frac{1}{\alpha}}\frac{\pi}{2} \int^{(1-s)\omega}_{0} \frac{d\eta}{2\pi} 
\frac{1}{\eta} \frac{1}{|\log \eta|^{\frac{3}{2}}} = \sqrt{\frac{1}{\alpha}}\frac{1}{2} 
\frac{1}{|\log \omega|^{\frac{1}{2}}} + \cdots. \label{limit2b}
\end{align}
For small $\omega$, Eq.~(\ref{limit2b}) clearly dominates over Eq.~(\ref{divide}). Thus, the DOS 
takes the following scaling form;
\begin{align}
\rho_{+}(\omega) \propto \frac{k^2_F}{c \sqrt{\alpha}} \frac{1}{|\log \omega|^{\frac{1}{2}}} + \cdots. 
\label{dos-1}
\end{align}  
 Eq.~(\ref{dos-1}) shows how the multiple-band Fermi system acquires a band 
gap when the boson system undergoes the quantum phase transition (Fig.~\ref{sfig:4}). 
The result shows that the density of states at the Fermi level 
$E_F$ ($\omega=0$) is zero at the QCP (`pseudo-gap' behaviour). 

When the temperature $T$ is finite, $\omega$ in Eq.~(\ref{dos-1}) is replaced by $T$ and 
the DOS at the Fermi level becomes finite;
\begin{align}
\rho_{+}(\omega \simeq 0 ,T \ne 0)  
\propto \frac{k^2_F}{c \sqrt{\alpha}} \frac{1}{|\log T|^{\frac{1}{2}}} + \cdots. 
\label{dos-2}
\end{align}
One can see this straightforwardly, by taking the momentum integral of the spectral function 
at $T\ne 0$; Eq.~(\ref{finite-T-spectral-weight-0}). 
 
\section{conclusion} 
In this paper, we have studied a Fermi system with a pair of electron and hole Fermi surfaces (FS), 
whose coupling is mediated by a critical U(1) $\phi^4$ boson field. Due to the presence of the 
finite volume of the FS, low-energy kinematics of the fermionic excitation and that of the 
bosonic excitations are quite distinct from each other in the momentum space~\cite{yamamoto10}. 
The low-energy fermionic excitations are constrained around the FS, while the low-energy bosonic excitations 
are restricted around the zero-momentum point in the momentum space. Due to this geometrical distinction of 
the two low-energy kinematics, a naive momentum-shell Wilsonian  
renormalization group (RG) analysis could overlook important renormalization effects in the 
fermion-boson coupled model. 

To avoid this difficulty, we employ in this paper a 
field-theoretical renormalization group analysis, where the UV cutoff dependence 
(UV divergent behaviour) in the vertex functions are systematically included into 
renormalizations of the field operator amplitudes and physical quantities. Thereby, 
the renormalized vertex function is given as a solution of the homogeneous Callan-Symanzik (CS) equation. 
By solving the CS equation in favor for the two-point fermion's Green function, we obtain the low-energy 
asymptotic form of the fermion's spectral function and density of states (DOS) at the quantum critical 
point (QCP). The analysis reveals that at the QCP, the Fermi velocities of the two FSs are renormalized 
into a same critical velocity as a boson velocity, and the fermion's DOS shows a pseudo-gap 
behaviour with the logarithmic energy dependence; the DOS at the QCP vanishes 
toward $E=E_F$ with $1/|\log |E-E_F||^{\frac{1}{2}}$. Since $|E-E_F|$ and the temperature $T$ 
has the same scaling dimension around the $T=0$ QCP, the fermion's DOS at the Fermi level will be 
also scaled as $1/|\log T|^{\frac{1}{2}}$ in a finite-$T$ side of the QCP. We expect that this log-$T$ dependence 
of the DOS manifest itself in various physical quantities in the quantum critical regime.  
Such are the electronic contribution of the 
specific heat and magnetic susceptibility. Namely, being proportional to the DOS at 
the Fermi level, the electronic contribution of the specific heat and Pauli paramagnetic susceptibility 
must be also scaled by $1/|\log T|^{\frac{1}{2}}$ in the high-$T$ region of the QCP.

\section*{ACKNOWLEDGEMENTS}
The work was supported by the National Basic Research Programs 
of China (No. 2019YFA0308401) and the National Natural Science Foundation 
of China (Grant No. is 11674011 and 12074008). 

\appendix 
\section{Callan-Symanzik equation for the boson Green's function and boson's spectral function at 
QCP}
In this appendix, the CS equation for the two-point boson Green's function will be derived 
at the quantum critical point (QCP) and it is solved in favor for the boson Green's function. Out 
of the solution, the boson's spectral function is obtained at the QCP. 
As in Sec.~VA, let us begin with the relation between the bare and renormalized function;
\begin{align}
& Z^{-1}_{\phi}(m^2=0,c^2,\cdots,\lambda,\kappa,\Lambda) 
\overline{G}^{-1}_{\phi}(q;c^2,\cdots,\lambda,\kappa) \nonumber \\
& \ \ \hspace{1cm} = G^{-1}_{\phi}(q;m^2_0,c^2_0,\cdots,\lambda_0,\Lambda).  
\label{start-boson}
\end{align} 
The $\kappa$-derivative of Eq.~(\ref{kappa-deri}) leads to the following inhomogeneous equation,
\begin{align}
&\Big\{\frac{\partial}{\partial \ln \kappa} 
+ \beta_c \frac{\partial} {\partial c} + \sum_{\sigma=\pm} \beta_{v_{\sigma}} \frac{\partial}{\partial v_{\sigma}} 
+ \beta_{g} \frac{\partial}{\partial g}  + \beta_{\lambda} \frac{\partial}{\partial \lambda} \nonumber \\ 
&  - \gamma_{\phi} \Big\} \overline{G}^{-1}_{\phi}(q;m^2=0,c^2,\cdot,\lambda,\kappa)  
= \frac{\partial m^2_0}{\partial \ln \kappa} \frac{\partial G^{-1}_{\phi}} {\partial m^2_0}. \label{CS-1-b}
\end{align}
Note that in the leading order in $g^2$ and $\lambda$, the $\gamma$ function for the boson Green's function is zero; 
$\gamma_{\phi}={\cal O}(g^4,g^2 \lambda,\lambda^2)$. Meanwhile, the right hand side of Eq.~(\ref{CS-1-b}) has a 
finite leading-order contribution, because $\partial G^{-1}_{\phi}/\partial m^2_0 = 1 + {\cal O}(g^2,\lambda)$. Thus, 
the CS equation takes an inhomogeneous form,
\begin{align}
& \Big\{\frac{\partial}{\partial \ln \kappa} + \sum_{\sigma=\pm }\beta_{v_{\sigma}} \frac{\partial}{\partial v_{\sigma}}  
+ \beta_{g} \frac{\partial}{\partial g} \Big\} \overline{G}^{-1}_{\phi}(q,c^2,\cdot,\kappa)  \nonumber \\ 
& \ \  \  \ \hspace{0.5cm} =  -\frac{g^2 k^2_F}{\pi^2 (v_{+}+v_{-})} + {\cal O}(g^4,g^2\lambda,\lambda^2). 
\end{align} 
Here we also omitted $\partial_{\lambda}$ in the left hand side, which leads to terms on the order of 
${\cal O}(\lambda^2)$; $\beta_{\lambda} \propto {\cal O}(\lambda^2)$.  
In the following, this equation will be solved in favor for the two-point boson Green's function. 

With $v_{\pm}=c$, the CS equation for the two-point boson vertex function is given by
\begin{align}
\Big\{\frac{\partial}{\partial \ln \kappa} +
 \beta_{g}(g) \frac{\partial}{\partial g}\Big\} \overline{G}^{-1}_{\phi}(q;g,\kappa)  
= - \frac{g^2 k^2_F}{2\pi^2 c}.  \label{CS-3-b}
\end{align}
Henceforth, we recover the implicit $k_F$-dependences in $\overline{G}_{\phi}$. According 
to Eq.~(\ref{ex-g}), we will expand $\overline{G}^{-1}_{\phi}$ in the powers of $k^2_F$;
\begin{align}
\overline{G}^{-1}_{\phi}(q;g,\kappa) = &\overline{G}^{-1}_{\phi,(0)}(q;g,\kappa) + k^2_F 
\overline{G}^{-1}_{\phi,(1)}(q;g,\kappa) \nonumber \\
&+ k^4_F \overline{G}^{-1}_{\phi,(2)}(q;g,\kappa) + \cdots. \label{ex-G-b}  
\end{align}
The dimensional analysis dictates that $\overline{G}_{\phi,(n)}$ has a different scaling form 
for different $n$;
\begin{align}
&\overline{G}^{-1}_{\phi,(0)}(q;g,\kappa) = r^2 g^{-1}_{(0)}(r/\kappa,\theta;g), \nonumber \\
&\overline{G}^{-1}_{\phi,(1)}(q;g,\kappa) = g^{-1}_{(1)}(r/\kappa,\theta;g), 
\nonumber \\  
&\overline{G}^{-1}_{\phi,(2)}(q;g,\kappa) = r^{-2} g^{-1}_{(2)}(r/\kappa,\theta;g), \cdots \label{scaling-form}
\end{align}
with $q \equiv (\varepsilon,{\bm q})$, 
$c|{\bm q}| \equiv r\sin\theta$, $\varepsilon \equiv r\cos\theta$. Since $\beta_g$ 
has no $k_F$ dependence, the CS equation can be decomposed into equations at every order in $k^2_F$;
\begin{eqnarray}
\left\{\begin{array}{l} 
\Big\{\frac{\partial}{\partial \ln \kappa} +
 \beta_{g}(g) \frac{\partial}{\partial g}\Big\} \big(r^2 g^{-1}_{(0)}(r/\kappa,\theta;g)\big) = 0,  \\
\Big\{\frac{\partial}{\partial \ln \kappa} +
 \beta_{g}(g) \frac{\partial}{\partial g}\Big\} \big(g^{-1}_{(1)}(r/\kappa,\theta;g)\big) = - \frac{g^2}{2\pi^2 c}, \\
\Big\{\frac{\partial}{\partial \ln \kappa} +
 \beta_{g}(g) \frac{\partial}{\partial g}\Big\} \big(r^{-2} g^{-1}_{(2)}(r/\kappa,\theta;g)\big) = 0,  \\
\hspace{2.0cm} \cdots. \\ 
\end{array}\right. 
\end{eqnarray}
In terms of Eq.~(\ref{gbar}), the CS equation at every order in $k^2_F$ can be solved in the leading order 
in $g^2$. The solutions are 
\begin{eqnarray}
\left\{\begin{array}{l}
g^{-1}_{(0)}(r/\kappa,\theta;g) = h_{(0)}(\theta;\overline{g}), \\
g^{-1}_{(1)}(r/\kappa,\theta;g) = - \frac{\overline{g}}{2\pi^2 c} \big(-t  + h_1(\theta;\overline{g})\big), \\
g^{-1}_{(2)}(r/\kappa,\theta;g) = h_{(2)}(\theta;\overline{g}), \\
\hspace{2cm} \cdots, \\
\end{array}\right. 
\end{eqnarray} 
with $t \equiv \ln(r/\kappa)$. $\overline{g}$ in the right hand side is 
given as a function of $t$ and $g$ by Eq.~(\ref{gbar}). 
This gives the two-point boson vertex function as follows, 
\begin{align}
&\overline{G}^{-1}_{\phi}(r,\theta;g,\kappa) = r^2  h_{(0)}(\theta;\overline{g}) 
\nonumber \\
&\ \ 
- \frac{\overline{g} k^2_F}{2\pi^2 c} \big(-t  + h_1(\theta;\overline{g})\big) 
+  r^{-2} k^4_F h_{(2)}(\theta;\overline{g}) + \cdots. \label{G-phi-2}
\end{align}
\begin{figure}[t]
	\centering
	\includegraphics[width=0.95\linewidth]{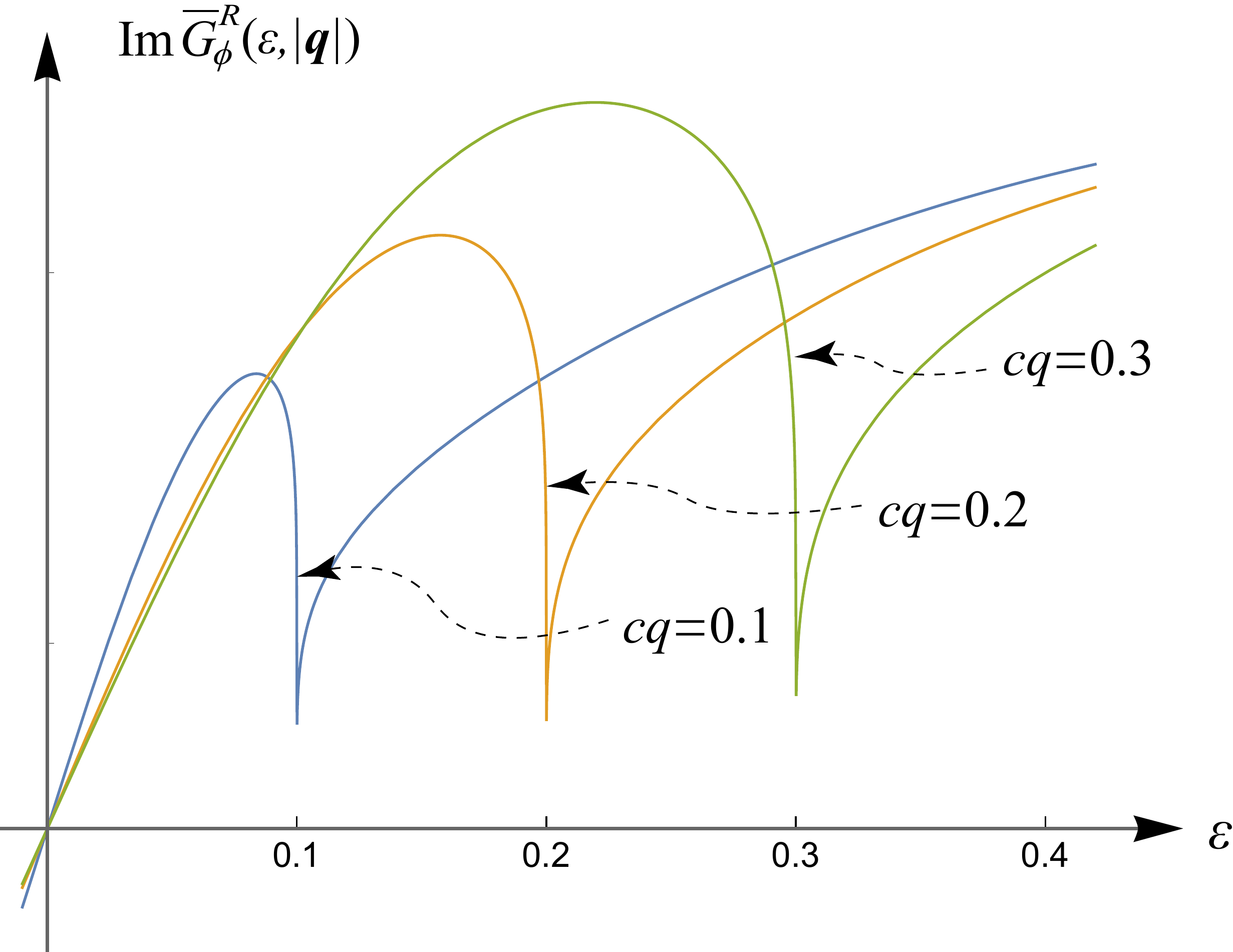}
	\caption{Bosonic spectral function ${\rm Im}\overline{G}_{\phi}(\varepsilon,|{\bm q}|)$ as a function of the frequency 
$\varepsilon$ for several $|{\bm q}|$. Eq.~(\ref{ImG-phi}) is plotted for $c|{\bm q}|=0.1$ (blue), $c|{\bm q}|=0.2$ (purple), 
$c|{\bm q}|=0.3$ (ocher). $\kappa=1$,  
$\alpha=0.1$, $\gamma=1$, and the horizontal axis is $\varepsilon$.}
	\label{sfig:3}
\end{figure}
According to Eqs.~(\ref{G-phi-2},\ref{gbar}), $\overline{G}_{\phi}(r,\theta;g,\kappa)$ in the IR limit 
($r \ll \kappa$) is determined by forms of $h_{(0)}(\theta;g)$, $h_{(1)}(\theta;g)$ and $h_2(\theta;g)$ for 
tiny $g$. To determine them for such small $g$, let us take $r=\kappa$  in Eq.~(\ref{G-phi-2}) 
and expand its right hand side in small $\overline{g}(t=0;g)=g$. 
For small $g$, $\overline{G}_{\phi}(r=\kappa,\theta;g,\kappa)$ in the 
left hand side can be calculated by the second order perturbation as;
\begin{align}
\overline{G}^{-1}_{\phi}(r=\kappa,\theta;g,\kappa) = 
& \kappa^2 - \frac{g^2 k^2_F}{2\pi^2 c} 
\Big(1 - \frac{\theta}{\tan\theta}\Big) \nonumber \\
& + {\cal O}(g^4,g^2\lambda,\lambda^2).  \label{G-phi-2a}
\end{align} 
Here Eqs.~(\ref{gphi0},\ref{int2},\ref{36}) with $v_{\pm}=c$ are substituted 
into Eq.(\ref{1-loop-Gphi}). A comparison between Eq.~(\ref{G-phi-2a}) and Eq.~(\ref{G-phi-2}) at $r=\kappa$ 
determines $h_{(0)}(\theta;g)$, $h_{(1)}(\theta;g)$ and $h_2(\theta;g)$ for tiny $g$;
\begin{align}
&h_{(0)}(\theta;g) = 1, \ \  h_{(1)}(\theta;g) = 1 - \frac{\theta}{\tan\theta}, 
\nonumber  \\ 
&h_{(2)}(\theta;g) = 0, \cdots.  
\end{align} 
This leads to an expected result for the two-point boson vertex function in the IR limit ($r\ll \kappa$);
\begin{align}
\overline{G}^{-1}_{\phi}(q;g,\kappa) = r^2 - \frac{\overline{g}^2 k^2_F}{2\pi^2 c}  
\Big(-t  + 1 - \frac{\theta}{\tan\theta}\Big), \label{G-phi-3}
\end{align}
with $q=(\varepsilon,{\bm q})$, $\varepsilon=r\cos\theta$, $c|{\bm q}|=r\sin\theta$, 
$t\equiv \ln(r/\kappa)$. $\overline{g}$ in the right hand side is given by $t$ and $g$ by Eq.~(\ref{gbar}). By 
comparing Eq.~(\ref{G-phi-3}) with Eqs.~(\ref{1-loop-Gphi},\ref{gphi0},\ref{int2},\ref{int3}), one can see that 
the $\Lambda^2_B$ term in the naive perturbation result is absorbed into the 
renormalization of the boson mass. ${\rm Log}\!\ \overline{\Lambda}_F$ and the Yukawa 
coupling $g$ in Eq.~(\ref{1-loop-Gphi}) 
are replaced by $\log \kappa$ and the IR-limit value of $g$, $\overline{g}$, respectively.   

By an analytic continuation $i\varepsilon \rightarrow \varepsilon + i\delta$, 
the imaginary part of the retarded boson Green's function 
(boson spectral function) is obtained as,
\begin{align}
\overline{G}^R_{\phi}(\varepsilon,{\bm q}) &\equiv \overline{G}_{\phi}(\varepsilon,{\bm q})
\Big|_{i\varepsilon \rightarrow \varepsilon+i\delta}, \nonumber \\
{\rm Im} \overline{G}^{R}_{\phi}(\varepsilon,{\bm q}) &= 
\frac{\gamma \overline{F}}{(-\varepsilon^2+c^2 {\bm q}^2 - \gamma \overline{E})^2 
+ \gamma^2 \overline{F}^2}, \label{ImG-phi}
\end{align}
where $\overline{E}$, $\overline{F}$ and $\gamma$ are given as follows,  
\begin{eqnarray}
\left\{\begin{array}{l}
\overline{E} = \frac{(1-2\alpha A)(1-A) + 2\alpha B^2 + 2\alpha B B^{\prime}}{(1-2\alpha A)^2 + (2\alpha B)^2}, \\  
\overline{F} = - \frac{(1-2\alpha A)B^{\prime} + (1-2\alpha) B}{(1-2\alpha A)^2 + (2\alpha B)^2}, \\
 \alpha \equiv \frac{g^2}{8\pi^2 c^3}, \ \gamma \equiv \frac{g^2 k^2_F}{2\pi^2 c}, \\
A \equiv \log \big[\frac{\sqrt{|(c|{\bm q}|)^2 - \varepsilon^2|}}{\kappa}\big],  \\  
B \equiv \frac{\pi}{2} \big[\theta(-\varepsilon-c|{\bm q}|) - \theta(\varepsilon-c|{\bm q}|)\big], \\ 
B^{\prime} \equiv -\frac{\varepsilon}{c|{\bm q}|} \frac{\pi}{2} 
\theta(c|{\bm q}|-|\varepsilon|).  \\
\end{array}\right. 
\end{eqnarray}
Note that the spectral weight is positive for $\varepsilon>0$ and negative for $\varepsilon<0$, 
for the low-energy boson, $|c^2 {\bm q}^2 - \varepsilon^2|\ll \kappa^2$, and for those Yukawa coupling smaller 
than a critical value ($\alpha<\frac{1}{2}$). Especially when 
$\varepsilon \rightarrow c|{\bm q}| + \eta$, `$A$' becomes negatively large and diverges logarithmically. 
Due to this logarithmic divergence, the spectral weight has a `valley' at $\varepsilon = c|{\bm q}|$ (Fig.~\ref{sfig:3}), 
around which $\overline{E}$ and $\overline{F}$ take the following asymptotic forms, 
\begin{align}
&\lim_{\varepsilon \rightarrow c|{\bm q}|} \overline{E} = \frac{1}{2\alpha}, \ \ \lim_{\varepsilon \rightarrow c|{\bm q}|-} 
\overline{F} = \frac{B^{\prime}}{2\alpha A} = \frac{\pi}{2} \frac{1}{\alpha |\log |\eta||}, \nonumber \\ 
& \lim_{\varepsilon \rightarrow c|{\bm q}|+} 
\overline{F} = -\frac{(1-2\alpha)B}{4\alpha^2 A^2} = \frac{\pi}{2} \frac{1-2\alpha}{\alpha^2 |\log |\eta||^2}.  
\end{align} 
The result shows that the delta-function peak at $\varepsilon=c|{\bm q}|$ for $g=0$ is wiped out 
completely at small but finite $g$,
and is replaced by the valley. The weight has broad peaks next to the valley. Toward the bottom of 
the valley, the spectral weight vanishes as
\begin{eqnarray}
{\rm Im}\overline{G}^{R}_{\phi}(\varepsilon,{\bm q}) = \left\{\begin{array}{cc} \frac{2\pi}{\gamma} 
\frac{\alpha}{|\log \eta|} & (\varepsilon = c|{\bm q}| - \eta),  \\
\frac{2\pi}{\gamma} \frac{1-2\alpha}{|\log \eta|^2} &  (\varepsilon = c|{\bm q}| + \eta). \\ 
\end{array}\right. \label{asy-boson2}
\end{eqnarray}

\bibliography{paper}

\end{document}